\documentclass[a4paper]{article}

\usepackage[a4paper,margin=2cm]{geometry}
\usepackage{amsmath}
\usepackage{graphicx}   
\usepackage[american]{babel}
\usepackage[utf8]{inputenc}
\usepackage[T1]{fontenc}
\usepackage{hyperref}
\usepackage{afterpage}
\usepackage{color}		
\usepackage{epsfig}
\usepackage{makecell}
\usepackage{subcaption}  
\usepackage[font={footnotesize}]{caption}

\def\hyp{{\hbox{-}}}

\begin{document}

\title{The Kinetic Basis of the Genetic Code}

\author{Hemachander Subramanian\\
\normalsize{Department of Physics}\\
\normalsize{National Institute of Technology, Durgapur, West Bengal, India} \\
\normalsize{hsubramanian.phy@nitdgp.ac.in} } 
\date{\today}
\maketitle

\begin{abstract}
The origin of the genetic code is an enigma, requiring explanations for the emergence of degeneracy, the organizational structure of the code, and the correspondence between codons and amino acids. Here, we use a recently-proposed asymmetric cooperativity model to show that these features arise from asymmetric cooperativity in codon–anticodon association process. We first provide mechanistic and evolutionary accounts of wobble-base-pair degeneracy, without providing any preferential treatment to the third position. The model, with four tunable parameters, reproduces the organizational structure of the genetic code, and provides theoretical codon-anticodon association times. Although ribosomal kinetic data are not used in parameter estimation, these theoretical association times correlate with experimentally determined ribosomal residence times ($r=0.73$), providing an independent empirical correspondence. The resulting timescales, extending over four decades, also exhibit correlations with diverse physicochemical properties of amino acids, at multiple timescales, motivating a nonequilibrium picture in which energy released during codon–anticodon association is transferred to its cognate amino acid to be dissipated into water. Finally, without incorporating information about aminoacyl-tRNA synthetases or further parameter adjustment, the model recovers the fundamental division of amino acids associated with the two aminoacyl-tRNA synthetase classes, with a combined segregation score of 0.92 ($ p\sim10^{-8}$). Together, these results suggest that wobble degeneracy, genetic-code organization, translation kinetics, amino-acid physicochemistry, and the two-class organization of aminoacyl-tRNA synthetases may reflect a common underlying kinetic constraint. We propose that non-equilibrium, asymmetrically cooperative kinetics, that possibly underlie DNA/RNA replication process, also provided a physical basis upon which the genetic code and its subsequent translation machinery evolved.
\end{abstract}

\section*{Synopsis}

The genetic code, a table that maps three-letter nucleic acid sequence families to various amino acids, is degenerate. In the section titled ``Wobble position degeneracy'', we probe both the evolutionary origin and the mechanistic underpinnings of this degeneracy, using a recently proposed kinetics-based model to simulate the association/dissociation kinetics of codon-anticodon base-pairing. We show, through clustering analysis of association times of codon-anticodon base-pairing, that the model reproduces the fourfold degeneracy of the wobble position, without providing any preferential treatment to the wobble position. This degeneracy is shown to be a consequence of directionally asymmetric kinetics that governs the unidirectional DNA/RNA templated strand construction.

In the section titled ``Quantitative reproduction of the codon table structure and its experimental validation'', we show that a simplified organizational structure of the genetic code can be quantitatively reproduced using the model introduced in the previous section that is designed to evaluate the kinetics of tRNA-mRNA association process, by extending the definition of asymmetric cooperativity to include sequence dependence. By organizational structure, we mean the dependence of the number of codons that code for a single amino acid on the class of nucleotides (strong/weak and purine/pyrimidine) present in the three positions of the codon sequence. Our model correctly clusters together the fourfold degenerate (synonymous) codons when C is in the second position, and twofold degenerate codons when the middle codon is A. It also reproduces the dependence of degeneracies of middle G and U codons on the strength of the first codon base-pair, with a strong first base mapping to fourfold degenerate, and weak first base to twofold degenerate codon assignments. This is achieved by treating the four model parameters as optimization variables, which are optimized to match the above simplified organizational structure of the genetic code. Clustering of the residence times of the codon-anticodon base-pairing process in the intermediate states leads to the genetic code structure reproduction. The optimized codon-anticodon association times are compared with the experimentally observed residence times of ribosomes on all $61$ codons. Remarkably, significant correlation $(\approx 0.73)$ between the model output and the residence times is found to be present mostly at the E site and at the edges of the ribosome, without any fitting involved. The model distinguishes the effects of two antibiotics on the ribosome residence times, discriminates between two ribosomal translational conformations, and is capable of identifying wobble base pairing. Translation kinetics therefore appear to be partly determined by the genetic code.

In the third section titled ``The non-equilibrium origins'', we identify correlations between, on one hand, the theoretically calculated residence times in the intermediate states of the codon-anticodon association process, and, on the other, some biophysical properties of individual amino acids, such as heat relaxation times, polarizabilities, softnesses, vibrational frequencies, dipole orientations, number of hydrogen bond acceptors, and side chain lengths. Assuming causation behind these correlations, we suggest that electronic energy transfer from the codon-anticodon association process to individual amino acids in aqueous medium as a possible mechanism for the emergence of the genetic code. These correlations suggest that the codon-amino acid assignments emerged, and are maintained, through matching of energy supply times and dissipation times, at the codon-anticodon and the amino acid ends respectively, at multiple timescales, from sub-femtoseconds to picoseconds. This impedance matching happens when the system is evolutionarily selected to minimize the energy transfer time, possibly to protect the mRNA from damage and prolong its lifetime. The identified correlations broadly reproduce the multilevel organizational structure and the amino acid assignments of the genetic code. This also provides a natural explanation for the well-observed correlation between hydropathy of amino acids and their genetic code assignments. The proposed non-equilibrium origin of the genetic code is consistent with the early explanations for the emergence and utility of life, where living systems are described as agents of efficient transduction of free energy, which preserves the agents' persistence. 

Aminoacyl-tRNA synthetases catalyze the charging of tRNAs with their cognate amino acids. They form two distinct classes with little sequence, structural or evolutionary homology between the classes, and each class is responsible for charging ten amino acids. The evolutionary rationale for the existence of two classes of Aminoacyl-tRNA synthetases is not known. In the last section titled ``Emergence of two classes of aminoacyl tRNA synthetases'', we provide a plausible answer to this question using the Asymmetric Cooperativity model to reproduce, validate and explain the evolutionary basis for the emergence of the genetic code. We show that the theoretically calculated codon-anticodon association kinetics can be used to correctly segregate into two sets seventeen of the twenty amino acids, aminoacylated by the two classes of tRNA synthetases. This result is emergent, since no information about synthetases entered our calculations. We use the experimentally calculated ribosomal residence times obtained from ribosomal footprinting (riboseq) experiments to verify this classification. We argue that the evolutionary rationale for the introduction of a second class of synthetases is to circumvent the effects of certain degeneracies observed in the codon-anticodon association kinetics, which possibly increased misacylation probabilities. We suggest that these degeneracies are removed by utilizing the distinct picosecond dissipation dynamics resulting from the amino acids being attached to the 2'-OH and 3'-OH ends of the ribose ring of the tRNA accceptor stem adenosine. Using the mapping between codon-anticodon association dynamics and amino acid dissipation dynamics demonstrated in the previous section, we argue that the bootstrapping problem of the evolution of translation, wherein proteins are needed for construction of proteins, can be resolved. 

\clearpage			
\begin{center}
    {\Large\bfseries I: Wobble position degeneracy} \\[0.5cm]
\end{center}

\section*{Introduction}
Living systems employ two types of heteropolymers: DNA, to store information, and proteins, to catalyze reactions of energy production and cellular maintenance. The information stored in the DNA heteropolymer sequence is translated, mediated by RNA, into an amino acid heteropolymer sequence, that folds to create proteins. The genetic code enables translation of the information encoded in nucleic acid sequences into amino acid sequences, mapping all possible three-letter nucleic acid codes (codons) to twenty or so amino acids (and start and stop signals). This mapping is degenerate, since RNA employs four types of nucleic acids (A,U,G and C in short), any of which can occupy the three positions of a codon, resulting in $64$ possible codons, all of which are mapped to twenty standard amino acids, leading to a many-to-one mapping.  

The genetic code, as the map is referred to, was initially suspected to be a frozen accident \cite{Crick1968,koonin2017frozen}, but was later shown to have significant symmetries and biochemical regularities that pointed to early evolutionary optimization \cite{taylor1989code, knight1999selection, wong1975co, yarus2005origins, copley2005mechanism, Koonin2009}. It was shown to have a high degree of error tolerance, with mutations generally substituting biochemically similar amino acids, thereby avoiding significant alterations in the protein structure \cite{woese1965evolution,Freeland1998Genetic}. Although the evolutionary origin of the genetic code is generally accepted, the precise nature of the evolutionary forces that led to the code is still unclear and is vigorously debated. 

Significant organizational structure is observed in the genetic code. All codons that have a 'C' in their second position are four-fold degenerate, with the codon families $ACN$, $GCN$, $CCN$, and $UCN$ each coding for an amino acid, where N denotes any of the four nucleotides A, U, G and C. All codons with an 'A' in the second position are doubly degenerate; for e.g, $GAY$ and $GAR$ code for an amino acid each, where Y denotes a pyrimidine (C or U) and R denotes a purine (A or G). If the middle codon position is a 'G' or 'U', then the first position nucleotide determines the degree of degeneracy: If the first position is a strong bond (G or C), then codons bearing 'G' or 'U' in their second position are four-fold degenerate, and if the first position is a weak bond (A or U), then they are doubly-degenerate, ignoring for the moment the modifications introduced by start and stop codons. For example, $GUN$ family codes for an amino acid (Valine), while $AUR$ and $AUY$ families code for an amino acid each (Leucine and Phenylalanine, respectively), in the vertebrate mitochondrial genetic code, the most symmetric of all the genetic codes \cite{Delarue2007}.

The presence of structure such as above suggests that the codon degeneracies are physicochemically determined. As explained above, the degeneracies are primarily due to the third codon position, called the wobble position, so named by Francis Crick to indicate possible, weak, non-Watson-Crick base-pairing between nucleotides at the first position of anti-codon on the tRNA and the third position of codon on the mRNA. Base-pairing between nucleotides such as G and U are allowed in the third codon position, due to the structural leniency shown by the ribosome for hydrogen bonding specifically at this position, proving Crick right \cite{murphy2004structure}.  However, this simple, structure-based explanation for the genetic code degeneracy raises deeper questions: Why degeneracy is imposed on the third codon position, and not on first or second positions?  What dictates the structured dependence of the degree of degeneracy (four-fold or two-fold) on the first and second codon positions? More importantly, it raises a chicken-and-egg problem: Which evolved earlier? Did the evolutionarily early genetic code degeneracy, evolved by as-yet unknown mechanism, force the ribosome to adopt a structure that implemented the degeneracy, or the evolution of the structure of the ribosome led to the genetic code degeneracy? Answering these questions require a deeper probing of the evolutionary forces that created the genetic code.


In this article, we intend to answer the following questions: I) What is the biophysical and evolutionary origin of wobble position degeneracy?  II) What is the biophysical origin of the organizational structure of the entire genetic code mentioned above? Can we quantitatively recreate the organizational structure using a biophysical model?  III) How are the codon-amino acid assignments made, and what is the biophysical mechanism that underlies the mapping?  IV) Why do two families of tRNA synthetases exist?.  As we show below, all the above questions can be answered by simulating the association/dissociation \textit{kinetics} of codon-anticodon base-pairs, with the inclusion of a new property which we call ``asymmetric cooperativity'', derived by maximizing the replication potential of primordial self-replicating heteropolymers \cite{subramanian2018evolutionary,subramanian2020evolutionary}. In this first section, we provide a brief introduction to the asymmetric cooperativity model and tackle the first question of wobble position degeneracy.

\subsection*{Symmetry breaking solves a primordial conundrum}
In order to answer the above questions, one must travel far back in time, even before the Last Universal Common Ancestor, to the time of free-floating, non-enzymatically self-replicating heteropolymers competing for monomers and energy supply, to identify the evolutionary forces that shaped the genetic code. In the primordial scenario, self-replicating heteropolymers that were proficient in both acquiring monomers and retaining them on the template long enough to enable intra-strand covalent bond formation between monomers, became evolutionarily successful. However, these two simple requirements for efficient self-replication cannot be simultaneously satisfied, since faster monomer-template bonding requires lower kinetic barrier, whereas, longer lifetimes of monomer-template bonds require higher kinetic barrier. Since self-replication should happen around the melting point of the monomer-template bond in order to enable both the construction of the daughter strand and its separation from the template after the construction is completed (as in Polymerase Chain Reaction), modification of thermodynamic contributions to the barrier will be counterproductive.

A solution that solves this conundrum is to introduce \textit{Asymmetric Cooperativity} between neighboring monomer-template hydrogen bonds of the heteropolymer \cite{subramanian2018evolutionary,subramanian2020evolutionary}. Asymmetric cooperativity is defined as unequal, non-reciprocal kinetic influence of a template-monomer bond on its left and right neighbors. For example, if a template-monomer hydrogen bond \textit{raises} the kinetic barrier for formation/dissociation of its neighboring template-monomer bond to the left, and \textit{lowers} the barrier for its right neighboring template-monomer bond, the self-replication process is asymmetrically cooperative. This would lead to unidirectional construction of daughter strands, from left to right, on heteropolymer templates. It is important to remember that the barrier raising/lowering happens only in the presence of a hydrogen bond, which makes the barriers vary dynamically with the construction of new bonds, as the daughter strand is constructed. Asymmetrically cooperative, self-replicating molecular species would have succeeded in the evolutionary competition with symmetrically cooperative/non-cooperative self-replicating molecular species, due to their ability to quickly acquire monomer substrates and energy supply, and due to their ability to retain monomers for covalent bond formation. This would have led evolution to choose asymmetric monomers that incorporated asymmetric kinetics in their interactions with their neighbors. The above evolution-based reasoning for the emergence of directional symmetry-broken self-replicators has been introduced in our earlier paper \cite{subramanian2018evolutionary}, where, we quantitatively demonstrated the evolutionary advantage of asymmetrically cooperative self-replicators over non-cooperative / symmetrically cooperative self-replicators. The above-proposed asymmetric cooperativity is of \textit{kinetic origin} and goes beyond the thermodynamic directional asymmetry, well-known from the experimentally determined nearest-neighbor free energies, enthalpies and entropies of pairs of base-pairs \cite{santalucia1998}. We would like to emphasize that this article assumes that asymmetric cooperativity is present in DNA and RNA, and it has to be accepted as the \textit{central premise} from which our results will be shown to follow.

\begin{figure}
\begin{center}
\includegraphics[width=0.9\textwidth]{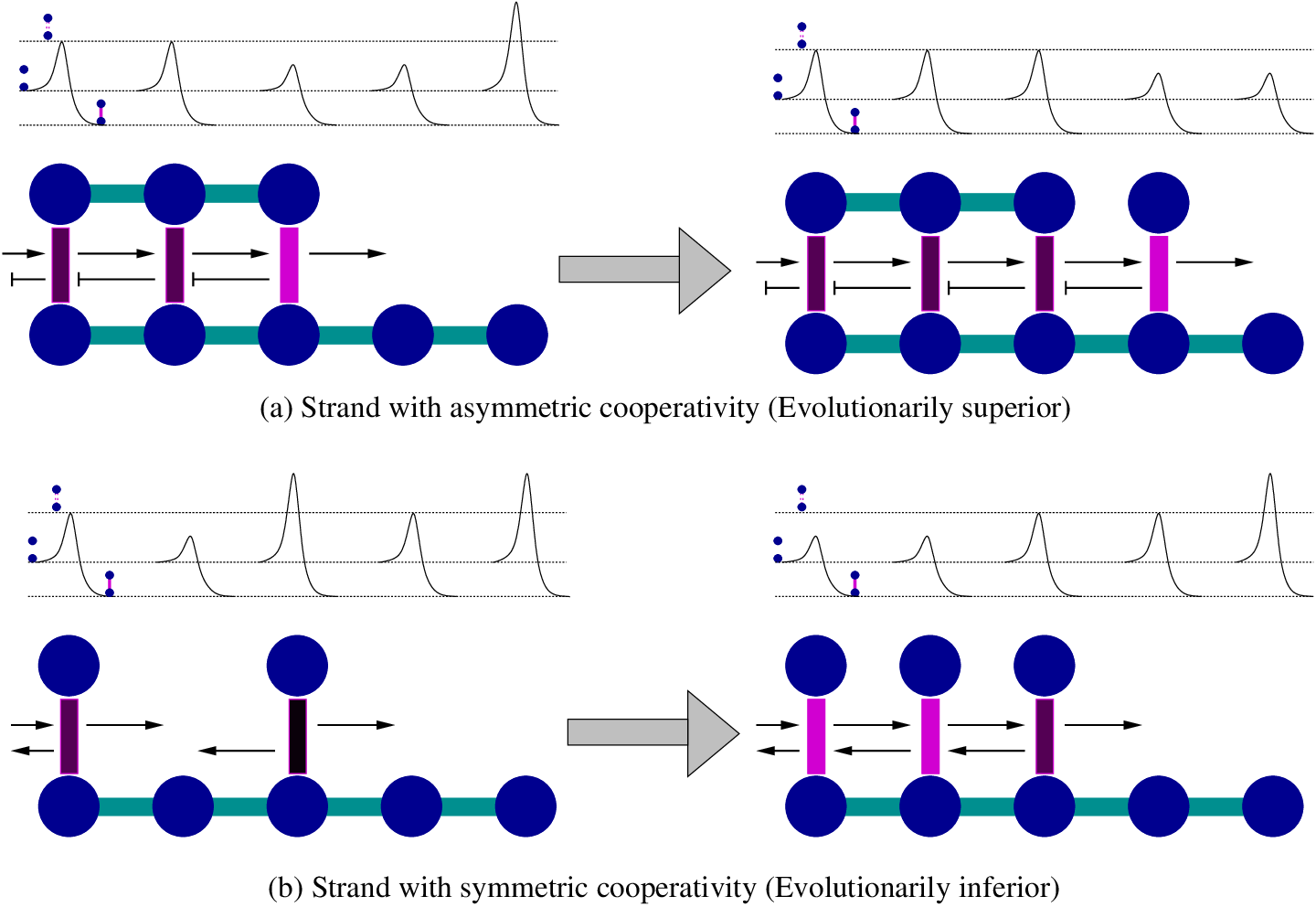}
\caption{Illustration of replica strand construction process of a circular autocatalytic polymer in the presence of (a) asymmetric and (b) symmetric nearest-neighbor hydrogen bond cooperativities (figure and caption reprinted from \cite{subramanian2018evolutionary} with permission from Elsevier). The circles represent monomers and the thick vertical lines connecting a pair represent the inter-strand hydrogen bonds. Horizontal lines connecting the monomers represent covalent bonds between them. The color of the hydrogen bonds represent the height of the kinetic barrier separating bonded and unbonded configurations, higher the barrier, darker the color. Hydrogen bonding energy diagram is shown above for both cases of cooperativity. The bottom line of the three lines in the energy diagram corresponds to the energy of the bonded configuration, and the middle line corresponds to that of the unbonded configuration. Only a section of the circular strand is shown here for convenience. (a) In a strand with asymmetric cooperativity, the catalytic influence on and from a hydrogen bond's left and right neighbors are unequal, simplified here to be catalysis and inhibition from left and right, respectively. Catalysis from a neighboring hydrogen bond is denoted by an arrow from the neighboring bond, and inhibition, with a bar-headed arrow. Such non-reciprocal, asymmetric catalytic influence among a pair of neighboring hydrogen bonds leads to low kinetic barrier for hydrogen bond formation/dissociation near the growth front, enabling faster monomer utilization for the replica strand construction. This also prevents the pre-existing hydrogen bonds behind the growth front from dissociation and provides them longer lifetime to facilitate covalent bond formation between the replica strand monomers. This asymmetrically cooperative hydrogen bonding behavior solves the conundrum of simultaneously requiring both low and high kinetic barriers for hydrogen bond formation/dissociation stemming from the conflicting needs of high monomer utilization and high covalent bond formation probability, but renders the strands directional. (b) In strands with symmetric cooperativity, two hydrogen bonds mutually catalyze each other's formation/dissociation, by reducing the kinetic barriers symmetrically. In this case, already formed hydrogen bonds in regions away from the growth front (second bond from left) have smaller kinetic barriers than the bonds at or near the growth front (fourth bond from left). This reduces the template's ability to attract monomers for replica strand elongation, and also its ability to keep the monomers bonded long enough to facilitate covalent bond formation, thereby reducing its replicative potential relative to templates with asymmetric cooperativity. Please note that the scales of energies in (a) and (b) are not the same.}
\label{symasym}
\end{center}
\end{figure}

In this article, through examination of \textit{association times of codon-anticodon base-pairing}, we show that the above-mentioned directional symmetry-breaking, that led to unidirectional replication of DNA and RNA, also leads to wobble position degeneracy.

\section*{The Asymmetric Cooperativity Model}
The mathematical model incorporating asymmetric cooperativity is constructed to enable us to evaluate association times, i.e., the time taken for base-pair formation, between codons of mRNA and anticodons of tRNA. We will show that the \textit{association times} of codons with their cognate anticodons with different wobble ($3'$-most) position nucleotides (for e.g., codons of the GGN family) are very similar, and hence cluster together.

In order to evaluate the association or zipping times of codon-anticodon base-pairing, we utilize a simple Continuous Time Markov Chain model \cite{van1992stochastic} with $8$ possible states, describing bonded/unbonded nature of each of the three codon positions $(2^3=8)$. The rates of transition between the states is given by the bonding / dissociation rates of the hydrogen bond that makes up a single base-pair between codon and its anticodon. For example, the state $000$ denotes a configuration where all the three base-pairing bonds are broken, and the state $111$ denoting all the three base-pairs between codon and anticodon formed and intact. The usual $5'-3'$ convention for writing DNA/RNA sequences is followed for the states as well. The transition rates between states, say $000$ to $001$, is $r(AU)$ or $r(GC)$, depending on the type of the nucleotide occupying the third codon position. The reverse transition from $001$ to $000$ is denoted $s(AU)$ or $s(GC)$. Here, $r(AU)$, $r(GC)$, $s(AU)$ and $s(GC)$, the thermodynamic parameters for base-pair formation and dissociation, are functions of nucleotide concentrations, temperature and pH, and are widely available in the literature.

Asymmetric cooperativity, defined as unequal and non-reciprocal kinetic influence on neighboring hydrogen bonds, is implemented in this mathematical scheme by modifying the rates of base-pair formation / dissociation, depending on whether its left neighboring hydrogen bond, right neighboring hydrogen bond, or both, are present. The bonding and dissociation rates, $\mathcal{R}$, of base-pairing interaction between a nucleotide in the codon and a nucleotide in the anticodon are 

\begin{equation}
\begin{aligned}
& \mathcal{R}\left ( 000 \rightarrow 010 \right ) = r_{bp}, \\
& \mathcal{R}\left ( 010 \rightarrow 011 \right ) = \alpha_R r_{bp}, \\
& \mathcal{R}\left ( 010 \rightarrow 110 \right ) = \alpha_L r_{bp} \qquad \text{and}\\
& \mathcal{R}\left ( 101 \rightarrow 111 \right ) = \alpha_R \alpha_L r_{bp}. \\
\label{ratesbonding}
\end{aligned}
\end{equation}

\begin{equation}
\begin{aligned}
& \mathcal{R}\left ( 010 \rightarrow 000 \right ) = s_{bp}, \\
& \mathcal{R}\left ( 011 \rightarrow 010 \right ) = \alpha_R s_{bp}, \\
& \mathcal{R}\left ( 110 \rightarrow 010 \right ) = \alpha_L s_{bp} \qquad \text{and}\\
& \mathcal{R}\left ( 111 \rightarrow 101 \right ) = \alpha_R \alpha_L s_{bp}. \\
\label{ratesunbonding}
\end{aligned}
\end{equation}

The equations \ref{ratesbonding} and \ref{ratesunbonding} are self-explanatory. If a base-pair's right neighboring basepair is present and left neighbor is absent, its association / dissociation rate is modified by $\alpha_R$, whereas, if the basepair's left neighbor is present and the right neighbor absent, its rate is modified by $\alpha_L$. If both neighbors are present, then the rate is modified by the product $\alpha_R \alpha_L$. 
Here $r_{bp} (s_{bp})$ refers to hydrogen bond formation (dissociation) rate between two nucleotides on the codon and anticodon, where $bp$ can be $AU$ or $GC$. These four thermodynamic parameters are widely available in literature for different temperatures, nucleotide concentrations and pH values. The hydrogen bond formation rates of $AU$ and $GC$, governed mostly by entropy, are assumed to be equal, and is taken to be $k_{on} = 0.1\: k_0$, where $k_0$ is of the order of $10^7 \: Mol^{-1}Sec^{-1}$ \cite{howorka2001kinetics}. Since we are modeling the association process between mRNA and tRNA within the tight environs of the ribosome, the hydrogen bond formation rate between the nucleotides, $k_0$, would be much higher than rates evaluated using simple reaction schemes. For now, we will keep the order of magnitude of $k_0$ \textit{undetermined}, but will later in the series find that it is of the order of inverse of picoseconds. The hydrogen bond dissociation rates of $AU$ and $GC$ are enthalpically determined, hence are distinct, and are calculated using the thermodynamic parameters governing RNA duplex formation, provided in \cite{xia1998thermodynamic}. Using the enthalpy and entropy changes for $AU$ and $GC$ base-pair formation, averaged over all nearest neighbor base-pairs, we calculate the dissociation rates to be $0.89 \: k_{on}$  and $0.42 \: k_{on}$ for $AU$ and $GC$ respectively. These values correspond to a temperature of $95^{\circ} C$, where the bonding and dissociation rates of \textit{single} base-pairs are nearly of the same magnitude, leading to better discrimination of association times of different codon-anticodon complexes in our model. On the other hand, the parameters $\alpha_L$ and $\alpha_R$ are entirely kinetic in nature, since they reduce or enhance both the formation and dissociation rates equally. We assume $\alpha_L = 1000$ and $\alpha_R = 0.1$, in the absence, within our knowledge, of studies that characterize polymerase-free, \textit{asymmetric}, strand construction kinetics of DNA/RNA single molecules, at single bond resolution. These parameter values can be motivated \textit{a posteriori} by pointing to the observation that the rate of zippering of double stranded RNA is about two to three orders of magnitude larger than the rate of the initial bond formation \cite{nagel2006structural}. It has to emphasized that the results below are almost completely insensitive to the precise values used for our parameters, with the exception of the constraint of kinetic asymmetry, that is,  $\alpha_L >> \alpha_R$.

Once the rate matrix is written down according to the equations \ref{ratesbonding} and \ref{ratesunbonding}, the mean first passage time (MFPT) or the hitting time to reach a final state $k$ can be calculated using the standard method of solving the following set of equations for non-negative $t_j$'s: 
\begin{equation}
\sum_j \mathcal{R}_{ij} t_j  = -1, \qquad i \ne k.
\label{hittingtimeeq}
\end{equation} 
 
In the above equation, $t_j$ represents the expected amount of time the Markov chain spends in state $j$ during its sojourn towards the absorbing state $k$, beginning at $i$. In our case, the final state $k=111$, the state corresponding to fully base-paired codon-anticodon complex, and the starting state $i=000$, the state of fully unzipped codon-anticodon complex. The time taken for the chain, beginning at state $000$, to reach $111$, is the time taken for the three basepairs between a codon and anticodon to form, henceforth simply called ``the association time''. This is simply the sum of the amounts of time spent by the chain at all the intermediate states between the starting state $000$ and the absorbing state $111$. That is, the association time $T_A=t_{000}+t_{001}+t_{010}+t_{011}+t_{100}+t_{101}+t_{110}$. Apart from using the association time $T_A$, we also use the seven-element residence time vector, $\mathbf{T_{res}} = \left [t_{000}, t_{001}, t_{010}, t_{011}, t_{100}, t_{101}, t_{110} \right ]$, to do cluster analysis. We will see in the next section that the residence time vector $\bf{T_{res}}$ can be different for different codons, even when their association times $T_A$ are similar. 

Obviously, the association time depends on the nucleotide composition of the codon, due to different dissociation rates for AU and GC basepairs. Due to the presence of asymmetric cooperativity, the association time also depends on the order of the nucleotides in the sequence. Below, we will elaborate on the dependence of association time on the codon sequence, and show that sequences with different third position nucleotides, grouped together as a codon family, have nearly the same association times. 

\section*{Results}
\subsubsection*{Clustering into four groups}
\begin{figure}
\begin{center}
\includegraphics[trim=20pt 20pt 20pt 20pt,width=0.95\textwidth]{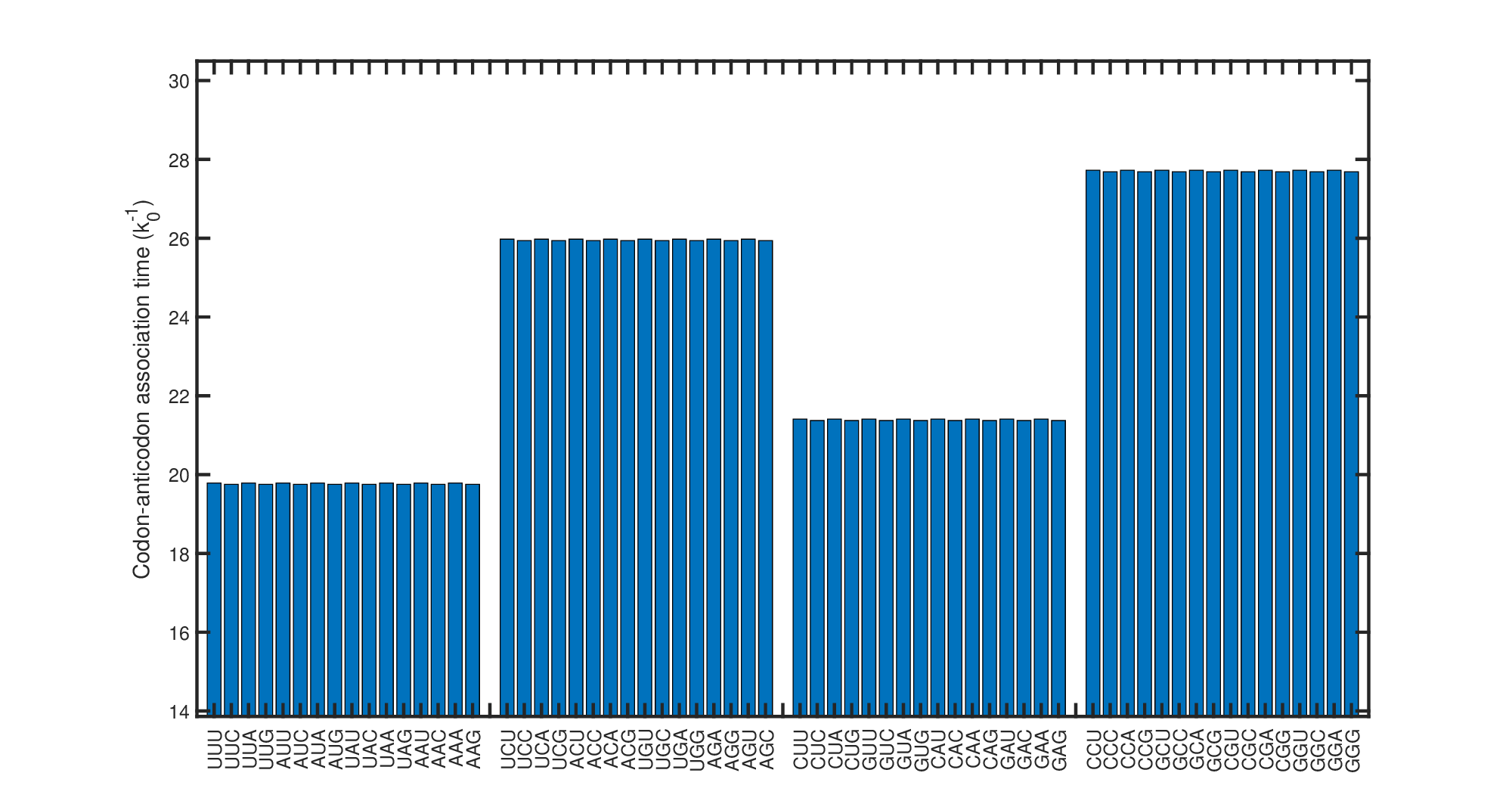}
\caption{Dependence of the codon-anticodon association time on the codon identity. Including only sequence-independent asymmetric cooperativity in the calculations, we observe that the association times are clustered into four groups with 16 elements in each. The codon groups are $WWN$, $WSN$, $SWN$, and $SSN$, where S and W denote strong and weak nucleotides respectively, and N for any nucleotide. The association times exhibit the interesting property of independence from the third codon identity, as observed in many degeneracies in the genetic code. It has to be emphasized that the third codon position is not treated in any special way in the model, and this independence or degeneracy emerges naturally from the asymmetric kinetic interactions between the base pairs of the codon-anticodon system.}
\label{clusters}
\end{center}
\end{figure}

Fig. \ref{clusters} shows the association time, the time it takes for the anticodon on the tRNA to fully bind with the codon on the mRNA, as a function of the codon identity. As can be seen, the association times form four distinct clusters, with each cluster containing four four-fold degenerate codons. The 64-fold degeneracy is split into four 16-fold nearly-degenerate clusters, $SSN$, $SWN$, $WSN$ and $WWN$, where $S$ denotes strong nucleotides $G$ and $C$, $W$, the weak nucleotides $A$ and $U$, and $N$ denoting any of the four nucleotides. It is important to emphasize here that the wobble position clustering is entirely due to the \textit{kinetic interactions} between the base-pairs of codon-anticodon duplex. Removing the asymmetrically cooperative interactions from the model destroys this clustering. The results show that the dependence of the association time on the third codon position is much weaker than the other two positions, with the association time varying very little when the third codon nucleotide is switched. This clustering is nearly insensitive to the values used for the above kinetic parameters, as long as the kinetic constraint $\alpha_L >> \alpha_R$ is satisfied, and is therefore a robust phenomenon.

\begin{figure}
\begin{center}
\includegraphics[trim=20pt 20pt 20pt 20pt,width=0.95\textwidth]{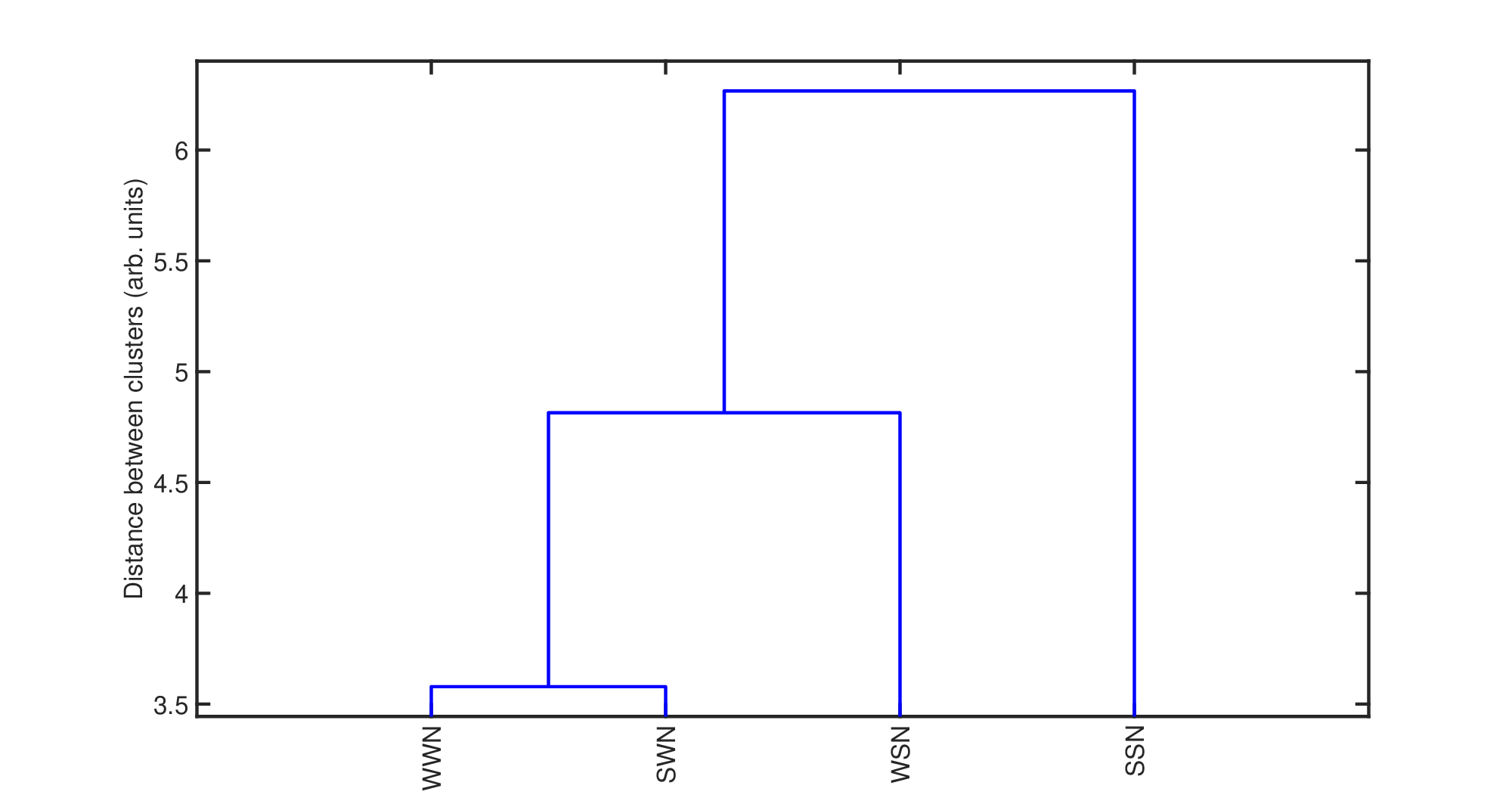}
\caption{Cluster analysis of the seven-element residence time vectors of all the 64 codon-anticodon systems. The residence time vectors cluster into four groups, $WWN$, $WSN$, $SWN$, and $SSN$, reproducing the third codon degeneracy observed in the genetic code. Note that the thermodynamically similar $SWN$ and $WSN$ clusters are kinetically segregated. This kind of clustering is maintained as long as the two free parameters of the model obey the constraint $\alpha_L >> \alpha_R$.}
\label{dendrogram}
\end{center}
\end{figure}

Apart from clustering using the association time $T_A$, we have clustered the codons using a 7-element residence time vector $\bf{T_{res}}$, distinct for different codons, with each element of the vector quantifying the average amount of time spent by the Markov chain at each of the seven states, beginning at $000$, before hitting the final fully zipped state $111$. The 7-element residence time vector is simply the first row of the inverse of the rate matrix $\mathcal{R}$ in eq.\ref{hittingtimeeq}. Clustering based on euclidean distance between the 7-element residence time vector for all codons results in the dendrogram of fig. \ref{dendrogram}. The four clusters in fig. \ref{dendrogram} are $WWN$, $WSN$, $SWN$ and $SSN$. The clustering segregates codons according to their first and second codon position, depending on their weak or strong nature of base-pairing, while being insensitive to the nature of the third codon position. Note that codons with distinct thermodynamic free energy differences during codon-anticodon association process, such as $GGA$ and $GGC$, are clustered together. On the other hand, the clusters $WSN$ and $SWN$ are segregated, despite having the same thermodynamic free energies for bonding/dissociation (we have not included stacking interactions between base-pairs in the model, for simplicity), due to the differences in the S and W base-pair lifetimes magnified and made position-dependent by the asymmetric kinetic interactions. Again, note that codons with similar thermodynamic free energy difference, such as $UGA$ and $GUA$, are discriminated in terms of the association time. This demonstrates that the thermodynamics of codon-anticodon association process cannot be the central determinant of the genetic code organizational structure. 


\subsubsection*{The mechanism behind wobble position degeneracy}
\begin{figure}[h!]
    \centering
    \begin{subfigure}[b]{0.45\textwidth}
        \includegraphics[width=\linewidth]{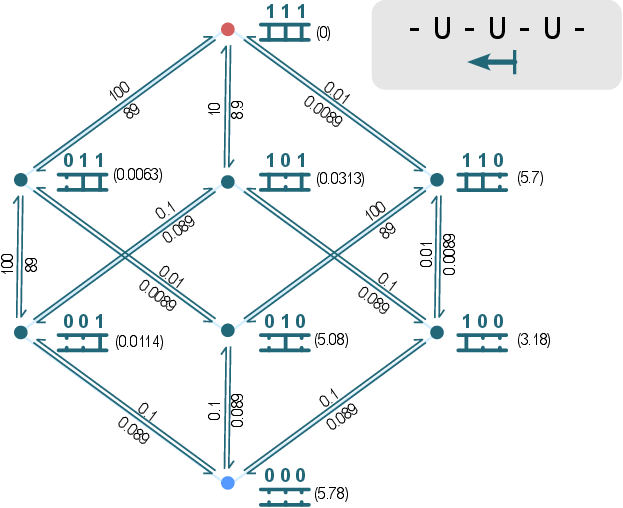}
        \caption{State transition graph for $UUU-AAA$ association}
        \label{states_UUU}
    \end{subfigure}
    \hspace{0.05\textwidth}  
    \begin{subfigure}[b]{0.45\textwidth}
        \includegraphics[width=\linewidth]{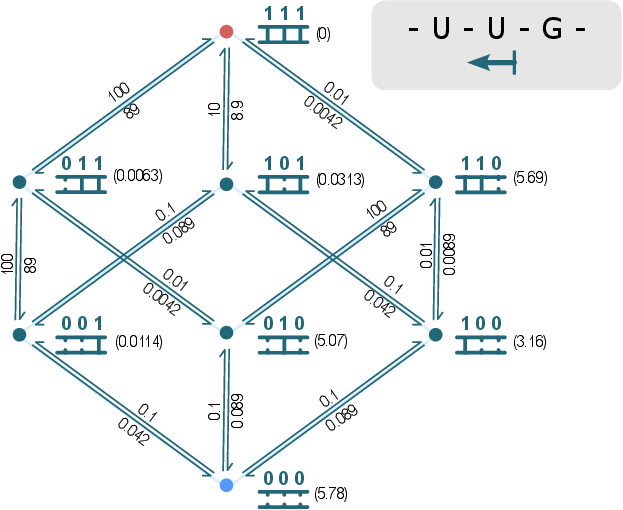}
        \caption{State transition graph for $UUG-CAA$ association}
        \label{states_UUG}
    \end{subfigure}
    \caption{Illustration of clustering of association times of codon-anticodon system with different third codon nucleotides in the genetic code. The state transition graphs of  two codon-anticodon systems that differ at the third codon position, $5'-UUU-3'/5'-AAA-3'$ and $5'-UUG-3'/5'-CAA-3'$ , are shown. The nodes of the graphs above denote states, represented by three binary digits, each of which denote the bonding state of the three  nucleotides of the codon and the anticodon. $000$ represent codon-anticodon system where none of the bonds exist between the three nucleotides of the codon and the anticodon, whereas, $111$ represent the state where the bonds between the three nucleotides of the codon and anticodon all exist. The edges of the graph represent transitions between the states, both forward and reverse. The two numbers near each edge are transition rates between the two states connected by the edge, with the larger number corresponding to the forward transition, a bond formation event, and the smaller one, the reverse transition rate. The residence time, the amount of time the Markov chain spends at any given state during its sojourn towards the absorbing state $111$, beginning at $000$, is given inside brackets. It can be observed that the residence times at the three intermediate states $001$, $011$ and $101$, where the third base pair exists between the codon and anticodon, are small, because of strong left asymmetric cooperativity that catalyzes the bond formation to the left of any bond present, resulting in fast state transitions from these states. The residence times are also nearly the same for both the $5'-UUU-3'/5'-AAA-3'$ and $5'-UUG-3'/5'-CAA-3'$ systems, because the probability for reverse transition from these low-residence time states is very low, and therefore, the discrimination becomes negligible. The residence times of the other three intermediate states lacking the third codon base pair, $010$, $100$ and $110$, are larger, because of the high barrier to bond formation to the right side, due to low $\alpha_R$. Their residence times also slightly differ between the $5'-UUU-3'/5'-AAA-3'$ and $5'-UUG-3'/5'-CAA-3'$ systems due to a small thermodynamically discriminatory back-flow from the $001$, $011$ and $101$ states. This similarity between the residence times of the $5'-UUU-3'/5'-AAA-3'$ and $5'-UUG-3'/5'-CAA-3'$  codon-anticodon systems lead to wobble base pair degeneracy.}
    \label{UUN_cluster}
\end{figure}

The mechanistic reason for the emergence of the wobble position or third codon position degeneracy can be explained using the state transition diagrams in figs. \ref{states_UUU} and \ref{states_UUG}, using $5'-UUU-3'/5'-AAA-3'$ and $5'-UUG-3'/5'-CAA-3'$ codon-anticodon systems as examples. The two figures show all the eight states that the Markov chain visits during the $5'-UUU-3'/5'-AAA-3'$ and $5'-UUG-3'/5'-CAA-3'$ codon-anticodon association processes respectively, the residence times of the chain at each of the eight states, and the transition rates between all pairs of connected states. The residence time at the absorbing state $111$ is undefined, since, once the Markov chain is absorbed, it never leaves the state, which is an assumption of the model. In both the figures, the residence times at the states $001$, $011$ and $101$, whose third/wobble positions are occupied by a base-pair, are much smaller compared to the residence times of the other states, whose third positions are all unoccupied. The reason for this significant difference in the residence times between ``wobble position occupied'' and ``wobble position unoccupied'' states is the presence of a very high rate transition from each of the three ``wobble position occupied'' states to the other states, i.e., from $001$ to $011$,  $011$ to $111$, and to a lesser extent, from $101$ to $111$, leading to a large out-flux, and therefore, smaller residence times. These high transition rates arise due to the presence of a large left asymmetric cooperativity parameter $\alpha_L (>> \alpha_R)$, which helps the wobble position base-pair catalyze its left neighboring base-pair formation, by reducing its kinetic barrier for formation/dissociation. Since the clustering of 64 codons into four groups involve calculating the euclidean distances between the seven-element residence time vectors between all possible pairs of codons, and since the residence times at the ``wobble position occupied'' states are considerably smaller than that of the ``wobble position unoccupied'' states, the euclidean distances are insensitive to the residence times of the ``wobble position occupied'' states, and is therefore insensitive to the identity of the wobble position nucleotide in the ``wobble position occupied'' states as well. This results in the clustering of codons with different wobble position nucleotides. Note that we have not provided any preferential treatment to the wobble position in our model.

The mechanism behind wobble base pair degeneracy can also be explained another way. The probability flux flowing through the $001$, $011$ and $101$ states are much higher than through the other ``wobble position unoccupied'' states, due to high transition rates, for the same reason above ($\alpha_L >> \alpha_R$). Therefore, the association process between a codon and an anticodon follows, with high probability, the path $000  \rightarrow 001 \rightarrow 011 \rightarrow 111$, which is a unidirectional zippering of anticodon beginning from the 3'-end of the codon and ending at its 5'-end. The reverse transition process is comparatively less probable, because of our initial choice of parameters that make bonding more favorable than dissociation. However, it is during the \textit{reverse transitions (dissociations)} that the discrimination between strong and weak base-pairs are possible, since the association rates for both these types of base-pairs are the same (Methods). The high \textit{non-discriminatory, forward} flux of probability through the $000  \rightarrow 001 \rightarrow 011 \rightarrow 111$ path, and the relatively low \textit{discriminatory, reverse} flux towards other ``wobble position unoccupied'' states from this high-probability path results in small differences in the state residence times between the $UUU$ and $UUG$ codons. The high forward flux through the ``wobble position occupied'' $000  \rightarrow 001 \rightarrow 011 \rightarrow 111$ path results in low residence times at the involved states, and these residence times are relatively insensitive to the identity of the nucleotide at the wobble position, as can be seen from the two figures \ref{states_UUU} and \ref{states_UUG}. This is because there is very little discrimination during the rapid forward transitions, i.e., base pair formations. Counter-intuitively, the small association time difference between the $5'-UUU-3'/5'-AAA-3'$ and $5'-UUG-3'/5'-CAA-3'$  codon-anticodon association processes comes from the residence times of the ``wobble position unoccupied'' states. This is due to the discriminatory ``back flow'' of flux from the states in the above high-probability flux path towards the ``wobble position unoccupied'' states. Since the residence times of the states in the high-probability flux path are low, the back-flow flux from these states towards the ``wobble position unoccupied'' states are also low, resulting in small difference between the residence times of these latter states between the association processes of $5'-UUU-3'/5'-AAA-3'$ and $5'-UUG-3'/5'-CAA-3'$  codon-anticodon systems. This explains the association time clustering of codons with different wobble position nucleotides.

`

\subsubsection*{Mechanism behind $WSN$ and $SWN$ segregation}
\begin{figure}[h!]
    \centering
    \begin{subfigure}[b]{0.45\textwidth}
        \includegraphics[width=\linewidth]{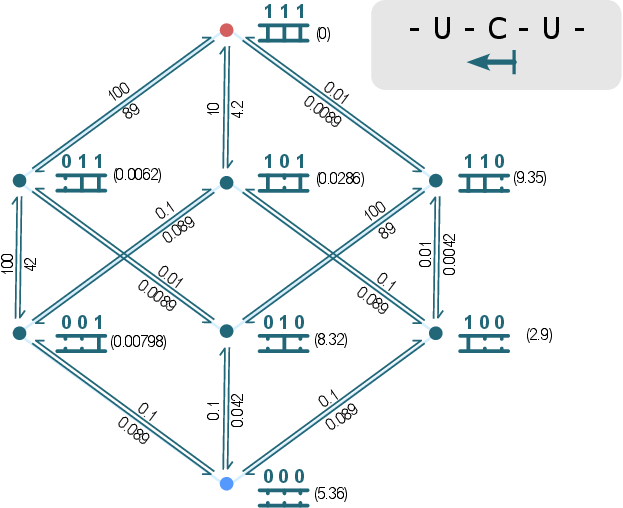}
        \caption{State transition graph for $UCU-AGA$ association}
        \label{states_UCU}
    \end{subfigure}
    \hspace{0.05\textwidth}  
    \begin{subfigure}[b]{0.45\textwidth}
        \includegraphics[width=\linewidth]{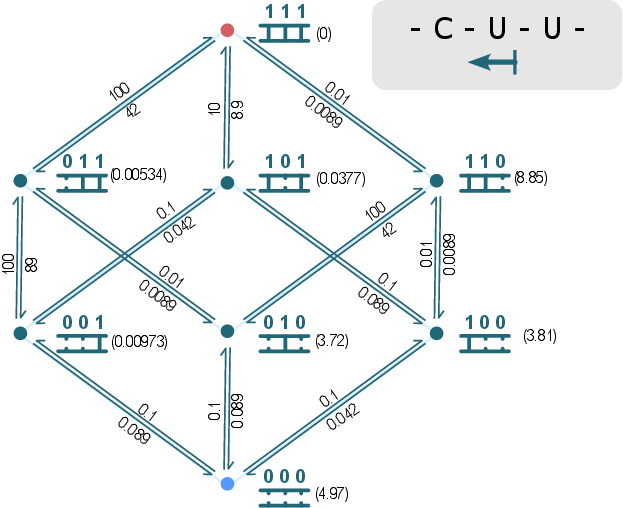}
        \caption{State transition graph for $CUU-AAG$ association}
        \label{states_CUU}
    \end{subfigure}
    \caption{Explanation for distinct residence times and association times for the thermodynamically identical $SWN$ and $WSN$ clusters, exemplified by $CUU$ and $UCU$ codons. Observe the difference in the residence time of the state $010$ in the two graphs above. This difference is due to the influx of probability from the $110$ state, due to difference in the transition rates from $110$ to $010$ between the two graphs. The higher residence times at $010$ and $110$ for the $WSN$ cluster, compared to the $SWN$ cluster, is because of the lowered barrier for the first base pair formation/dissociation, due to the presence of a stable, strong base pair at the second position, and the simultaneous lack of outflow from both these states, due to low transition rates towards states involving bond formation. The Markov chain therefore gets trapped in these two states, for $WSN$ codons, leading to higher residence times. }
    \label{SWN_WSN}
\end{figure}

The difference in the association times between the $WSN$ and $SWN$ clusters can similarly be explained by examining their state transition graphs (figs.  \ref{states_UCU} and \ref{states_CUU}). Here, we use the association process between $UCU$ and $CUU$ and their respective cognate anticodons for illustrating the $WSN-SWN$ segregation. Although the free energy change during the bond formation is the same for both the clusters within our model, their association times are distinct. This is again due to the asymmetrically cooperative interactions between the base pairs of the codon-anticodon system. A base pair in the codon-anticodon system catalyzes the formation/dissociation of its left neighboring base pair, while inhibiting the formation/dissociation of its right neighbor, according to the asymmetric cooperativity model. Since strong base pairs have longer lifetimes, their ability to catalyze the formation/dissociation of their left neighbors is larger. Therefore, in the $WSN$ cluster ($UCU$), the central strong base pair catalyzes the left neighboring weak base pair, further weakening it, leading to its instability. This results, counter-intuitively, in longer residence times in states that involve breaking or forming of this weak base pair, such as $010$ and $110$ in fig. \ref{states_UCU}, and leads to longer association times at these states for $WSN$ cluster. This is because of the rapid and continuous shuttling of the Markov chain between these two states, in the absence of edges from both these states with high transition rates towards the absorbing state. On the contrary, in the $SWN$ cluster, the strong base pair at the 5'-end inhibits the formation/dissociation of its right neighbor, resulting in a stabler codon-anticodon system, leading to shorter association times. The foregoing exemplifies the position-dependent effects of asymmetric cooperativity, rendering the kinetics of otherwise thermodynamically identical sequences drastically different.

\subsubsection*{Evolutionary reason for wobble position degeneracy}
The evolutionary reason, as opposed to the mechanistic reason above, for the emergence of wobble position degeneracy, is the evolutionary advantage gained by the unidirectionally replicating heteropolymers in both inducting and retaining monomers for self-replication, by decreasing kinetic barrier for new base-pair formation, and increasing or leaving untouched the barrier for already-formed base-pairs behind the growth front, thereby increasing the covalent bond formation probability. In our model above, a single hydrogen bond on a $5'-3'$ DNA/RNA strand catalyzes the formation of another hydrogen bond immediately to its left, due to high $\alpha_L$, which then catalyzes another to its left and so on, thereby creating a new daughter strand from the template's $3'$ end towards its $5'$ end. Thus, high $\alpha_L$, that leads to the unidirectional daughter strand construction on the DNA/RNA template strand, also results in similar residence times or association times for codons with different nucleotides at the third codon positions, leading to wobble position degeneracy.

\section*{Discussion}
The fourfold/twofold degeneracy of the wobble position underlies most of the organizational structure of the genetic code. We have shown that this degeneracy arises from the clustering of \textit{residence times} of the codon-anticodon base-pair in various states of bonding. Degenerate codons have similar residence times across all the seven available intermediate states of bonding, which minimizes the euclidean distance between them. The insensitivity of the residence time to the third codon nucleotide is due to the presence of left-asymmetric cooperativity in RNA zipping/unzipping kinetics. The presence of a base-pair at the third codon position catalyzes the formation of another base-pair to its left, which then catalyzes the formation of a base-pair to \textit{its} left, making the residence times of the states with a base-paired third codon position small, compared to residence times of states with non-base-paired third codon position. This results in the insensitivity of residence times to the identity of the nucleotide at the third codon position. We have shown that the evolutionary force that resulted in unidirectional daughter strand construction on DNA/RNA template strand has also led to the wobble position degeneracy.

A deeper question regarding the above results concerns the use of residence times to cluster codons into degenerate groups. Why use residence times of codon-anticodon base-pairing to cluster codon families, and what is the relationship of residence times to the decoding mechanism that operates within the tRNA synthetases, which enables accurate loading of appropriate amino acid to the tRNA carrying cognate codon?  The answer to this question will be provided in part-III and IV of this series.

Our approach here is top-down, as opposed to bottom-up, beginning with the identification of the evolutionary forces that shaped the molecules that constitute life today, and positing biophysical interactions (asymmetric cooperativity) within and among these molecules that would have benefited them evolutionarily, in the primordial scenario. A tacit assumption underlying this approach is that evolution was allowed to experiment with a very large pool of diverse molecules, until it found a small subset of specific molecules that served its purpose of continually catalyzing the conversion of free energy to entropy while maintaining its internal order and information across many self-replication cycles. This directly contradicts ``Frozen Accident'' explanations, which presume that evolution was constrained by the primordial availability of much smaller pool of appropriate molecules and biophysical pathways, and had to make do with the molecules provided by primordial geochemistry. We strongly interpret the statement ``Evolution is supervenient to Chemistry''. This point of view has helped us create a simple model that appears to capture the essential properties of DNA self-replication \cite{subramanian2018evolutionary}. Identification of the physicochemical origins of the posited interactions require the complementary bottom-up investigations, which is beyond the scope of our approach. Therefore, the posited interactions (asymmetric cooperativity) must be taken as axioms, whose veracity partly lies in its ability to explain fundamental observations regarding DNA/RNA replication and translation.

In summary, in this section, we have introduced a simple kinetics-based model that reproduces the wobble position degeneracy in the genetic code, and identified both the biophysical and evolutionary origins of the degeneracy. By identifying the biophysical determinants of the degeneracy, we have ruled out the ``Frozen Accident'' hypothesis of the genetic code evolution, and provide support for physicochemical theories of genetic code evolution. In the next section, we will show that, when the asymmetric cooperativity's dependence on the \textit{orientation} of a base-pair (which differentiates between 3'-A-5'/5'-T-3' from 3'-T-5'/5'-A-3') is included, the entire organizational structure of the genetic code, as explained in the introduction, is reproduced, and experimental support is provided.  In third section,  correlations between the codon-anticodon association times and amino acid energy dissipation times are established, and their consequences for the emergence of the genetic code are analyzed. In the final part-IV, a reason for the emergence of two types of tRNA synthetases is provided, and a classification system, using codon association times, is identified  that classifies codons into the two classes of synthetases. 

\clearpage			
\begin{center}
    {\Large\bfseries II: Quantitative reproduction of the codon table structure and its experimental validation} \\[0.5cm]
\end{center}

\subsection*{Background}

The biophysical origin of the genetic code, that enables translation of information stored in DNA sequences to amino acid sequences constituting proteins, is yet to be deciphered. The non-random, structured, many-to-one mapping of the $64$ codons to twenty or so amino acids suggests the presence of a biophysical basis for such an assignment \cite{woese1966molecular,wolfenden1979water,Koonin2009}.  A thorough understanding of the biophysical basis of the genetic code can revolutionize our ability to rationally engineer genetic information for specific purposes, such as production of antibiotics, and may clarify the conceptual underpinnings of life itself. 

In the first part of four sections, we have theoretically demonstrated that the evolutionary forces that underlie unidirectional growth of the daughter strand on DNA/RNA template also lead to wobble position degeneracy. Both the above phenomena are shown to be governed by a kinetic property we called  ``Asymmetric Cooperativity'', which we have derived earlier \cite{subramanian2018evolutionary} using evolutionary reasoning. In this part of the series, we modify asymmetric cooperativity to include sequence-dependence, as done earlier \cite{subramanian2020evolutionary}, and, optimizing just four parameters characterizing asymmetric cooperativity, we reproduce the clustering of the $64$ codons into $24$ degenerate groups, with each group containing either four or two codons, depending on the nature of the nucleotides at the three positions of the codon. 

It can be guessed from part-I that the clustering of codons into degenerate groups that code for a single amino acid is based on the \textit{kinetics} of codon-anticodon association process. The kinetics of the association process is quantified by the time taken for all the three hydrogen bonds between the three nucleotides of the codons and anticodons to form. As explained in part-I, we model, using a simple Markov chain method, the kinetics of the association and dissociation of bonds between the codon on mRNA with its anticodon on tRNA. We calculate the residence times, the average amount of time the system spends at each of the seven intermediate states during its sojourn, beginning with a completely detached state and ending in the fully-zipped double-stranded state. The sequence-dependence of kinetics of the codon-anticodon association process arises primarily from our postulate of sequence-dependence of asymmetric cooperativity, which will be elaborated in the next section.

\subsection*{The Asymmetric Cooperativity Model - Sequence Dependence}
\begin{figure}
\begin{center}
\includegraphics[width=0.6\textwidth]{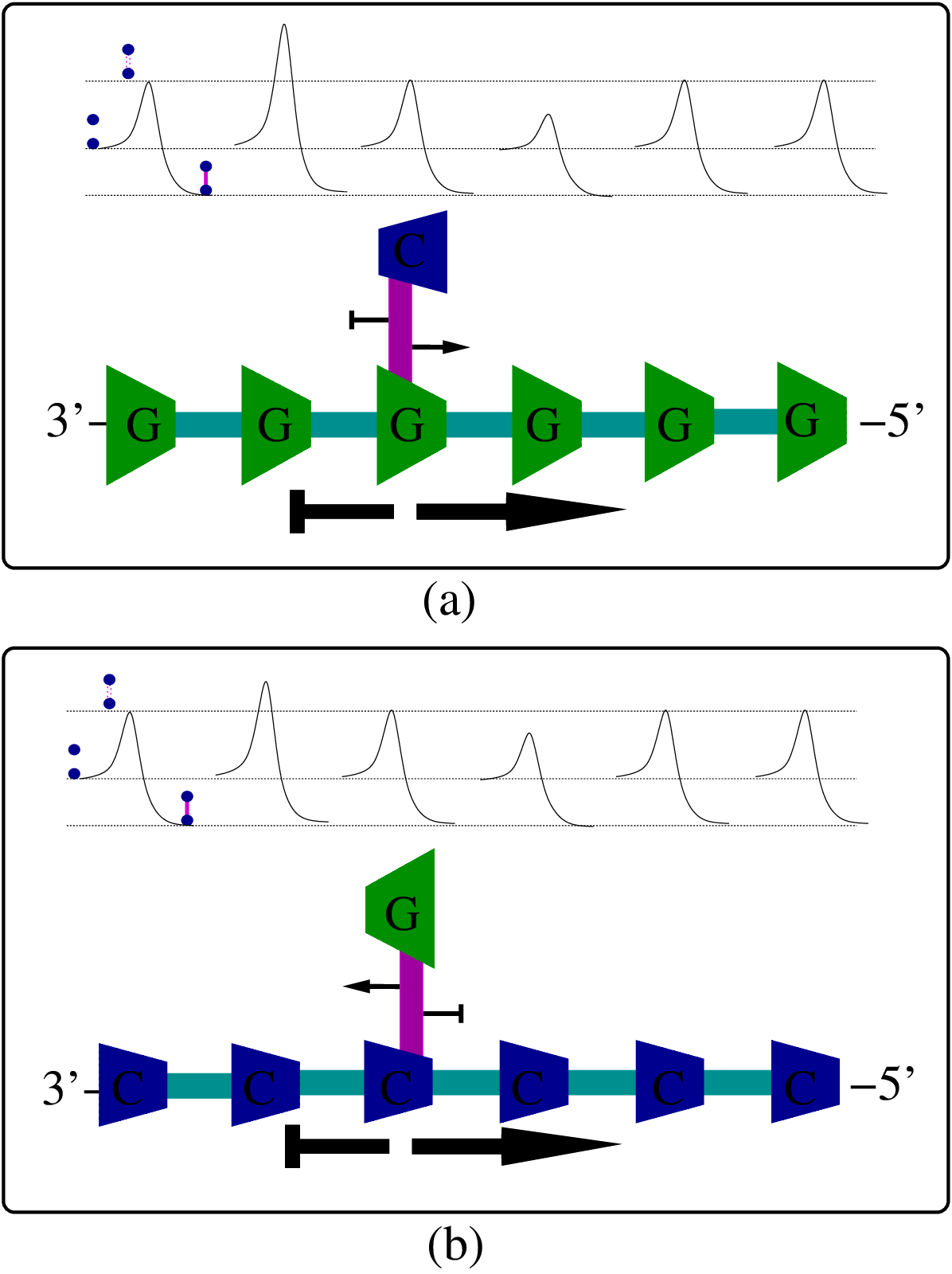}
\caption{Illustration of sequence-independent asymmetric cooperativity in DNA single strands. The asymmetric cooperativity mode of a single template strand is dictated by the $3'-5'$ directionality of the strand (figure and caption adapted from \cite{subramanian2020evolutionary} under a Creative Commons Attribution $4.0$ International License). (a) A hydrogen bond between a lone nucleotide $C$ and the template strand catalyzes the formation of another hydrogen bond to its right by reducing its kinetic barrier, while inhibiting the formation of its left neighbor by raising its barrier. The strength and the mode of sequence-independent asymmetric cooperativity dictated by the template strand is denoted by the thick black arrow below the template strand, pointing to the right. The thin arrows attached to the hydrogen bond denote the weaker sequence-dependent asymmetric cooperativity strength and mode. (b) Irrespective of the type of nucleotides occupying the template strand, the directionality of the template strand alone dictates the mode of sequence-independent asymmetric cooperativity. The thinner arrows on the hydrogen bond, denoting sequence-dependent part, though pointing in the opposite direction, does not alter the overall mode of asymmetric cooperativity due to the relative strength of sequence-independent part. This can be seen in the kinetic barrier diagrams above the bonds, where the barrier on the right of the hydrogen bond is lower in both (a) and (b), and the barrier on the left is higher. This assumption ensures that the daughter strand is always constructed from the $5'$ end of the daughter strand to its $3'$ end.}
\label{seqindependent}
\end{center}
\end{figure}

In \cite{subramanian2018evolutionary}, we introduced the asymmetric cooperativity model, justifying its introduction using evolutionary arguments, which we recapitulated in part-I. In \cite{subramanian2020evolutionary}, we theoretically factorized asymmetric cooperativity into a stronger sequence-independent part operative in DNA/RNA \textit{single} strands, and a weaker sequence-dependent part operative in the DNA/RNA \textit{double} strands, in order to demonstrate the evolutionary advantage of anti-parallel DNA duplex strands and to justify the counter-intuitive evolutionary choice of the complicated, piecemeal ``lagging strand'' replication mechanism. The part that was recapitulated in part-I is termed henceforth as sequence-independent, operative in DNA/RNA single strands, and is illustrated in fig. \ref{seqindependent}. Below, we introduce the weaker sequence-dependent part, operative in DNA/RNA double strands. Although the exposition below has already been covered in greater detail in our earlier article \cite{subramanian2020evolutionary}, we include it here to keep this series self-contained.

\subsubsection*{Sequence-Dependent Asymmetric Cooperativity}
\begin{figure}
\begin{center}
\includegraphics[width=0.6\textwidth]{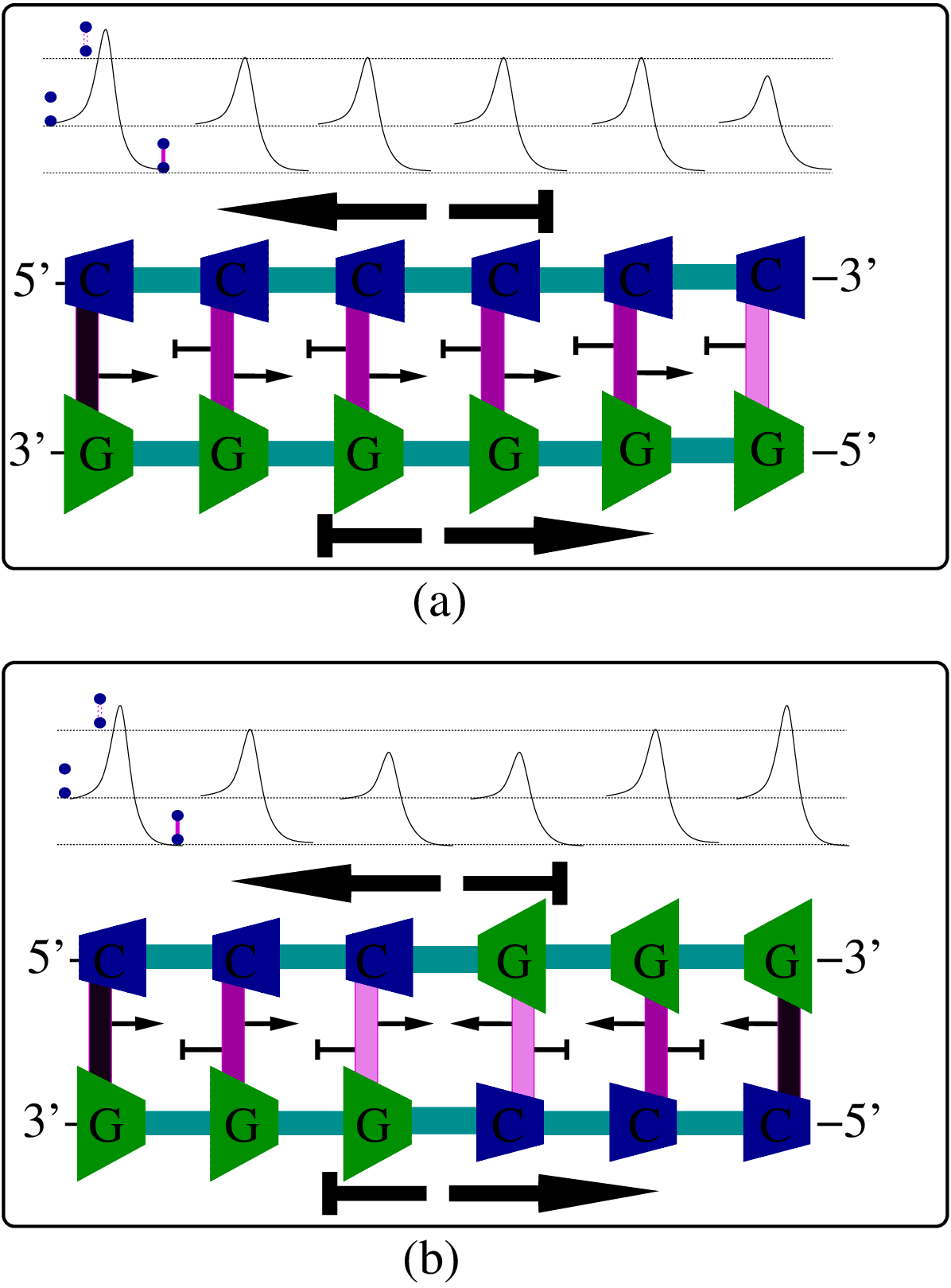}
\caption{Illustration of sequence-dependent asymmetric cooperativity (figure and caption adapted from \cite{subramanian2020evolutionary} under a Creative Commons Attribution $4.0$ International License). In a DNA double strand, the anti-parallel orientations of the two strands result in the cancellation of their respective opposing asymmetric cooperativity modes. If the nucleotides on both the strands are of the same type, the cancellation would be complete, due to symmetry. When the types of nucleotides on the two strands are different, say, with $C$ on the $3'-5'$ strand and $G$ on the $5'-3'$ (heteromolecular basepairing), the cancellation is not complete and the residual asymmetric cooperativity is dictated by the types of nucleotides, making the asymmetric cooperativity sequence-dependent. The thick arrows denote the sequence-independent asymmetric cooperativity dictated by the individual strands' directionality, whereas the thinner arrows attached to the hydrogen bonds denote the sequence-dependent asymmetric cooperativity that changes its mode depending on the orientation of the base-pair. $5' \hyp C \hyp 3'/3' \hyp G \hyp 5'$ base-pair orientation of the hydrogen bonds instantiates right asymmetric cooperativity, as shown in (a), whereas the $180^\circ$ -rotated $5' \hyp G \hyp 3'/3' \hyp C \hyp 5'$ instantiates left asymmetric cooperativity, as the last three bonds of (b) illustrates. The kinetic barrier diagrams above the strands in (a) and (b) are significantly different, illustrating the sequence-dependence of the unzipping behavior of DNA double strands. The unzipping of the strand in (a) would proceed sequentially from the rightmost end, whereas the middle two bonds would break and the strand will simultaneously unzip in both the directions in (b). }
\label{seqdependent}
\end{center}
\end{figure}

The weaker sequence-dependent asymmetric cooperativity dominates in DNA/RNA \textit{double} strands, since the stronger sequence-independent part, dictated by the $3'-5'$ directionality of the single strands, stands cancelled due to the anti-parallel orientation of the two strands in the duplex. The sequence-dependence of asymmetric cooperativity in duplex strands arises from the dependence of the \textit{mode} of asymmetric cooperativity on the ``orientation'' of the hydrogen-bonding base-pairs, which differentiates, for example, the base-pair $5' \hyp G \hyp 3'/3' \hyp C \hyp 5'$ from that of its $180^\circ$-rotated counterpart, $5' \hyp C \hyp 3'/3' \hyp G \hyp 5'$. Thus, the base-pair $5' \hyp G \hyp 3'/3' \hyp C \hyp 5'$ would lower the kinetic barrier for the formation/dissociation of its left hydrogen bond neighbor and raises the barrier for its right neighbor (left mode), whereas, the $180^\circ$-rotated base-pair $5' \hyp C \hyp 3'/3' \hyp G \hyp 5'$ catalyzes its right neighbor and inhibits its left neighbor (right mode). The effect of the presence of sequence-dependent asymmetric cooperativity on the kinetic barriers of hydrogen bonds in a DNA duplex is shown in fig. \ref{seqdependent}. The presence of sequence-dependent asymmetric cooperativity in DNA/RNA duplex strands renders the rate and the direction of unzipping of the duplex \textit{sequence-dependent}, thereby providing a crucial, additional degree of freedom to control the kinetics of DNA/RNA replication. This degree of freedom is exploited by evolution in creating origins of replication, by introducing substantial nucleotide skews around the potential origin, as we argued in \cite{sahu2024high}. Such skews are found in nearly all genomes, both prokaryotes and eukaryotes, studied thus far, and are widely used as a bioinformatic tool to find the origins of replication \cite{rocha2004,tillier2000contributions,dai2005dna,marsolier2012asymmetry,niu2003strand,bartholdy2015allele,touchon2005replication}. 

In case of the association process between codons and anticodons on mRNA and tRNA, the cancellation of sequence-independent asymmetric cooperativity is not complete, due to the structural/topological differences between the mRNA, with its linear strand, and tRNA, where the strand is bent as part of the anticodon loop. This is in contrast to our assumption of complete cancellation of sequence-independent asymmetric cooperativity in DNA double-strand unzipping \cite{subramanian2020evolutionary, sahu2024high}, where both the strands are topologically similar. Therefore, here, we assume that both sequence-independent and sequence-dependent asymmetric cooperativities are present in the codon-anticodon system. Importantly, we also assume that the two sequence-dependent asymmetric cooperativity parameters for both GC and AU basepairs are the same, and denote them as $\alpha_U$ and $\alpha_D$. This makes the sequence-dependent asymmetric cooperativity parameters independent of the strength of the basepairs. This is done primarily to keep the number of model parameters minimal. The basepair $5' \hyp C \hyp 3'/3' \hyp G \hyp 5'$  is right-asymmetrically cooperative, and modifies the kinetic barrier of its right neighbor by $\alpha_U$, and its left neighbor by $\alpha_D$. Whereas, its $180^\circ$-rotated counterpart $5' \hyp G \hyp 3'/3' \hyp C \hyp 5'$ modifies the kinetic barrier of its right neighbor by $\alpha_D$, and its left neighbor by $\alpha_U$. This results in sequence dependence of the kinetics of basepairing between the codons and anticodons. The same scheme applies to AU basepairs, with A behaving similar to G (purines, R), and U with C (pyrimidines, Y). The sequence-independent cooperativities are denoted, as in part I, as $\alpha_L$ and $\alpha_R$. The thermodynamically dictated bare rates of formation and dissociation of single hydrogen bond between two nucleotides on the codon and anticodon are denoted as $r_{bp}$ and $s_{bp}$, respectively, where $bp$ can be either GC or AU. These association/dissociation rates are kinetically modified due to the presence of neighboring hydrogen bonds, by multiplying them with the appropriate asymmetric cooperativity parameters, taking care to multiply both the association and the dissociation rates with the same asymmetric cooperativity parameters, thus altering only the kinetics and not thermodynamics. With these assumptions, the rate matrix for the Markov chain model to capture the kinetics of the association process, in terms of the four parameters above, is constructed as follows. We illustrate the construction of the rate matrix for some of the state transitions during the association process of  (i) $5'-AGC-3'$ codon with its $3'-UCG-5'$ anticodon, and (ii) $5'-UCG-3'$ codon with its $3'-AGC-5'$ anticodon. The generalization for other codons and for other transitions is straightforward.

\begin{equation}
\begin{aligned}
& \mathcal{R}_{AGC}\left ( 000 \rightarrow 010 \right ) = r_{GC}, \\
& \mathcal{R}_{AGC}\left ( 010 \rightarrow 011 \right ) = \alpha_R \alpha_D r_{GC}, \\
& \mathcal{R}_{AGC}\left ( 010 \rightarrow 110 \right ) = \alpha_L \alpha_U r_{AU} \qquad \text{and}\\
& \mathcal{R}_{AGC}\left ( 101 \rightarrow 111 \right ) = \left ( \alpha_R \alpha_D \right ) \left ( \alpha_L \alpha_D \right ) r_{GC}. \\
\label{ratesbondingI}
\end{aligned}
\end{equation}

\begin{equation}
\begin{aligned}
& \mathcal{R}_{UCG}\left ( 000 \rightarrow 010 \right ) = r_{GC}, \\
& \mathcal{R}_{UCG}\left ( 010 \rightarrow 011 \right ) = \alpha_R \alpha_U r_{GC}, \\
& \mathcal{R}_{UCG}\left ( 010 \rightarrow 110 \right ) = \alpha_L \alpha_D r_{AU} \qquad \text{and}\\
& \mathcal{R}_{UCG}\left ( 101 \rightarrow 111 \right ) = \left ( \alpha_R \alpha_U \right ) \left ( \alpha_L \alpha_U \right ) r_{GC}. \\
\label{ratesbondingII}
\end{aligned}
\end{equation}

The equations \ref{ratesbondingI} and  \ref{ratesbondingII} describe the rate assignment procedure for four different state transitions that involve a single hydrogen bond formation between $AGC$ or $UCG$ codons and their corresponding cognate anticodon nucleotides. The reverese transitions of \ref{ratesbondingI} and  \ref{ratesbondingII}, representing hydrogen bond dissociation, can be obtained by simply replacing $r_{bp}$ with $s_{bp}$. The states, such as $000$ and $010$, represent configurations where no hydrogen bond is present, and only one hydrogen bond is present between the middle codon nucleotides of codon and anticodon, respectively.  The $000 \rightarrow 010$ state transition is assigned bare rates, since there are no neighboring bonds, whereas, the  $010 \rightarrow 011$ transition involves third codon bond formation with a neighbor present, whose rate is therefore modified due to the presence of the middle codon bond. In $AGC$, the middle codon bond is $5' \hyp G \hyp 3'/3' \hyp C \hyp 5'$, and therefore, its right neighboring bond's association and dissociation rates are modified by the sequence-dependent asymmetric cooperativity factor $\alpha_D$ (apart from the sequence-independent factor $\alpha_R$), whereas, in $UCG$, the middle codon bond is $5' \hyp C \hyp 3'/3' \hyp G \hyp 5'$, and hence,  its right neighboring bond's rates are modified by $\alpha_U$ (and $\alpha_R$). Similarly, for the $101 \rightarrow 111$ transition, the orientations of the first and third codon basepairs dictate the kinetic barrier for the middle codon bond formation and dissociation. Since the orientations of the first and third codon hydrogen bonds are different for $AGC$ and $UCG$, the sequence-dependent asymmetric cooperativity factors multiplying the middle bond formation/dissociation rates are different for the two codons in equations \ref{ratesbondingI} and  \ref{ratesbondingII}. The  first codon $5' \hyp A \hyp 3'/3' \hyp U \hyp 5'$ and third codon $5' \hyp C \hyp 3'/3' \hyp G \hyp 5'$ in $AGC$ each provide an $\alpha_D$ for its right and left neighbor respectively. Similarly, $5' \hyp U \hyp 3'/3' \hyp A \hyp 5'$ and $5' \hyp G \hyp 3'/3' \hyp C \hyp 5'$ in $UCG$ each provide an $\alpha_U$ to modify the middle bond's formation/dissociation rates. 


The Markov chain rate matrices are constructed according to the above prescription for all the $64$ codon-anticodon association processes. The Mean First Passage Time (MFPT) is calculated for the Markov chain to travel from the fully-unzipped state $000$ to fully-zipped state $111$, traversing the six other intermediate states, by solving the following set of equations \cite{van1992stochastic}:  
\begin{equation}
\sum_j \mathcal{R}_{ij} t_j  = -1, \qquad i \ne k.
\label{hittingtimeeq}
\end{equation} 
Here, $t_j$ represents the expected amount of time the Markov chain spends in state $j$ during its sojourn towards the absorbing state $k$, beginning at $i$. Since our starting state is $1$, representing $000$ state (see table.\ref{stateindex}), the required mean first passage time, or the time taken for the codon-anticodon association process, is given by $\sum_{j=1}^{7}{t_j}$, calculated for each of the $64$ codons. The eqn.\ref{hittingtimeeq} is usually written as a matrix-vector product $\bf{\mathcal{R'}}.\bf{t} = \bf{-1}$, where, $\mathcal{R'}$ is derived from $\mathcal{R}$ by removing the row and column corresponding to the final state $111$. This equation can be used to solve for $t_j$'s, by inverting the reduced rate matrix $\mathcal{R'}$, and summing and negating the first row. The seven individual (negated) elements of the first row of $\mathcal{R'}^{-1}$ provide the \textit{residence times} of the Markov chain, defined as the amount of time the Markov chain spends in each of the seven states, from $000$ to $110$, during its travel from $000$ to $111$. The sum of the seven residence times is $\sum_{j=1}^{7}{t_j}$, the mean first passage time. This seven-element residence time vector, denoted henceforth by $\textbf{T}=\{ t_1 \ldots t_7\}$, calculated for each of the $64$ codons, is all we need for cluster analysis, and is thus the \textit{central quantity} provided by the theoretical model.   

\begin{table*}	
		\centering 
		\begin{tabular}{|c|c|}
			\hline
        State Index & State   \\ 
        \hline
        $1$   & $000$ \\ 
				$2$   & $001$ \\ 
				$3$   & $010$ \\ 
				$4$   & $011$ \\ 
				$5$   & $100$ \\ 
				$6$   & $101$ \\ 
				$7$   & $110$ \\ 
				$8$   & $111$ \\ 
        \hline
		\end{tabular}
		\caption{Assignment of an index to each of the eight Markov chain states}
	\label{stateindex}
\end{table*}

\subsubsection*{Optimization and Parametrization}
The seven-element residence time vector $\bf{T}$, calculated for each of the $64$ codons, are treated as functions of the four asymmetric cooperativity parameters, $\alpha_L$, $\alpha_R$, $\alpha_U$ and $\alpha_D$.  These residence time vectors are then hierarchically clustered into $24$ groups, first by calculating the euclidean distances between the $64$ vectors and clustering two groups of vectors according to the average distance between the elements of the two groups. Our choice of $24$ clusters  reflects the fact that there are $20$ or so proteinogenic amino acids used by extant living systems, and the nearest number of clusters that obey the symmetry (see below) found in the genetic code is $24$.  We use Genetic Algorithm procedure to find the optimum values for the four asymmetric cooperativity parameters that lead to the correct clustering found in the genetic code. It has to be noted that there are multiple optima available within the parameter space that satisfy the constraint of matching with the observed genetic code structure, which we have narrowed down by fixing the value of $\alpha_L$ and searching along the three remaining dimensions to find a unique optimum. We have experimented with a few sets of optimal parameters and found no significant difference in the results that follow in this series. We therefore decided to choose the following set, among all other possible optimal parameter sets:  $\alpha_U = 1.5196$, $\alpha_D = 9.4091$, $\alpha_R = 3.1793$, and $\alpha_L = 45.0000$. The thermodynamic parameters are the same as those provided in part-I, with $r = k_{on} = 0.1\: k_0$ for both $GC$ and $AU$, where $k_0$ is as yet undetermined, and $s = k_{off} = 0.89 \: k_{on}$  and $0.42 \: k_{on}$ for $AU$ and $GC$ respectively.  These parameters will remain unaltered for the rest of the sections, and all the results that follow are produced using the above parameter set. \textit{It has to be emphasized that no fitting is involved in producing the correlations between experimental data and the model output, shown below and hereafter in the series, beyond the optimization for the reproduction of the organizational structure of the genetic code, mentioned above}. 


\subsection*{Results}
\begin{figure}
\begin{center}
\includegraphics[trim=40pt 20pt 20pt 20pt,width=1.05\textwidth]{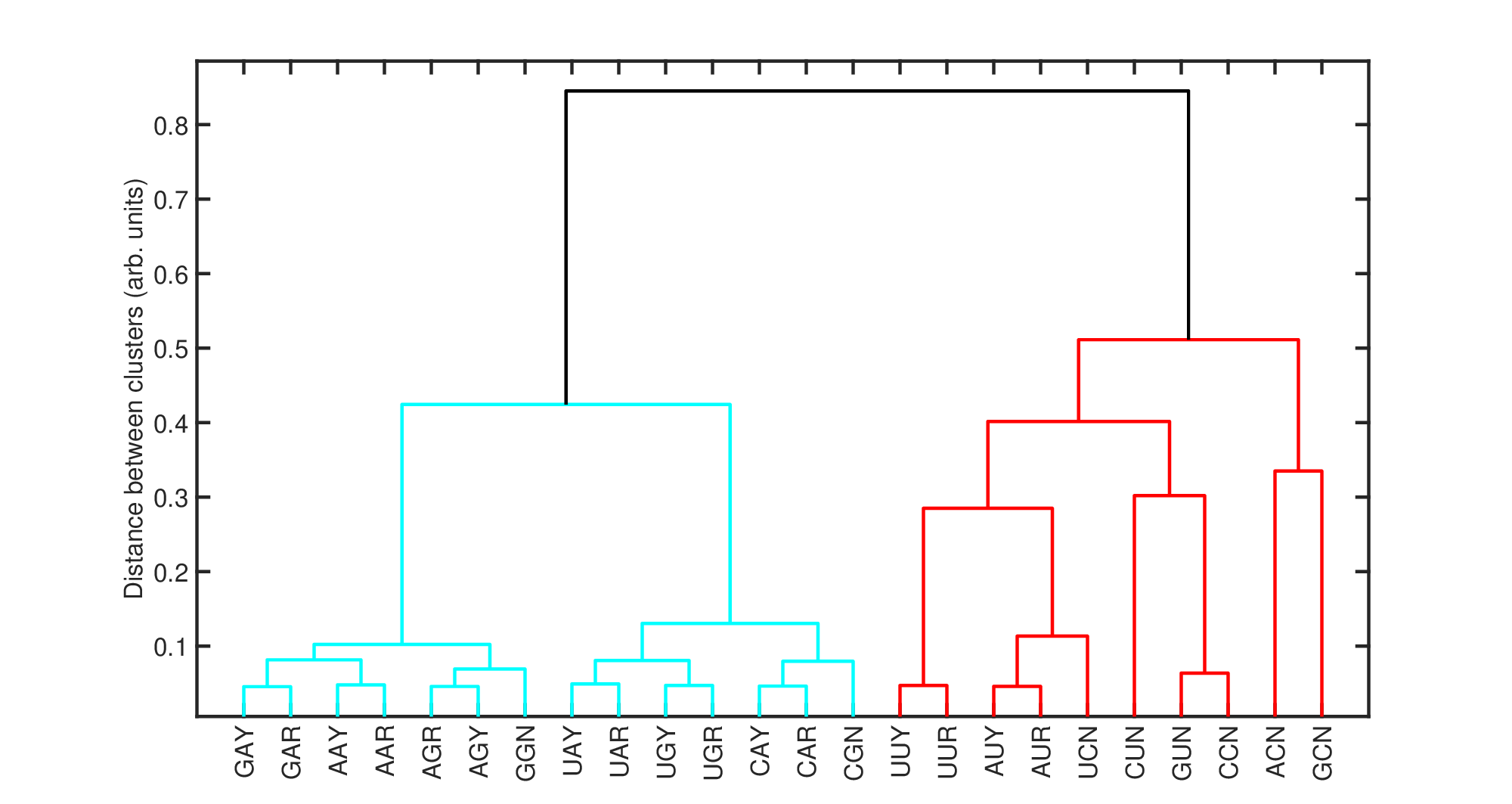}
\caption{A hierarchical tree structure clustering the 64 seven-element residence time vectors representing the 64 codons, $\bf{T_c}$, into 24 groups. The symmetry present in the simplified genetic code is apparent: All middle-C codons are clustered into quadruplets, $ACN$, $UCN$, $GCN$ and $CCN$. All middle-A codons are clustered into doublets, for example, $GAR$ and $GAY$. All middle-U and middle-G codons are clustered depending on their first position nucleotide identity: if the first position nucleotide is weak (A or U), these codons are clustered as doublets, and if it is strong (G or C), quadruplets. For example, $UGY$ and $UGR$ are clustered as doublets, whereas, $GGN$ is a quadruplet. This clustering closely corresponds to the vertebrate mitochondrial genetic code, which is the most symmetric of all genetic codes. This tree is obtained by treating the residence time vectors as functions of the four asymmetric cooperativity parameters, and optimizing the parameters such that the residence time vectors representing 64 codons cluster into 24 degenerate groups as in the genetic code. The tree also clearly segregates $NYN$ and $NRN$ codons.}
\label{clustering}
\end{center}
\end{figure}

The result of the above-described optimization procedure, a hierarchical tree structure that clusters the residence times of the association process of $64$ codons with their cognate anticodons into $24$ groups, is shown in fig.\ref{clustering}.  The symmetry of the genetic code is apparent. It is clear from the figure that (i) all codons with a middle C are clustered into groups with four codons each, i.e., as \textit{quadruplets}. These clusters are $ACN$, $GCN$, $CCN$ and $UCN$, where, $N$ denotes any of the four nucleotides. (ii) All codons with a middle A are clustered into groups with two codons each, i.e., as \textit{doublets}. For example, $CAY$ and $CAR$ are clustered into two groups, with two codons each, with Y denoting pyrimidines and R, purines. (iii) All codons with middle G or T are clustered into quadruplets or doublets depending on the strength of the first basepair; i.e., whether the first basepair is strong ($GC$) or weak ($AU$). If the first basepair is strong, the clusters are quadruplets and, if weak, doublets. For example, $GGN$ is clustered as a quadruplet, whereas, $AGY$ and $AGR$ are segregated as doublets. Thus we observe that the asymmetric cooperativity model neatly reproduces, with four independent parameters, the full simplified organizational structure of the genetic code. It has to be mentioned here that the clustering that reproduces the genetic code structure depends on the residence times ($t_j$'s) of the codon-anticodon association process at \textit{every individual intermediate step}, and not simply on the total zipping time, denoted by the mean first passage time, $\sum_{j=1}^{7}{t_j}$. Thus, information about the genetic code structure is embedded in the entire \textit{path} of  the Markov chain. We will later see that this leads to a conclusion that communication between the codon-anticodon system and amino acids happen at multiple, distinct timescales, due to the presence of multiple timescales in the residence time vector $\bf{T}$, as can be seen in the table of residence time vectors of all 64 codons, provided in the supplementary material, table \ref{residencetimestable}.

The splitting between $NNA$ and $NNG$, where these two codons are assigned to different amino acids (such as between stop codon $UGA$ and tryptophan $UGG$, and isoleucine $AUA$ and methionine $AUG$, both in the standard code), cannot be captured by the current model, because we assume that the asymmetric cooperativity parameters to be the same for both $AU$ and $GC$ basepairs. These splittings can be captured by breaking this symmetry between $AU$ and $GC$ basepair kinetic influences, but at the expense of two extra asymmetric cooperativity parameters, which we have chosen to avoid. The model can nevertheless be easily extended to include the differences in the asymmetric cooperativity parameters between $GC$ and $AT$ base pairs, and optimized individually to fit every genetic code variant.

One can observe, from fig.\ref{clustering}, that codons with middle pyrimidines and middle purines, $NYN$ and $NRN$, are clearly segregated, with the middle purine $NRN$ codons generally having smaller total residence times than $NYN$ codons (fig. \ref{clustering}).  This difference in the kinetics of association between $NRN$ and $NYN$ codons on mRNA with their cognate anticodons on tRNA arises entirely from the sequence-dependent asymmetric cooperativity, within our model. Sequence-dependent asymmetric cooperativity discriminates between the \textit{orientation} of  the base pairs, which specifies which nucleotide (R or Y) occupies which of the two associating RNA strands, as explained in the Model introduction above. Switching the orientation of the middle base pair, say from $5' \hyp G \hyp 3'/3' \hyp C \hyp 5'$ to $5' \hyp C \hyp 3'/3' \hyp G \hyp 5'$, which takes $NRN$ codons to $NYN$ codons, switches the asymmetric cooperativity direction from catalyzing to inhibiting (or less catalyzing), thereby producing significant differences in the association kinetics of $NRN$ and $NYN$ codons, and segregating them. A scree plot for the above clustering procedure, to estimate the optimum number of clusters, is provided in fig. \ref{screeplot}. The optimum number of clusters appears to be 12, from the plot.

A table (table \ref{residencetimestable}) of  the seven-element residence time vector, $\bf{T}$, for all the 64 codons, is provided in the supplementary material. From the table, one can observe that the order of the individual elements of the residence time vector of any given codon ranges from $10^{-5}$ (state 6) to $1$ (state 1). This rather large range of timescales in the residence time vector of a given codon arises from the presence or absence of base pairs at the three locations of a particular state, and the catalytic influence these base pairs experience due to, and provide to, their neighboring base pairs. For example, the low residence times at states 2 ($001$), 4 ($011$) and 6 ($101$), for all codons, is due to the presence of a base pair at the third position, which strongly catalyzes the formation of the middle base pair, leading to the Markov chain leaving these states immediately after entering them. In part-I, we identified these low residence times as the cause of wobble position degeneracy. Similarly, the large residence time at state 1 ($000$) is due to the non-availability of catalytic assistance from any neighboring base pair. We will later see, in part III, that this large range of timescales of the association process is important for the accurate assignment of codons to their corresponding amino acids.

Although the thermodynamic parameters that influence the bonding between a codon-anticodon system and its palindromic counterpart, such as $AAU$ and $AUU$, are the same, their kinetics and hence their residence time vectors are different. This is due to the presence of sequence-independent asymmetric cooperativity, which maintains its directional effects irrespective of whether a codon or its palindromic counterpart occupies mRNA, whereas, sequence-dependent asymmetric cooperativity switches direction, due to altered orientation of the basepairs. If it were not for sequence-independent asymmetric cooperativity, the kinetics of a codon-anticodon system and its palindromic counterpart system would be the same, and the genetic code would be two-fold degenerate, and unable to distinguish between a codon and its palindromic counterpart. This differentiation ultimately arises from the topological difference between (locally) linear mRNA and curved tRNA anticodon loop, which motivated us to add sequence-independent asymmetric cooperativity to the codon-anticodon system in the first place.

It has to be remembered that the thermodynamic parameters for bond formation between $GC$ and $AU$ are the same, and hence do not distinguish between the two basepairs, in our model. The thermodynamic discrimination between the two basepairs comes from hydrogen bond dissociation. If we evaluate the time taken for codon-anticodon \textit{association} process, involving bond formation, at very high thermodynamic drive ($r >> s$), the discrimination between different codons become minimal, and all the $64$ $\bf{T}$ vectors cluster together, because of the small probability for bond dissociation. Moreover, near thermodynamic equilibrium, ($r \approx s$), the time taken for association becomes very large, counterproductive for the translation process. This implies that the thermodynamic and kinetic parameters needed to be evolutionarily optimized, to balance discrimination or accuracy, and speed, as typically observed in many biological error correction processes.  

It is important to note that the optimization of the four free parameters was to simply cluster the codons into degenerate groups, and no constraint was imposed on the magnitude of the state residence times of individual codons. Therefore, no ordering of codons in terms of their residence times was imposed. Yet, we find that the model has ordered the residence times appropriately  (table \ref{residencetimestable}) to result in significant correlations with the experimentally measured residence times of the ribosome on each codon, as we will see below. The model assigns the shortest residence time for the smallest of all amino acids, glycine, and a large residence time for phenylalanine, one of the largest amino acids. If the genetic code assignment is dictated by thermodynamics, the stability values calculated from calorimetry should be able to distinguish between codons such as Gly ($GGG$) and Pro ($CCC$). However, these values are nearly same \cite{klump2020energy,breslauer1986predicting}, suggesting that thermodynamics plays a subdued role in determining the genetic code structure. Worse still, these thermodynamic values vary significantly with the third codon nucleotide identity, exhibiting no hint of degeneracy observed in the genetic code. The introduction of direction-dependent kinetic influence of a basepair on its neighbors (sequence-independent and sequence-dependent asymmetric cooperativities), motivated by evolutionary reasoning, resolves these issues and results in appropriate clustering of the codon-anticodon association times, and segregates codons with similar thermodynamic stabilities, such as $GGG$ and $CCC$, kinetically.

\subsection*{Experimental Support}
To validate the theoretically evaluated association times of tRNA-mRNA bonding process, we make use of publicly available data generated from a technique developed to interrogate the kinetics of the translation process, called ``Ribosome profiling'' or ``Riboseq'' \cite{ingolia2009genome}. The technique, briefly, involves isolating biochemically/biophysically stalled ribosomes on mRNA by digesting away the exposed mRNA and then sequencing the ribosome-protected mRNA fragments and aligning them to known gene sequences of the organism considered. This technique is capable of measuring relative residence times of ribosomes on different codons at the three functional sites E, P and A, of the ribosome. This is done by measuring the frequency of appearance of mRNA fragments with a specific codon at a specific ribosome site in the sequenced data, and associating it with the residence time of the ribosome on that codon at that site. We utilize the data-normalization technique used by Gardin \textit{et al} \cite{gardin2014measurement}, since it has been demonstrated to help evaluate accurately the relative residence times of ribosomes on different codons at distinct sites of the ribosome, by averaging away the pervasive and significant site-site correlations present in the sequenced Riboseq data. By averaging away the probability distribution of codons that is common to all sites of the ribosome that leads to the above correlations, this normalization brings to the fore the site-dependent variation of the relative residence times of the ribosome. 

These site-dependent residence times of all codons can help understand the kinetics of distinct processes of translation at distinct ribosomal sites, such as waiting for the cognate aminoacylated tRNA molecule to bind with A-site codon, peptide bond formation times and the exit tunnel residence times of the nascent polypeptide at the P-site, and the mRNA-tRNA binding at the E-site, which has been proposed to be connected to the next tRNA reception at the A-site \cite{nierhaus1993solution}. Since the experimentally calculated residence times of the ribosome is a complicated combination of all these probability distributions, disentangling the relative contribution of each of these processes is necessary to understand the site-by-site dynamics of the ribosome and its dependence on a specific codon. The normalization technique, introduced by Gardin \textit{et al} \cite{gardin2014measurement}, is an attempt to achieve this. Roughly, it takes the total number of sequences with a specific codon $c$ at a specific ribosomal site $i$, $N_{ic}$, yielded by Riboseq, and sums them over all available sites within the ribosome, $N_c = \sum_i{N_{ic}}$. The relative residence time of a given codon is then calculated by dividing the occurrence frequency $N_{ic}$ by the sum: $T_{ic}^{exp} = N_{ic}/N_c$. This procedure eliminates the information about the relative occupancies of the $61$ codons to be present within all the sites of the ribosome, and only retains information about each codon's relative occupancy of each of the ten or so sites within the ribosome. The sum of the relative occupancies of any given codon across all ribosomal sites is then $1$, due to this normalization. This procedure has been shown to be capable of deciphering the relative times of events, such as tRNA waiting times, within individual sites (A-site) of the ribosome, relatively unaffected by the time taken for other events at other ribosomal sites, such as E and P sites.

We have extracted the preprocessed and pre-validated publicly available data from the publications of Gardin \textit{et al} \cite{gardin2014measurement}, Stein \textit{et al} \cite{stein2022ageing} and Lareau \textit{et al} \cite{lareau2014distinct}. We utilized $41$ datasets from a total of $31$ samples, four from Gardin \textit{et al}, ten samples containing both long and short sequences from Lareau \textit{et al} (for a total of twenty datasets), and twelve from Stein \textit{et al}, all belonging to Yeast (\textit{S. cerevisiae}), and five from Stein \textit{et al} belonging to worms (\textit{C. elegans}). The long footprints are approximately 28-30 nt long and short footprints, 20-22 nt long, are produced using different elongation inhibitors, and report on different structural conformations of the translating ribosomes \cite{lareau2014distinct}. We have not eliminated any sample from consideration from these three publications. The data were aligned to the format of Gardin \textit{et al}, with the same normalization procedure employed by them, with the difference that a total of 17 sites were considered for the samples of Lareau \textit{et al} and Stein \textit{et al}, whereas, Gardin \textit{et al} provide data for just ten sites (thirty nucleotides). We calculate the relative residence times of ribosomes on all the three functional codon sites, E, P and A, of the ribosome, apart from seven sites to the left of E, and seven to the right of A, for a total of $17$. This wide window enables us to validate if the model's data truly captures what is going on inside the ribosome within a window of ten codons or less, or randomly correlating with dynamics unconnected with tRNA-mRNA interactions outside this window.

Our model allows us to theoretically calculate the association times of anticodons on tRNA with codons on mRNA, while the experimentally-derived relative residence times of ribosomes on a specific codon ``c'', $T_{ic}^{exp}$ above, allows one to investigate the kinetics of  tRNA-mRNA interactions at a specific site ``i'' of the ribosome. We speculated that the above approach to estimate relative residence times of ribosomes on all $61$ codons on all sites could be utilized to test the theoretical model's predictions on tRNA-mRNA association kinetics, specifically because this approach allows us to disentangle the contributions of other processes, such as tRNA waiting times, peptide bond formation time etc., from the tRNA-mRNA association times, since these processes happen at distinct sites. We first theoretically calculate the codon-anticodon association time, which is obtained by summing the residence time vector $\textbf{T}=\{ T_1 \ldots T_7\}$ over all the seven Markov chain intermediate states, to yield $T_c^{th}$, for every codon ``c''. This is the total time taken for an anticodon to go from completely dissociated state to completely-bonded state with the codon. We proceeded to correlate the theoretically calculated residence times of the $61$ codons $T_c^{th}$, with the experimentally obtained relative residence times of the $61$ codons at different sites ``i'' within the ribosome $T_{i}^{exp}$, for all the $41$ datasets from three publications.

\begin{figure}
\begin{center}
\includegraphics[trim=30pt 20pt 30pt 20pt,width=1.05\textwidth]{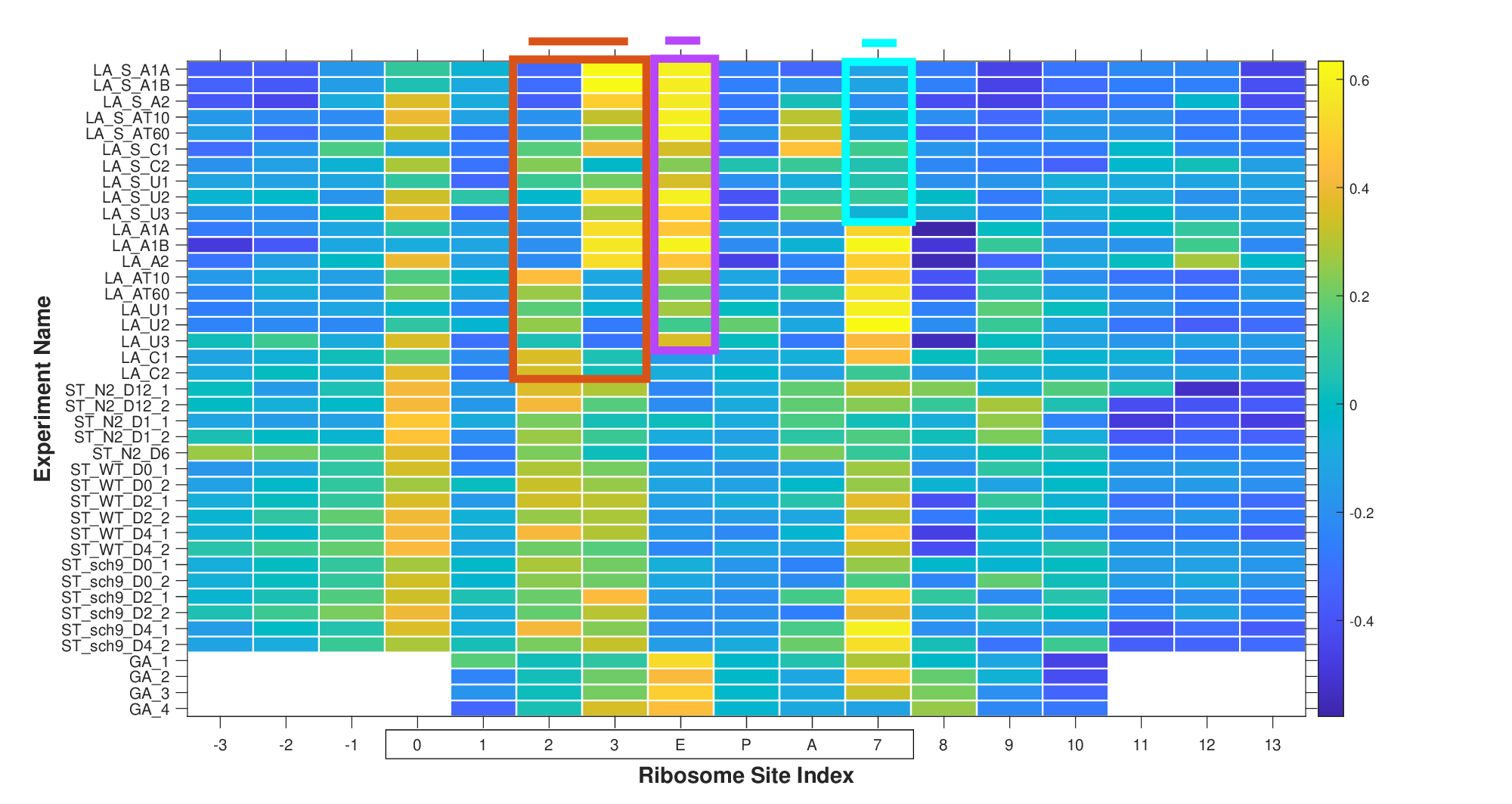}
\caption{Correlation between theoretically determined mRNA-tRNA association times and ribosomal residence times at sites (x-axis) within and near the ribosome, for 41 experiments, listed along the y-axis. Experimental riboseq data from three publications are used: 20 experiments from Lareau \textit{et al} (which include 10 experiments isolating short fragments, and 10 experiments of long fragments), 10 from Stein \textit{et al}, and 4 from Gardin \textit{et al}. These experiments are labeled as LA\_S\_ (Lareau \textit{et al}'s short fragment experiments), LA\_ (Lareau \textit{et al}'s long fragment experiments), ST\_ (Stein \textit{et al}), and GA\_ (Gardin \textit{et al}), along the y-axis. Seventeen ribosome sites are included in the correlation analysis, seven to the left of Exit (E) site, seven to the right of Aminoacyl site (A), and the E, P and A sites themselves, totaling 17. Gardin \textit{et al} do not provide data for sites other than 1 to 10. The site numbering follows Gardin \textit{et al}'s convention. Correlations are observed at site 0, at the leftmost $5'$ edge of the ribosome for many experiments. Short fragments and Anisomycin-treated long fragments show correlation at site 3, whereas untreated and cycloheximide-treated long fragments show correlation at site 2, in Lareau \textit{et al}'s data, which segregate short and long fragments. This is highlighted by the red box. Other experiments do not do so, and the correlations at these two sites are non-discriminatory. These correlations therefore appear to report on translocation dynamics. Exit (E) site shows correlation for all experiments other than the ones that used cycloheximide, which binds to the E site and probably destroys the correlation \cite{schneider2010inhibition}. This distinction is highlighted by the purple box. Site 7 shows no correlation for short fragments, whose lengths extend only up to the A site, highlighted by cyan box. Long fragments show correlations at this site. Sites 0-7 are usually inside the ribosome where correlations are seen, and is highlighted by a box in the x-axis.}
\label{correlationmatrix}
\end{center}
\end{figure}

The figure \ref{correlationmatrix} shows the correlation of $T^{th}$ with $T_{i}^{exp}$ for all the $41$ datasets, for all 61 codons and $i=\{-3,-2, \ldots 13\}$ sites for $37$ datasets of Stein et al and Lareau et al, and $i=\{1 \ldots 10\}$ for the four datasets of Gardin et al. The site numbering convention is chosen to align with Gardin \textit{et al}'s numbering system, where, site 4 is the exit (E) site, site 5, Peptidyl (P) and site 6, Aminoacyl (A) site, and are denoted as such in the figure. Each row corresponds to a single experiment, whereas, each column corresponds to a specific site in or near the ribosome.  The bright (yellow and orange) colors denote high positive correlation between theoretical and experimental residence times of all $61$ codons, and dark blue colors denote high negative correlation. Fig. \ref{correlationmatrixpvalue} highlights ribosomal sites and experiments where the p-value of the correlation between $T^{th}$ and $T_{i}^{exp}$ is below 0.05, and when the absolute correlation is above 0.3. A number of observations can be made from this figure.

(1) Significant positive correlations between experimental and theoretical residence times exist only at the sites that are either inside or at the edge of the ribosome (from 0 to 7). Sites outside this range show mostly negative correlations, that are produced due to the normalization procedure followed, because the residence times at these sites are normalized with (divided by) the sum of all sites' residence times that include sites with strong positive correlations. This suggests that the model captures processes happening only inside the ribosome.

(2) Significant positive correlations are seen at the E site, where we expect to observe codon-anticodon binding. Although A site residence times include codon-anticodon binding , this signal is corrupted by the waiting times for cognate tRNAs, and thus might have degraded the signal. At the E site, the experiments can be cleanly divided into two groups, ones showing significant correlation, and the ones that do not. Experiments that used cycloheximide to inhibit elongation of the nascent polypeptide show almost no correlation with the theoretical data at the E site, whereas, experiments that didn't use this inhibitor show strong correlations. Two datasets of Lareau \textit{et al} and all the datasets of Stein \textit{et al} used cycloheximide to stabilize long mRNA fragments, and thus do not show any correlations. All other datasets of Lareau \textit{et al}, and Gardin \textit{et al} (who flash-froze the samples before adding cycloheximide) show significant correlations, including all the short-footprint data of Lareau \textit{et al} (including two samples treated with cycloheximide, whose correlations are relatively weaker). It is known that cycloheximide binds to the E site of the ribosome \cite{schneider2010inhibition} and inhibits tRNA release, blocking subsequent processes. The lack of correlation for these datasets probably arises from this inhibitory activity of cycloheximide. It is surprising to find that the theoretical data could distinguish clearly between these two groups of experiments.

(3) We observe significant correlations at site 7, immediately to the right of A site (towards $3'$), near the ribosomal right edge. Again, we observe two clearly separated groups of experiments, with short footprint datasets (20-22 nt) isolated by Lareau \textit{et al} showing minimal correlations, and most other experiments that included long footprints (28-30 nt) showing significant correlations. Since the short footprints are of 22 nt maximum length, the number of sites occupied by these fragments within the ribosome cannot be more than seven. Assuming site 0 to be inside the ribosome, since significant correlations are seen at this site for most experiments, the mRNA fragment would only extend up to the A site (site 6), with site 7 sitting outside the ribosome for short footprints, which therefore produces no correlations. On the other hand, long fragments do show correlations at this site.

(4) The P site shows no correlations for all experiments, since the processes determining the P site residence times, such as peptide bond formation, and exit tunnel interactions, have little to do with codon-anticodon interactions. The A site residence times, although includes codon-anticodon interactions, have tRNA waiting times overlapped on them, thereby making the correlations minimal.

(5) Sites 2 and 3, to the left (5') of E site, show interesting variation, specifically for the experiments of Lareau \textit{et al}. For these experiments, which separate out short and long footprints, the correlations between the experiments and theory appear complementary at sites 2 and 3. For most of the short footprint and anisomycin-treated long footprint datasets of Lareau \textit{et al}, site 3 shows higher correlation, whereas, the other samples not treated with anisomycin show higher correlation at site  2.  Since it has been shown that anisomycin protects short footprints embedded within post-translocational conformation of the ribosome, whereas, cycloheximide protects long footprints within pre-translocational ribosomal conformation \cite{lareau2014distinct}, post-translocational, short-footprints (and anisomycin-treated long footprints) show correlations at a site (site 3) that is one codon ahead of the pre-translocational (site 2) long footprints generated by cycloheximide. Therefore, correlations at sites 2 and 3 appear to report on translocation dynamics of the ribosome. It is unclear why correlations at other sites are unaffected by translocation. Data from the other two experiments do not show such a clear separation between the two sites since they do not segregate short and long footprints. The theoretical data thus appears to capture the ribosomal translocation dynamics, provided the footprints are segregated by length. The origin of this correlation, although, is not obvious, and needs further bottom-up investigations.

(6) Site 0, sitting at the left-most (5') edge of the ribosome, shows positive correlation with the theoretical residence times for almost all the datasets (except for Gardin \textit{et al} datasets, where this information is not available). In the absence of correlations with E site, for all the cycloheximide-treated sample datasets (total $19$), we utilize the correlations at this site and site 7 to calculate average relative residence times across all $19$ datasets. The correlations seen at this site, and sites 2,3 and 7, are unexpected, since these does not conform to our current textbook understanding of ribosome function, where, all the translation activity happens within the three sites E, P and A. However, our study is not the first to notice the importance of sites other than E, P and A in determining the ribosomal residence times on mRNA. O\' Connor \textit{et al} \cite{o2016comparative}, while analyzing $30$ riboseq datasets (including 3 used here) from multiple publications, have found that ribosome residence times are predominantly dictated by, apart from the A site codon, codons at the $5'$ and the $3'$ ends of the ribosome-protected mRNA fragments. Pavlov \textit{et al} \cite{pavlov2021estimation} state that ``ribosomal profiling spectra are ultra-sensitive to codon identity near ribosome edges''. Tunney \textit{et al} \cite{tunney2018accurate} trained a neural network to predict ribosome residence times in terms of codons within the embedded mRNA fragments, and found that ``models trained on these data learned disconcertingly high weights for both the –5 and+3 codon positions. The  $–5$ codon, i.e., the $5'$ end of a footprint, was the single strongest predictor of footprint counts, exceeding even those of the A site''. However, these observations have been attributed to ``sequence-specificity of enzymes used during library construction'' \cite{o2016comparative}. Here, we find that tRNA-mRNA interactions might be happening at these sites as well, resulting in the observed correlations, although, structural analysis of the ribosome do not support this.

(7) We have initially suspected that the correlations found at sites other than E site may have been produced due to random thermal movements of ribosome \textit{after} getting inhibited by antibiotics. This motion is capable of carrying residence time information from one site to another. However, the uniform lack of correlation at certain sites, such as the P-site and site 1, militate against this possibility. More over, Gardin \textit{et al}'s flash-freezing method should have decreased the possibility of random motion and hence the appearance of correlations at sites other than E, P and A, which it didn't. These suggest that the observed correlations are real and are produced by yet to be explored events within the ribosome.

(8) The $41$ datasets include five from Stein \textit{et al} belonging to worms (C. elegans). We do not see any significant difference in the residence time correlation across sites, between the yeast and worm datasets. This suggests that the ribosome dynamics uncovered by the correlations in the figure are not specific to yeast alone, and may be extended to other eukaryotes as well.

\begin{figure}
\begin{center}
\includegraphics[trim=50pt 20pt 20pt 20pt,width=1.05\textwidth]{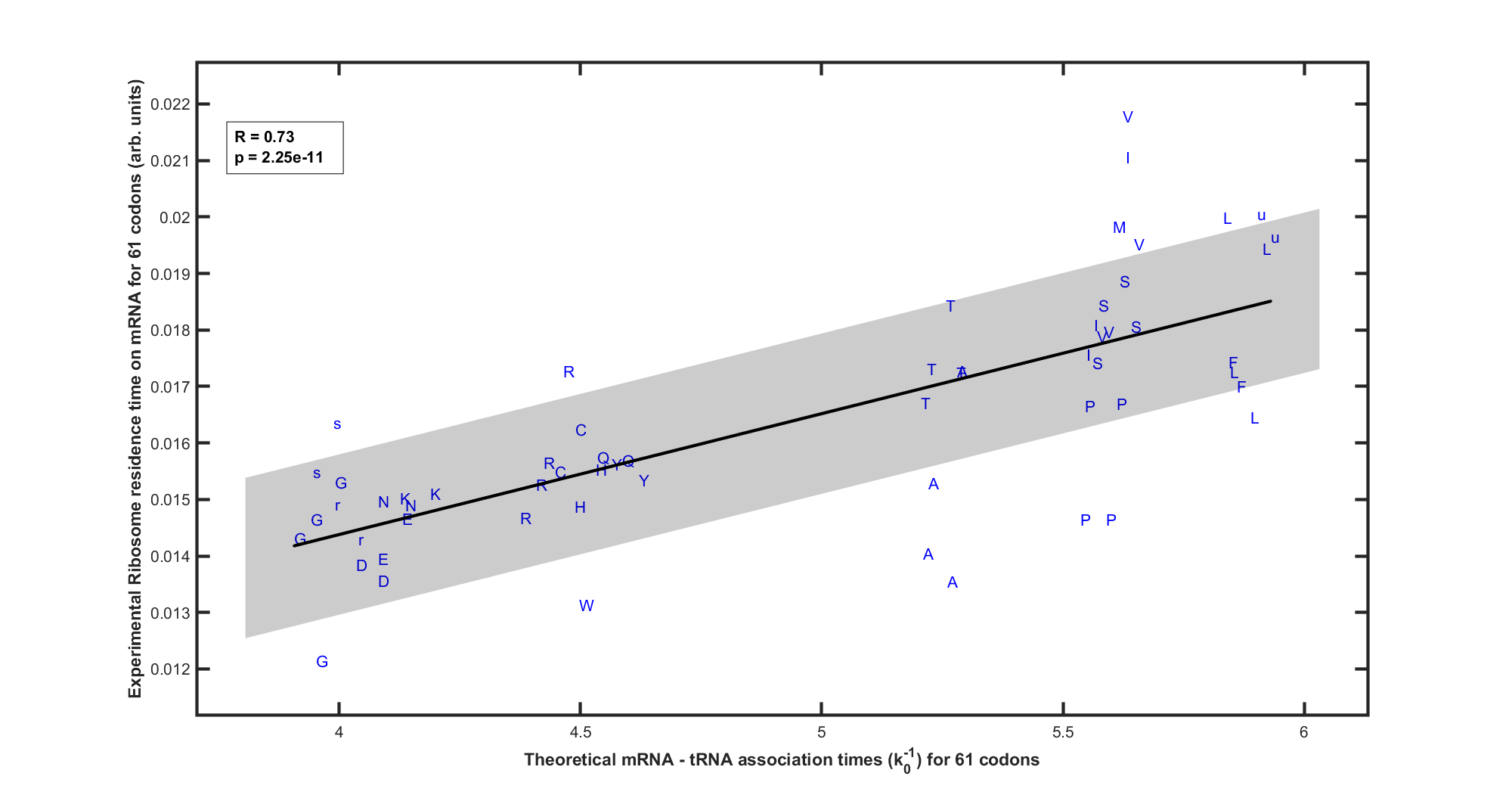}
\caption{Scatter plot between theoretically determined mRNA-tRNA association times and average ribosomal residence times calculated from data of 41 experiments. The residence times at E site are used to calculate the average, and for data from cycloheximide-treated samples (two of Lareau \textit{et al}'s and all of Stein \textit{et al}'s), where E site correlations are not present, the sum of site 0 and site 7 residence times are used. Amino acid single letter names corresponding to all the 61 codons are used as data markers in the scatter plot. For the three amino acids that code for both a quadruplet and a doublet, Arg, Leu and Ser, we use upper case letters R, L and S for the quadruplets, and lower case r, u and s for the doublets, respectively. The shaded area around the line of best fit extends one standard deviation on either side. It can be observed that the spread in the experimental data (y-axis), specifically for synonymous codons, is larger than the insignificant spread in the theoretical data (x-axis). We argue that this spread is due to wobble base pairing. Since the model clusters the genetic code into 24 clusters with certain symmetries, it does not segregate Ile ($AUA$) from Met or start codon ($AUG$), or Trp ($UGG$) from stop codon ($UGA$). This could be the reason for all these three codons appearing as outliers.  }
\label{correlation_scatter}
\end{center}
\end{figure}

The foregoing demonstration of the correlations between theory and experiments provides us sufficient confidence to suggest that the asymmetric cooperativity model, however simplistic, captures certain fundamental aspects of tRNA-mRNA interactions. In order to visualize the calculated correlations, it is instructive to look at the scatter plot of the averaged experimental residence times from ribosomal footprinting experiments versus theoretically calculated residence times. Figure \ref{correlation_scatter} shows a plot between the theoretically calculated tRNA-mRNA association times $T_c^{th}$ and the average relative residence times of ribosomes $T_{c}^{exp}$ , calculated over the E sites of $22$ datasets (non-cycloheximide-treated), and sites 0 and 7 for the other $19$ datasets (cycloheximide-treated), where we see pervasive correlations. The unweighted (weighted using standard deviation) correlation coefficient is 0.73 (0.75), with $p<10^{-11}$.

As can be seen in the figure, the spread of the experimental residence times of codons coding for the same amino acid is larger (along y-axis), compared to the spread in the theoretically predicted residence times (along x-axis). To identify the origin of this dependence of the experimental residence time on the identity of the nucleotide at the third codon position, we segregate the $61$ codons into four groups, each ending in one of the four nucleotides, A, U, G and C. The correlations between theoretical and experimental residence times of these four groups are shown in the fig. \ref{splitATGC}. It can be seen that the spread in the experimental residence times of G-ending codons are larger than that of the other three groups. To identify the origin of this spread, we isolate the outliers in all the four plots, by calculating the residuals, i.e., the difference between the experimental and the theoretically predicted residence times, and identifying the codons whose residuals lie outside \textit{one standard deviation} in the four plots. Please note that the standard deviation is calculated individually for the four plots. We identify 14 codons as outliers.

\begin{figure}[htbp]
    \centering
    
    \begin{subfigure}[b]{0.99\textwidth}
        \includegraphics[trim=40pt 20pt 20pt 20pt,width=\textwidth]{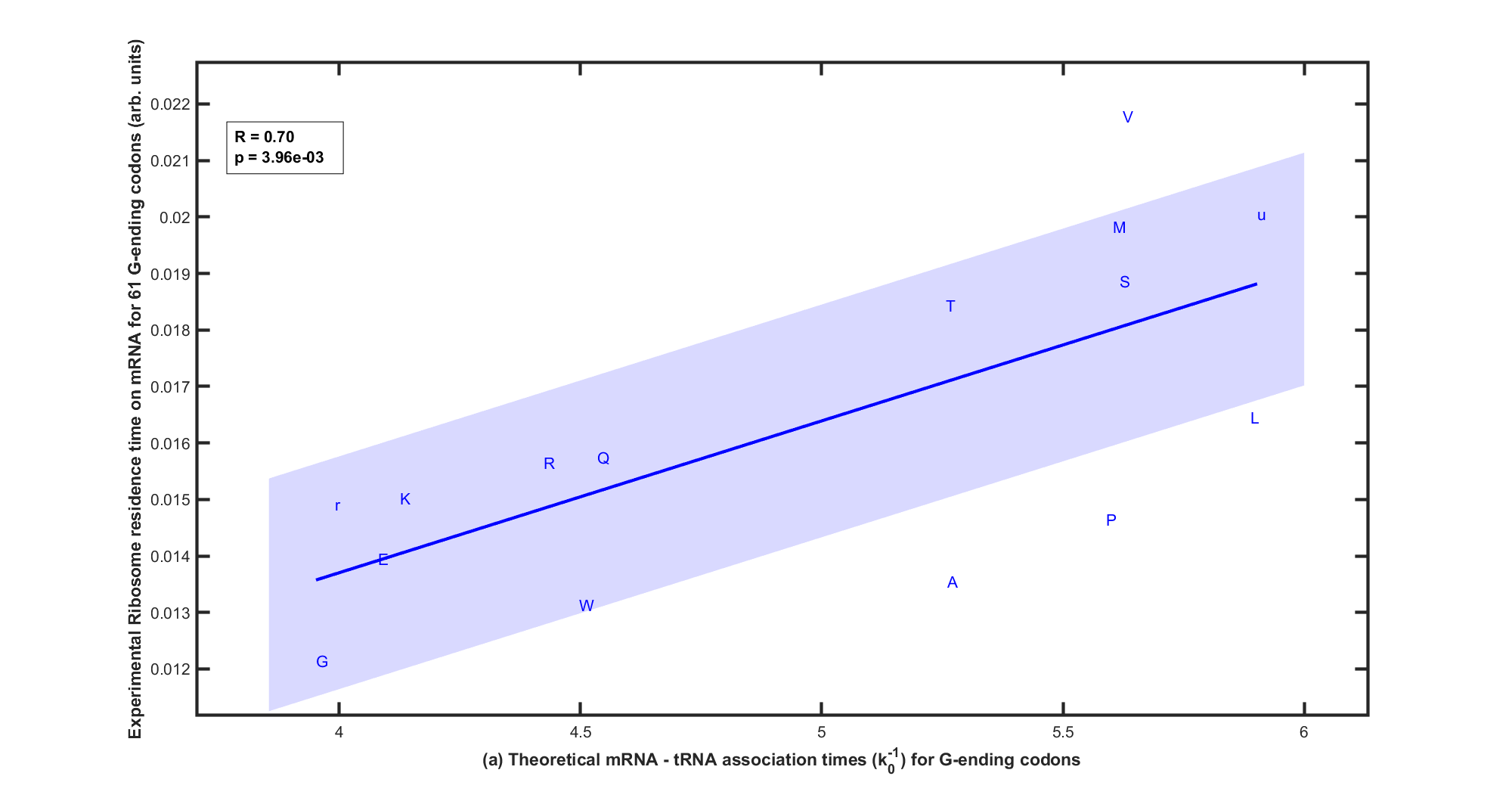}
       \caption{}
        \label{splitG}
    \end{subfigure}
    \hfill
    \begin{subfigure}[b]{0.99\textwidth}
        \includegraphics[trim=40pt 20pt 20pt 20pt,width=\textwidth]{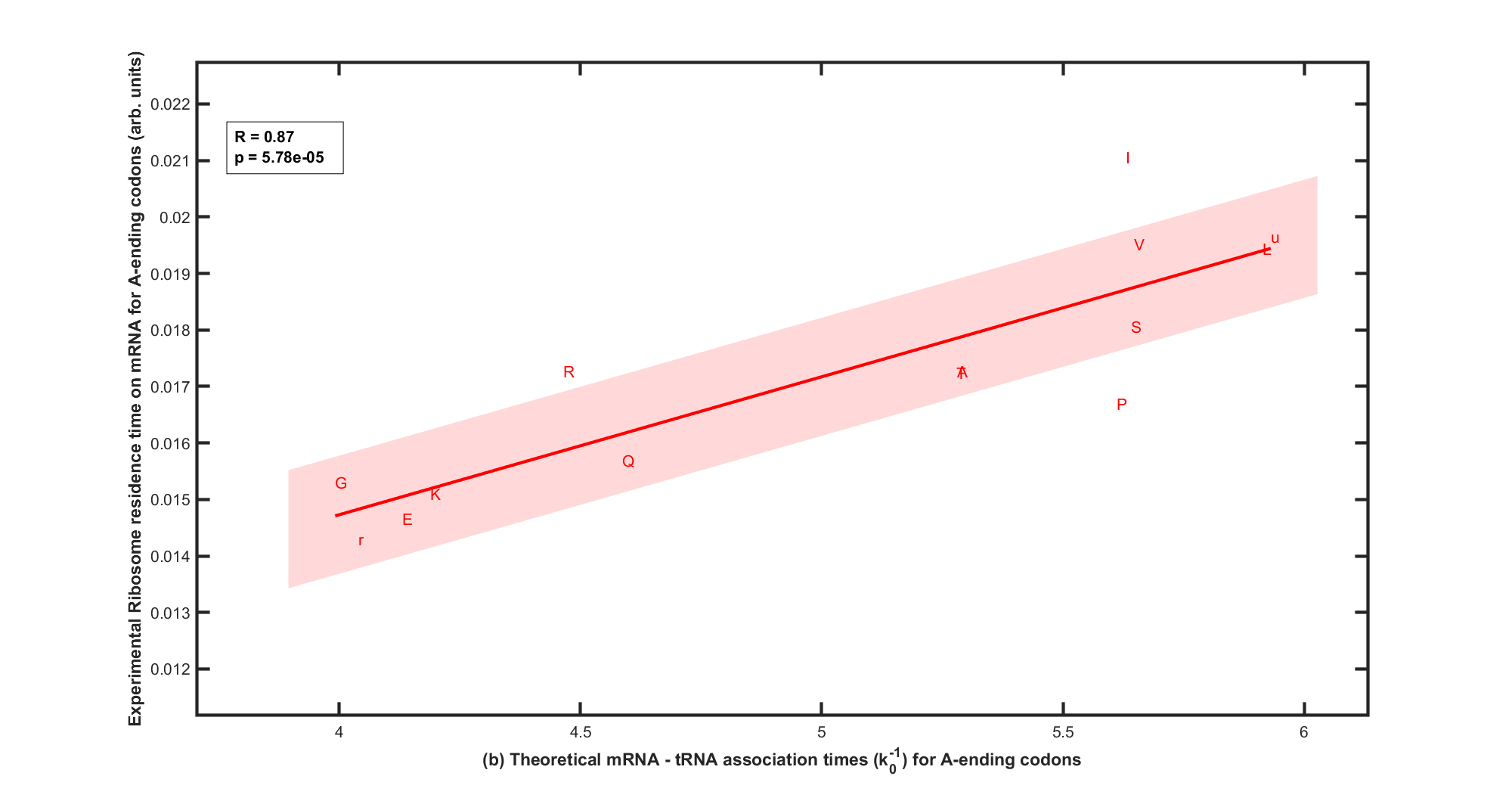}
        \caption{}
        \label{splitA}
    \end{subfigure}
   \end{figure}
		
		\begin{figure}
		\ContinuedFloat
		 \centering
    \begin{subfigure}[b]{0.99\textwidth}
        \includegraphics[trim=40pt 20pt 20pt 20pt,width=\textwidth]{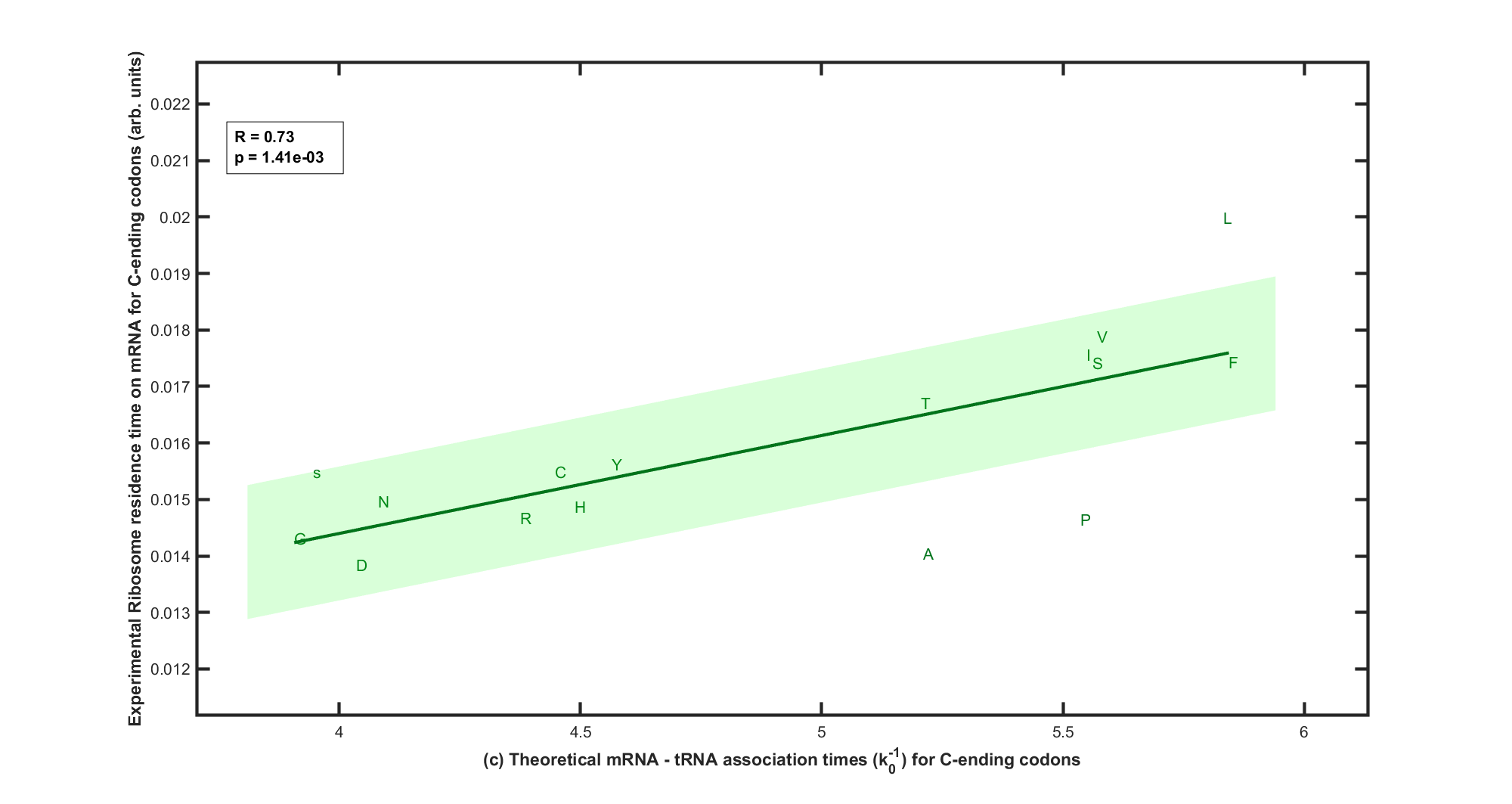}
        \caption{}
        \label{splitC}
    \end{subfigure}
    \hfill
    \begin{subfigure}[b]{0.99\textwidth}
        \includegraphics[trim=40pt 20pt 20pt 20pt,width=\textwidth]{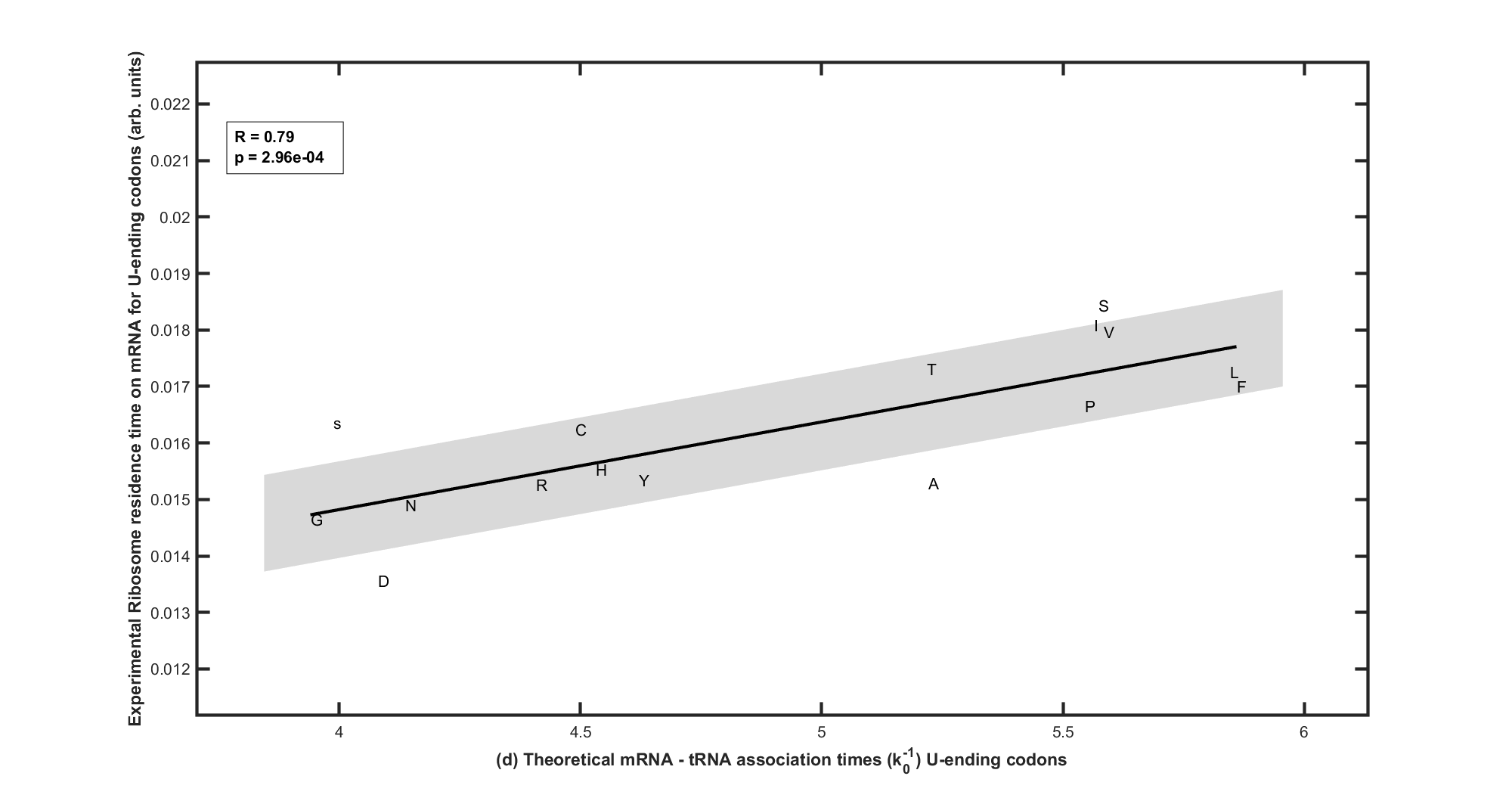}
        \caption{}
        \label{splitU}
    \end{subfigure}
    
    \caption{Correlation between experimental ribosomal residence times and theoretical mRNA-tRNA association times of all codons with the same wobble position base pair, in four subplots. (a) The spread in the data is larger for G-ending codons, compared to the other three sets. This large spread is primarily due to the anticodon U : codon G base pairing, resulting in large deviations ($>1 SD$) from the theoretical expectation for Ala, Leu, Pro and Val, of which the first three amino acids lack cognate tRNA isoacceptors to decode their corresponding codons, in Yeast. Gly, decoded both by U:G and a cognate isoacceptor, can be seen as a near-outlier as well. Trp and Met are also near-outliers, possibly because $AUG$ and $UGG$ are clubbed with $AUA$ and $UGA$ respectively in our model, to preserve the doublet-quadruplet symmetry. Only a subset of U:G wobble base pairing codons are identified as outliers, with most of the remaining codons decoded by cognate isoacceptors. (b) Arg, Ile and Pro are seen as outliers among the A-ending codons, whose third position involves wobble base pairing of codon U with I, $\Psi$ and A anticodon nucleotides respectively. Among these three amino acids, only Pro has a cognate isoacceptor. (c) Among C-ending codons, Ala, Pro and Leu are outliers, which base pair with I, U and U anticodon nucleotides respectively, with Leu having a cognate isoacceptor. (d) As for U-ending codons, four codons, two of Ser, Ala and Asp are identified as outliers. Only two codons of the four do not have cognate isoacceptors, one of Ser and Ala, which are decoded by I anticodon nucleotide, while, for the other two codons, Ser ($AGU$) and Asp ($GAU$), we do not have an explanation. Of the 14 outliers, 12 are observed to be wobble-base paired. Of the 29 possible wobble base pairs in Yeast, the model identifies 12 as outliers. Most of the wobble base pairs not identified as outliers are decoded by anticodons containing I. }
    \label{splitATGC}
\end{figure}

\begin{table}
{\footnotesize
\begin{tabular}{|c|c|c|c|c|c|}
\hline
\textbf{Codon} &  \makecell{\textbf{Wobble basepair}\\\textbf{Anticodon:codon} \cite{roth2012decoding,hani1998trna} } & \makecell{\textbf{Expt-theory}\\\textbf{Outlier ($>1 SD$)?}} & \textbf{Residual} & \makecell{\textbf{Frameshifting}\\\textbf{propensity ($>5\%$)}} \cite{vimaladithan1994special} & \makecell{\textbf{Cognate tRNA} \\ \textbf{isoacceptor available?} \cite{roth2012decoding} } \\ 
\hline
GCG (Ala) & U:G & Yes & \bf{-0.0036} & Yes & No (U:G) \\ 
\hline
GGG (Gly) & U:G & No & -0.0014 & Yes & Yes \\ 
\hline
\makecell{CUG, \\ CUC, \\ CUU (Leu)} & U:G/C/U & \makecell{Yes, \\ Yes, \\ No} & \makecell{\bf{-0.0023}, \\ \bf{0.0024}, \\ -0.0004 } & \makecell{Yes, \\ Yes, \\ Yes} & \makecell{No (U:G), \\ Yes, \\ No (G:U)} \\ 
\hline
\makecell{CCG, \\ CCC, \\ CCU (Pro)} & U:G/C/U & \makecell{Yes, \\ Yes, \\ No} & \makecell{\bf{-0.0033}, \\ \bf{-0.0024}, \\ -0.0006 } & \makecell{Yes, \\ Yes, \\ Yes} & \makecell{No (U:G), \\ No (U:C), \\ Yes} \\ 
\hline
GUG (Val) & U:G & Yes & \bf{0.0037} & Yes & Yes \\ 
\hline
ACG (Thr) & U:G & No & 0.0013 & No & Yes \\ 
\hline
AGG (Arg) & U:G & No & 0.0012 & No & Yes \\ 
\hline
CAG (Gln) & U:G & No & 0.0006 & No & Yes \\ 
\hline
UCG (Ser) & U:G & No & 0.0008 & No & Yes \\ 
\hline
CCA (Pro) & A:A & Yes	& \bf{-0.0020}	& No	& Yes \\
\hline
\makecell{GCU, \\ GCC (Ala)} & I:U/C & \makecell{Yes, \\ Yes} & \makecell{\bf{-0.0014}, \\ \bf{-0.0025}} & \makecell{No, \\ No} & \makecell{No (I:U), \\ No (I:C)} \\ 
\hline
\makecell{CGU, \\ CGC, \\ CGA (Arg)} & I:U/C/A & \makecell{No, \\ No, \\ Yes} & \makecell{-0.0002, \\ -0.0004, \\ \bf{0.0014}} & \makecell{No, \\ No, \\ Yes} & \makecell{No (I:U), \\ No (I:C), \\ No (I:A)} \\ 
\hline
\makecell{AUU, \\ AUC, \\ AUA (Ile)} & \makecell{I:U/C \\ \\ $\Psi$:A} & \makecell{No, \\ No, \\ Yes} & \makecell{0.0008, \\ 0.0005, \\ \bf{0.0023}} & \makecell{No, \\ No, \\ No} & \makecell{No (I:U), \\ No (I:C), \\ No ($\Psi$:A)} \\ 
\hline
\makecell{UCU, \\ UCC, \\ UCA (Ser)} & I:U/C/A & \makecell{Yes, \\ No, \\ No} & \makecell{\bf{0.0012}, \\ 0.0003, \\ -0.0007} & \makecell{No, \\ No, \\ No} & \makecell{No (I:U), \\ No (I:C), \\ Yes} \\ 
\hline
\makecell{ACU, \\ ACC (Thr)} & I:U/C & \makecell{No, \\ No} & \makecell{0.0006 , \\ 0.0002} & \makecell{No, \\ No} & \makecell{No (I:U), \\ No (I:C)} \\ 
\hline
\makecell{GUU, \\ GUC (Val)} & I:U/C & \makecell{No, \\ No} & \makecell{0.0007, \\ 0.0008} & \makecell{No, \\ No} & \makecell{No (I:U), \\ No (I:C)} \\ 
\hline
\end{tabular}
}
\caption{ Association between wobble base pairing codons in Yeast and their outlier status in the theory-experiment correlation plot (fig. \ref{splitATGC}). Anticodon U : codon G/C/U, anticodon I : codon U/C/A, A:A and $\Psi:A$ are considered wobble base pairs here. Outlier codons in fig. \ref{splitATGC} are the ones that lay $>1$ standard deviation away from the theory-experiment fit. Of the 14 codons identified as outliers in fig. \ref{splitATGC}, 12 are seen here as being wobble base paired, irrespective of whether cognate tRNA isoacceptors are available for decoding the codons or not. The column titled Residual provides the deviation of the experimentally determined ribosmal residence times of codons from the linear fit between experimental and theoretical residence times, in arbitrary units. The codons identified as outliers from the fit are also compared with codons identified as frameshift-inducing, with 7 out of 10 frameshift-inducing codons identified as outliers in fig. \ref{splitATGC}. }
\label{wobblecodons}
\end{table}

The first column of Table \ref{wobblecodons} lists all codons (29 in total) that are decoded using wobble basepairing, in yeast, irrespective of whether cognate tRNA isoacceptors are available for these codons or not. This wobble base pairing list is obtained from \cite{roth2012decoding,hani1998trna} . Information on whether cognate tRNA isoacceptors are available or not for a given codon, listed in the last column of the table \ref{wobblecodons}, was obtained from table 8 of \cite{roth2012decoding}. We identify anticodon U: codon G/C/U, anticodon I: codon U/C/A, anticodon A: codon A, and one case of anticodon $\Psi$ : codon A as wobble basepairs. I denotes inosine, the deaminated form of adenosine, A. Here we denote all uridines, whether post-transcriptionally modified or not, as simply U, and Pseudouridine  as $\Psi$, which is used in Ile tRNA anticodon cognate to $AUA$ codon, to avoid recognizing methionine $AUG$ \cite{roth2012decoding}. Note that we have not included anticodon G: codon U in the list, since this basepairing is accepted readily and is ubiquitous, whereas, U:G is universally disfavoured and this tendency appears intrinsic to the genetic code \cite{percudani2001restricted}.   Since only five of the 41 data sets were from worms (C. Elegans), we did not analyze wobble base pairing in worms, although the averaged data includes that of worms.

In column 3 of table \ref{wobblecodons}, we specify if the codons of column 1 are present as outliers in the experiment-theory comparison plots, fig. \ref{splitATGC}. Of the 14 codons isolated as outliers, 12 are seen to be wobble basepaired, with or without cognate tRNA isoacceptors, with two exceptions: Asp ($GAU$), and Ser ($AGU$). All the codons that are decoded by U:G base pairs and are without a cognate tRNA isoacceptor present in yeast, Ala ($GCG$),  Leu ($CUG$), Pro ($CCG$), are identified as outliers. Val ($GUG$) is decoded both by U:G base pair and a cognate isoacceptor, and yet it is identified as an outlier. Only four codons do not have cognate tRNA isoacceptors to decode them, in yeast:  Ala ($GCG$),  Leu ($CUG$), Pro ($CCG$), and Arg ($CGA$) \cite{roth2012decoding,hani1998trna}, and all of them are identified as outliers. Arg ($CGA$) is decoded by I:A wobble basepair, and is identified as an outlier as well in the figure. Similarly, Pro ($CCA$) is decoded by A:A wobble base pair, along with its cognate U:A base pair. Of the 16 codons that use I:U/C/A, A:A or $\psi$:A basepairing, the model identifies 6 as outliers. The case of Ile ($AUA$) needs more elaboration. Since the model is designed to cluster Ile ($AUA$) and Met ($AUG$) codons into a single doublet (see The Model above), this genetic code simplification might also have led to Ile ($AUA$) being identified as an outlier. Column 4 lists the residuals, the deviation of experimentally observed residence times from the theoretically calculated residence times for all outliers. Given the variety of post-transcriptional modifications of tRNA anticodon loop (table 2 of \cite{roth2012decoding}), which can alter the codon-anticodon association kinetics, we cannot expect the model to perfectly identify all the wobble basepairing codons as outliers.

The case of Serine doublet ($AGU$, $AGC$) is special: Of the three amino acids, Arg, Leu and Ser, that are decoded by six codons (a quadruplet and a doublet), Arg and Leu have the same nucleotide in the middle codon position in the quadruplet and the doublet, U and G, respectively. Therefore, the difference in the kinetics of codon-anticodon association between the quadruplet and the doublet is minimal. However, Ser is coded by $UCN$ quadruplet and $AGY$ doublet, whose middle nucleotides are significantly different, the first, pyrimidine and the second being purine. Since the association kinetics primarily differentiate between middle pyrimidines and middle purines, the kinetics of the $UCN$ quadruplet and the $AGY$ doublet are very different, and the doublet appears to attempt to match the kinetics of the quadruplet, making $AGU$ an outright outlier and $AGC$ a near-outlier. Thus, the only outlier, out of 14, that is left without an explanation is Asp ($GAU$).

It is gratifying to note that 12 out of 14 codons ($>85\%$) identified as outliers are wobble basepaired, which mostly explains the scatter seen in the experimental  residence times in fig. \ref{correlation_scatter}. Since we have not included wobble basepairing in the model, codons that wobble basepair with their near-cognate tRNA anticodons appear as outliers in the correlation plot, because of their altered kinetics. It has to be noted that our ability to identify a subset of all the wobble basepairs in Yeast arises from our choice to \textit{not fit} the experimental residence times by altering the parameters in our theoretical model. The parameters were optimized simply to reproduce the genetic code degeneracy structure, and the residence time data thus produced by the model was compared with the experimental residence times. If we had chosen to fit the model to the experimental residence times from riboseq instead, the model's correlation with experiments might have improved, but we would have lost the information about the outliers, produced by wobble basepairing and embedded in the experimental residence times.

In order to cross-validate our results, we have compared the list of outlier codons, extracted from the correlation plot between experimental and theoretical residence times in fig. \ref{splitATGC}, and listed as present in column 3, with the list of codons experimentally shown to induce frameshifting during translation (KK240 strain in Table 1 in \cite{vimaladithan1994special}). Since frameshifting has been shown to be caused by wobble basepairing \cite{farabaugh2000translational,lim1997analysis,bjork1999transfer}, such a comparison can help us validate our results above. The frameshifting-prone codons, identified in the experiment \cite{vimaladithan1994special}, with frameshift efficiency greater than $5\%$, are Leu ($CUU$, $CUC$, $CUG$), Val ($GUG$), Pro ($CCG$, $CCC$, $CCU$), Ala ($GCG$), Arg ($CGA$), and Gly ($GGG$).  It is noteworthy that there is a significant overlap between codons experimentally observed to promote frameshifting in Yeast and codons that appear as outliers in the above plots, particularly codons that are U:G wobble basepaired. Of the 10 frameshifting-prone codons, 7 are found as outliers in the plots \ref{splitATGC}.  These codons are identified in column 5 of table \ref{wobblecodons}. It is not clear why the rest of the wobble basepaired codons, most of them involving basepairing with inosine (I), do not induce frameshifting.

\subsection*{Summary}
In this section, we have shown that the genetic code structure (specifically, vertebrate mitochondrial code structure, which is the most symmetric of all codes) is dictated by codon-anticodon association kinetics. We quantify this kinetics by theoretically evaluating the time taken for the association process to go from fully dissociated state to fully hydrogen-bonded state, and the time the process spends in all the intermediate stages of the association process. We have used a model we have constructed earlier \cite{subramanian2018evolutionary,subramanian2020evolutionary} to make sense of unidirectional DNA strand construction and bidirectional replication fork progression, the asymmetric cooperativity model, to evaluate the residence times of the codon-anticodon association process. The model posits the existence of nearest-neighbor \textit{asymmetric} kinetic influence of a basepair, where, the basepair influences the association/dissociation kinetics of its nearest neighbors to the left and right, unequally. This asymmetry leads to the $5'$-end of the codon treated differently from the $3'$-end, and results in the wobble position degeneracy, as shown in part-I of this series.

The 64 residence time vectors (each vector storing the residence times at all seven intermediate states of a codon-anticodon association process) were subjected to cluster analysis. The four asymmetric cooperativity parameters, which quantify the sequence-independent and sequence-dependent kinetic influence of a basepair on its neighbors to the left and right, were optimized to reproduce the clustering of codons observed in the genetic code, capturing the full symmetry of the genetic code. The fig. \ref{clustering} shows the result of a successful optimization, with the 24 clusters exhibiting the full symmetry observed in the simplified genetic code. There are infinitely many optimized parameter sets that reproduce the genetic code structure, requiring us to arbitrarily fix one of the four parameters to obtain a unique optimal set.

In order to validate our theoretical results, we have compared (not fit) the theoretically evaluated residence time data (the time taken for codon on mRNA and anticodon on tRNA to basepair) with experimentally obtained residence times of 61 codons (the relative times that ribosomes spend on each codon on each site inside the ribosome), from 41 datasets obtained from 31 experiments. We have shown that the correlation between the theoretical and experimental residence times are significant, particularly at the E site of the ribosome, where confounding processes such as tRNA waiting times and peptide bond formation times are absent (fig. \ref{correlationmatrix}). We also find significant correlations near the edges of the ribosome, whose origin is unclear. We have explained how experimental conditions, specifically the use of antibiotics such as cycloheximide and anisomycin, affect the residence times, whose signatures are evident in the correlations. We have shown that the outlier codons in the correlation plot between theoretical and experimental residence times are predominantly wobble basepaired, thereby providing us a way to identify wobble basepairing from riboseq data for other organisms.   

\subsection*{Discussion}
The genetic code, argued to have been present in the Last Universal Common Ancestor, is used by every single life form on Earth, albeit with minor variations. Its universality and persistence suggest that its purpose of serving as a map between two sets of informational heteropolymers is central to life. Although the mapping between the two systems, nucleotides and amino acids, has been deciphered 60 years ago, the biophysical origin of the map itself hasn't been deciphered. Any hope of purposefully altering the genetic code rests on our ability to thoroughly understand the biophysical basis of the codon-amino acid assignments at a fundamental level.

Here, we have used the previously introduced Asymmetric Cooperativity model to recreate the degeneracy structure of the genetic code. It is surprising that a simple kinetics-based model, created to understand the evolutionary need for unidirectional replica strand construction of DNA/RNA and bidirectional replication fork progression, could reproduce the degeneracy structure of the genetic code, with just four free parameters. More remarkable is the fact that the residence times derived theoretically from the optimization procedure to reproduce the degeneracy structure of the genetic code, matches significantly, without fitting, with the experimentally observed residence times from sophisticated ribosome profiling experiments, and even enables identification of wobble base pairing and the effects of different antibiotics on ribosome function.

By reproducing the genetic code structure without assuming the presence of  ribosomes, we have effectively eliminated the contribution of ribosome to the emergence of genetic code structure, and moved its emergence back to a time when the interactions between the nucleic acids and the amino acids were possibly mediated by simpler physical forces. We can conclude, with some confidence, that the ribosome is an interpreter of the genetic code, not its creator. This conclusion helps resolve the chicken-and-egg paradox of which came earlier, the ribosome or the genetic code. The ribosome was invented by evolution possibly to provide a predictable environment within which these simpler forces can operate without the blurring effects of environmental interference on the code. The correlation demonstrated between the theoretically derived association times  between mRNA and tRNA and the experimentally observed ribosomal residence times suggest that the fundamental processes governing the kinetics of the translation machinery, at specific ribosomal sites, are governed by the codon-anticodon binding, which then dictate the ribosomal dynamics at these sites at lower spatiotemporal resolutions.

A fruitful by-product of the observed correlations between theoretical and experimental residence times is that the theoretical residence times (association times) can serve as a cheap, experiment-independent benchmark against which future riboseq experiments can be validated, at least as a crude first step. More over, due to the ability of the theoretical data to differentiate between the actions of different antibiotics on the ribosome residence times at distinct sites, as demonstrated above, these data can also be used for examining the actions of various other antibiotics that are designed to interfere with the ribosomal functions at specific sites. Since the correlations of residence times at two ribosomal sites (sites 2 and 3) with the theoretical residence times appear to report on ribosomal translocation dynamics, these correlations can be used for interrogating the translocation dynamics as well. The correlations can be used to partially identify wobble base pairing in different organisms, as demonstrated above for yeast. More over, the data can help identify the correct reading frame of translation and programmed frameshifts in riboseq experiments as well. This can be done by identifying the reading frame that produces the highest correlation with the theoretical data.

\subsection*{Open Questions}

First, why is the association \textit{time} between the codons and anticodons appear as the organizing variable to recreate the genetic code structure? What is the biophysical mechanism governing the interaction between nucleic acids and amino acids where time appears as the central variable of consequence? Energy transfer between these two molecular systems may be a possible answer, since the donor lifetimes determine the efficiency of energy transfer. In part -III, we argue, with some experimental evidence, that this could be the case. However, a direct experimental proof of such a phenomenon is lacking, and is crucial to test the model's veracity.

Second, there appears a mismatch between the timescales of the codon-anticodon association and the timescales of ribosome residence on mRNA. Although we have avoided providing a magnitude for $k_0^{-1}$, the timescale of codon-anticodon association time, it is clear \cite{nibbering2004ultrafast,asbury2004dynamics} that the hydrogen bond formation and dissociation, which the model simulates, happens at timescales of picoseconds, whereas, ribosomal movement on mRNA happens at timescales of milliseconds to seconds \cite{Sharma2019}. The possibility of energy transfer between nucleic and amino acids also hint at picosecond timescales for codon-anticodon association, above which solvation and local dissipation dominates, for these systems \cite{bagchi2010solvation}. What process(es) bridges these two highly disparate timescales?  In part-III, we argue that the multifractal nature of ribosomes, and proteins in general, may provide an answer.

Third, the appearance of correlation of experimental and theoretical residence times at the ribosomal edges is intriguing. Although this is not the first article to note the importance of ribosomal edges in determining its residence times, we have provided specific evidence that mRNA-tRNA interactions might be happening at the edges as well. This however contradicts the current understanding of ribosome function. Further bottom-up investigations of ribosome dynamics are needed to shed more light on this intriguing observation.

\clearpage
\begin{center}
    {\Large\bfseries III: Non-equilibrium Origin} \\[0.5cm]
\end{center}

\section*{Background}
The origin of the genetic code, or the biophysical basis behind the mapping between 64 codons and twenty or so amino acids, remains unsolvable, despite considerable attention having been paid to the problem since its discovery in the 1950s. Multiple hypotheses have been put forward to solve this ``forbiddingly difficult'' \cite{koonin2017frozen,Koonin2009} problem: Steriochemical hypothesis, asserting that the codon assignments reflect the chemical affinities between amino acids and their cognate (anti)codons \cite{Yarus2009,Johnson2010}, Coevolution theory, positing that the code’s structure reflects biosynthetic relationships among amino acids \cite{Wong2005}, error-minimization hypothesis, resting on the observation that the standard code is much better than random codes at minimizing the phenotypic impact of single-base mutations or mistranslations \cite{Freeland1998Genetic}, and the frozen-accident hypothesis \cite{Crick1968}. Each of these hypotheses appear to explain certain, limited facets of the genetic code puzzle, such as the colocation of (anti)codons and their cognate amino acids in ribosomes, the observed precursor-product relationships between amino acids and their related codon assignments, the code's superior error tolerance, and the universality of the genetic code. However, the narrow scope and insufficient data supporting these hypotheses prevent them from being used to confidently construct a model of early life or to rationally alter or expand the codon-amino acid assignments for medicinal purposes.

A reasonable requirement of a testable hypothesis may be stated as its ability to explain both the mechanistic (Aristotle's material cause) and evolutionary (the final cause) underpinnings of a biological phenomenon, since, without the latter, the former loses its purpose to be present in living systems. Although the above hypotheses of the emergence of genetic code attempt to explain the mechanistic basis of the codon-amino acid assignments, such as the preferential affinity of an amino acid for its cognate anticodon, they do not explain why such assignment was ultimately, evolutionarily useful for the organism. The induction of amino acids into the predominantly RNA-based early lifeforms is currently explained as due to the superior catalytic abilities of proteins over RNA, wrongly attributing foresight to a process that is blind to the future. A causally reasonable explanation of genetic code should demonstrate immediate accrual of benefit to the organism upon induction of amino acids, even before machinery catalyzing peptide bond formation between amino acids were invented by evolution.  


Below, we provide a mechanistic explanation for the emergence of the genetic code that additionally identifies an evolutionary rationale for the recruitment of amino acids by RNA-based living systems. Our approach to identify the biophysical basis of genetic code is to utilize a model constructed earlier to justify certain molecular properties of RNA/DNA, such as their unidirectional replication and anti-parallel orientation, as arising from the evolutionary force for high replication rates \cite{subramanian2018evolutionary,subramanian2020evolutionary}. Since the model assumptions allow for the replication rates (specifically, (un)zipping rates of the double strands) to vary with the DNA/RNA sequence, information becomes important for evolution even before the emergence of the genetic code. We speculated that the sequence characteristics-driven (un)zipping dynamics of DNA/RNA double strands might also underlie the emergence of the genetic code. Utilizing a simplified version of this kinetics-based asymmetric cooperativity model, in part I of this series (Title: Kinetic basis of the genetic code: I - wobble position degeneracy), we have shown that the wobble position degeneracy, wherein codon sequences that differ only in their third position generally code for the same amino acid, is the result of clustering of \textit{association times} of codon-anticodon basepairing of all such degenerate codons. We showed that the 64 codons clusters into four groups: SSN, WWN, SWN and WSN, where W stands for weak base-pair between A and T/U, S for strong base-pair between G and C, and N denoting any of the four nucleotides, without providing any preferential treatment to the wobble position. This surprising result pointed to the possibility that the wobble position degeneracy, and perhaps the entire organizational structure of the genetic code, is based on the \textit{kinetics} of the codon-anticodon base-pairing process.

Making use of the full asymmetric cooperativity model, which additionally included sequence-dependent cooperativity, we have demonstrated in part-II of this series that the clustering of codon-anticodon association times can be matched with that of the degeneracies observed in the genetic code by tuning \textit{four} asymmetric cooperativity parameters, thereby theoretically reproducing the entire organizational structure of the highly symmetric vertebrate mitochondrial genetic code (Title: Kinetic basis of the genetic code: II - Quantitative reproduction of the codon table structure and its experimental validation). We have also shown that significant correlations between the theoretically measured codon-anticodon association times and experimentally observed codon residence times within ribosomes in ribosome profiling (riboseq) experiments in \textit{Yeast}, specifically at the E site of the ribosome, where codon-anticodon interactions, unencumbered by other interactions, are expected to happen. This demonstration makes the relationship between the kinetics of codon-anticodon association process and the genetic code concrete.


In this third part, by identifying correlations between various biophysical properties and their associated relaxation times of individual amino acids on one hand, and the theoretically-derived residence times at various states of their corresponding cognate codon-anticodon association process, on the other, we provide tentative evidence that \textit{electronic energy transfer} happens between these two entities, at multiple timescales, at the fastest possible rates. We argue with evidence that the matching of timescales of energy supply from the codon-anticodon association event and the energy dissipation event at the cognate amino acid end suggest that evolution chose cognate amino acids in order to maximize the rate of energy transfer, akin to the critical damping regime in simple harmonic motion, where the system returns to equilibrium after excitation in the shortest possible time. Through these associations, we highlight the importance of non-equilibrium processes in the emergence and maintenance of spatiotemporal processes and structures associated with living systems.

\section*{Methods}
The modeling of the codon-anticodon association process, in the presence of asymmetric cooperativity, is done using a Markov chain model, as elaborated in part II. The association process of the codon-anticodon system going from completely unzipped state to fully zipped state is characterized by the residence times of the Markov chain, the average time that the chain spends in each of the seven intermediate states corresponding to partially zipped states. These seven residence times for each of the 64 codon-anticodon basepairs, provided in part II, constitute all the data needed in this part for the codon-anticodon system. On the amino acid end, various physicochemical properties of the 20 amino acids constitute our data set. These are collated from extant literature, which are referenced at appropriate locations. It has to be emphasized that no fine-tuning of parameters or fitting is involved in generating the correlations below, and the parameters used to cluster the residence times to reproduce the genetic code structure in part II are not altered. Throughout the section, a p-value less than 0.05 is taken to be significant.

Since this section is about finding the biophysical mechanism behind the mapping between these 64 codon-anticodon basepairs and 20 amino acids, we need to pick a specific genetic code among the many available genetic codes that differ from each other in only minor details. We pick the vertebrate mitochondrial genetic code \cite{Delarue2007}, since this appears to be the most symmetric of all alternate codes available. In this code, two amino acids, Ser (S) and Leu (L), are coded by six codons. We ignore the doublet codons (AGY, UUR) for these amino acids and consider only the four-fold degenerate codons (UCN, CUN) for the analysis below.

Confounding factor analysis was not done for the correlations we report below for the following reason: Our primary aim is to correlate the theoretically calculated residence times in various states of codon-anticodon association process with various biophysical properties of the 20 proteinogenic amino acids, in order to hypothesize mechanisms behind the origins of the genetic code. Since many of these amino acids' biophysical properties are correlated due to functional connections between the biophysical variables, such as between volume and polarizability, heat relaxation times and hydropathy \cite{Hamzi2022}, HOMO-LUMO gaps and softness \cite{Pearson2005}, molecular size and dipole moments, vibrational times and molecular weight etc, identifying the biophysical variable that causes the correlation with the residence times is not straight-forward. In cases where we suspect the presence of a confounding variable, we pick the biophysical variable that has dynamical relevance, at the appropriate timescale, for the mapping with the codon-anticodon residence times (say, polarizability or heat relaxation times), while ignoring the inert variable (say, volume or hydropathy).



\section*{Results}

\begin{figure}
\centering
\includegraphics[trim=80pt 20pt 20pt 20pt,width=1.05\textwidth]{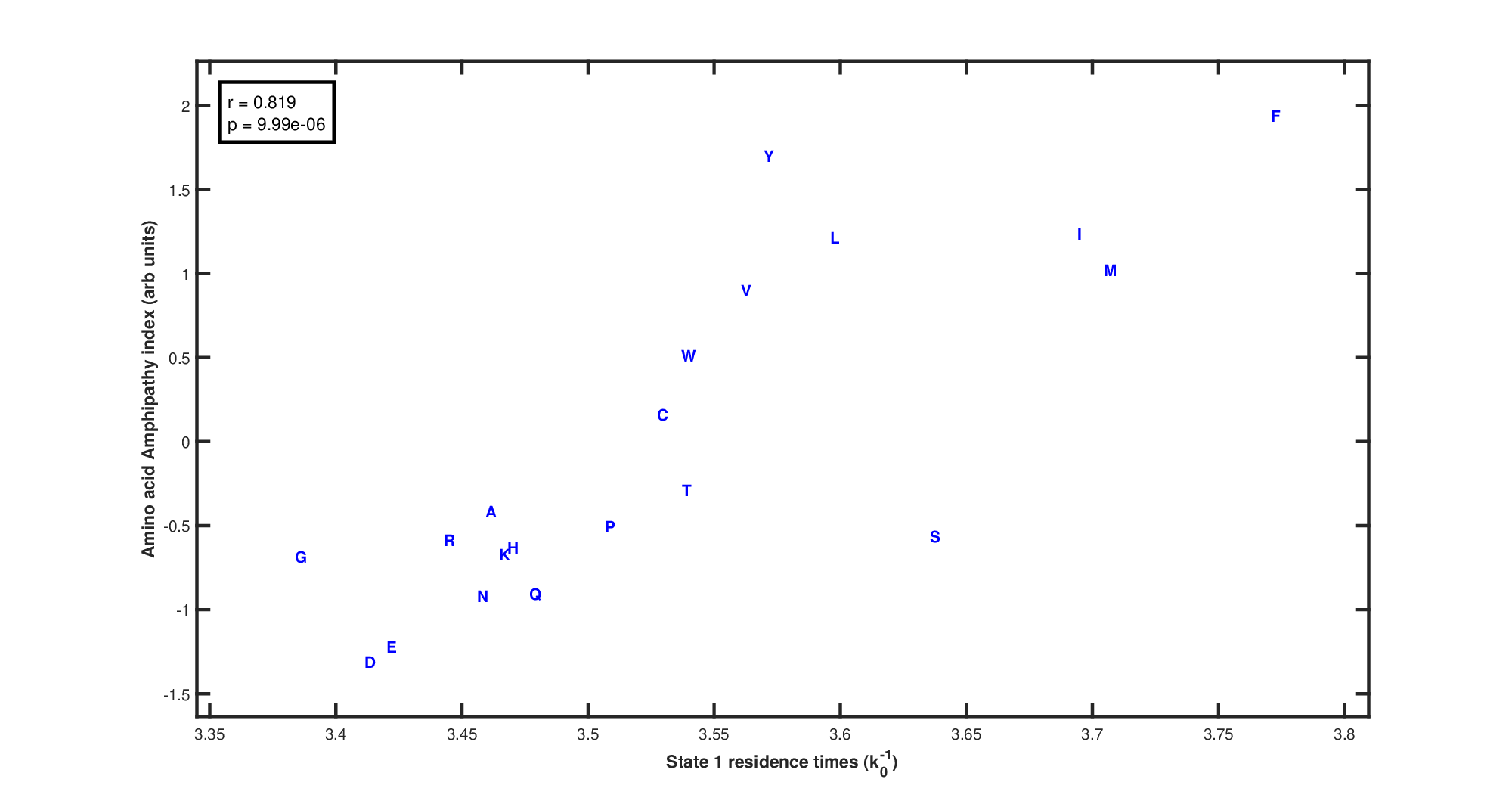}
\caption{Correlation between amino acid amphipathy index \cite{Cornette1987} and average state 1 residence times of corresponding cognate codon groups. Such correlations between amino acid properties and codon sequence structures have been widely reported since 1970s \cite{Woese1973,Taylor1989}, although not explicitly quantitatively as here. Our model corroborates such observations; however, it does not rely on them, since such indices constructed through complicated analyses of protein structures do not help point us towards a biophysical mechanism or evolutionary rationale for the emergence of the genetic code, during a prebiotic era when protein construction has not yet evolved. The Pearson correlation r and the p-value are given in the figure. The correlation is seen to be significant.}
\label{amphipathy}
\end{figure}

At the outset, we would like to note that, since the aim of this work is to identify the biophysical mechanism behind the mapping of codons and amino acids, we concentrate only on the properties of amino acids that provide mechanistic insight into the coupling between the amino acids and codons that possibly led to the genetic code mapping. Specifically, since we are aiming to identify amino acid properties that correlate with the \textit{residence times} of codon-anticodon association process, we narrow our search for amino acid properties that are either \textit{dynamical} or directly influence the electronic or vibrational dynamics of amino acids. Therefore, although we find strong correlations of the codon-anticodon residence times with amino acid properties such as hydropathies (fig.\ref{amphipathy} \cite{Cornette1987}) or volume, extracted from well-known databases such as AAindex (\url{https://www.genome.jp/ftp/db/community/aaindex/aaindex1}), we do not concentrate on them because they provide little insight into the dynamical processes that underlie the origin of the genetic code. These correlations have been widely reported in the literature since the 1970s \cite{Woese1973,Taylor1989}, without leading to much significant insight into the origin of the genetic code later. Although it is gratifying to observe that the model robustly reproduces these oft-noted correlations with amino acid properties such as hydropathies, we limit ourselves below to examining microscopically meaningful amino acid properties that we deem to lie causally upstream of these previously reported correlations. For the same reason, we avoid using statistical techniques such as Principal Component Analysis to identify linear combinations of amino acid properties that correlate with the codon-anticodon residence times, since such combinations are difficult to interpret mechanistically. Our choice of relevant amino acid properties below are guided more by physical intuition and parsimony of explanation for the emergence of the genetic code, than by identification of properties that maximize statistical correlations, devoid of mechanistic insights.  

\begin{figure}
\centering
\includegraphics[width=0.9\textwidth]{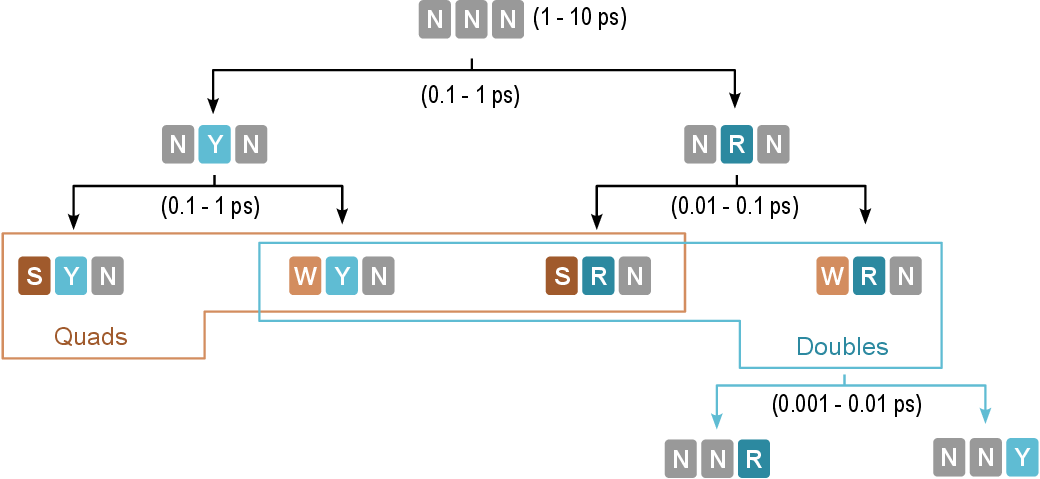}
\caption{A schematic, temporal tree structure of the codon table organization, according to the residence times during the codon-anticodon association process, theoretically calculated using the asymmetric cooperativity model. At the root of the tree, at level 1, lies NNN, representing all 64 codons. This level corresponds to the timescale of 1 - 10 picoseconds, the timescale of the total residence times of the association process. This timescale is arrived at from the correlations observed between the total residence times of the codons and heat relaxation times and vibrational timescales of the cognate amino acids (see below). The processes that underlie the segregation between NYN and NRN codon clusters, at level 2, happens at the timescale of 0.1 - 1 picoseconds, according to the model. Level 3 corresponds to the segregation of NYN into SYN and WYN clusters, which also has the same timescale of 0.1 - 1 picoseconds. NRN splits into SRN and WRN at timescales of 0.01-0.1 picoseconds, at level 4. We introduce a mezzanine level 5M, for instructional clarity, segregating all the codons in the four clusters SYN, WYN, SRN and WRN into two clusters, quadruplets and doublets. This level does not have a corresponding timescale. At the final level 5, the doublets are split into NNR and NNY, with the shortest timescale of 0.0001 - 0.01 picoseconds. Below, we identify various physicochemical properties of amino acids that correlate with the residence times at different intermediate states, corresponding to different timescales.}
\label{codon_org}
\end{figure}

The organization of this section follows the organizational structure of the genetic code, which is depicted in the fig. \ref {codon_org}. In order to emphasize the centrality of \textit{time} in dictating the structure of the genetic code, we begin by concentrating on the slowest time scale processes involved in the codon - amino acid mapping, and then proceed down to faster time scales. As shown in the figure, for the purposes of this section, the genetic code is organized as a hierarchical structure of five levels, with four distinct ranges of timescales. Below, we first identify the timescale of each level, and find one or more physicochemical properties of amino acids that correlate with the residence times of the corresponding cognate codon-anticodon association process. We then speculate, assuming that there is causation behind the observed correlations, on the biophysical processes that may lead to such correlations.

\subsubsection*{First level: NNN, 1 - 10 picoseconds}

\begin{figure}
\centering
\includegraphics[trim=80pt 20pt 20pt 20pt,width=1.05\textwidth]{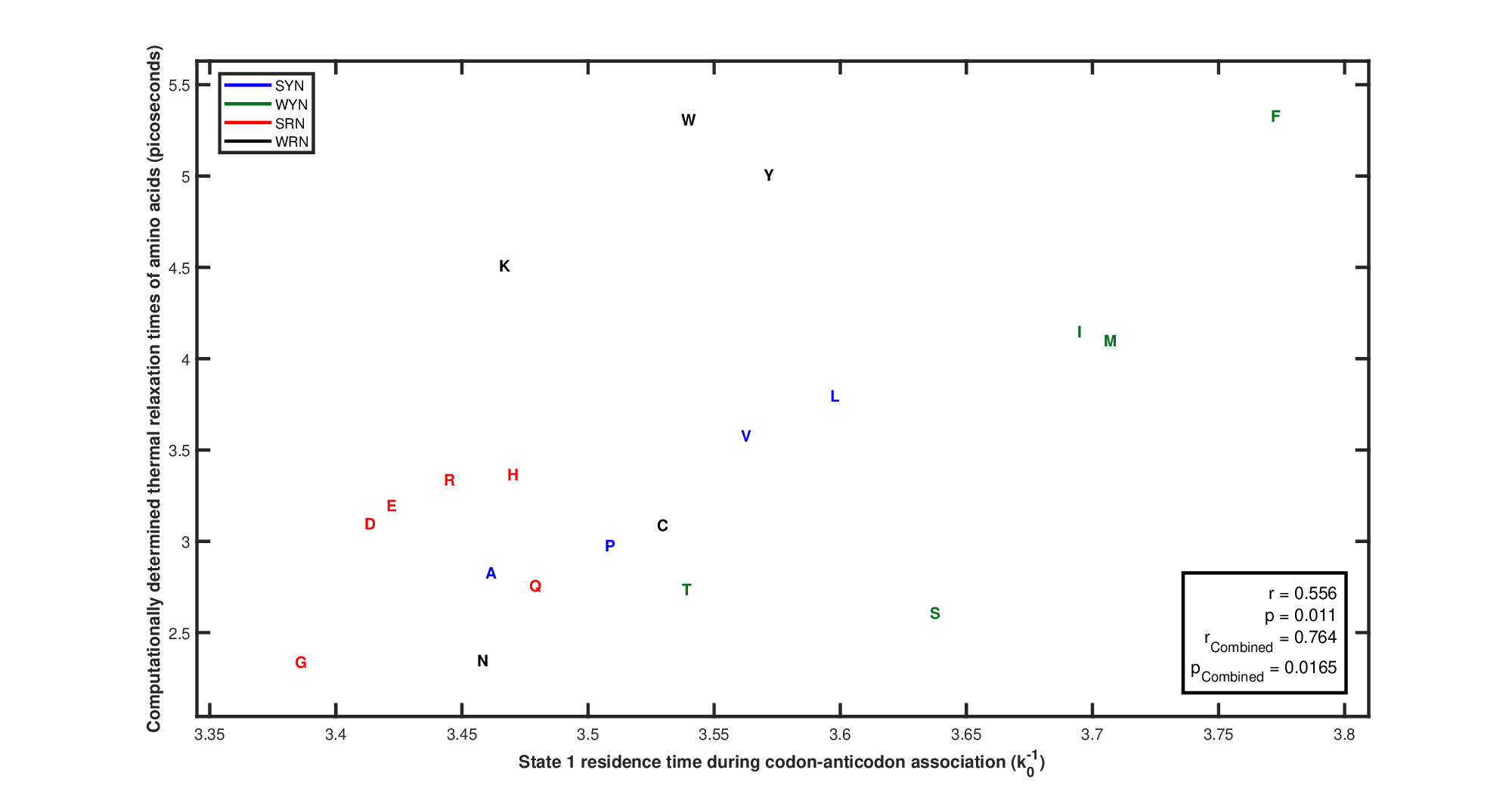}
\caption{Computationally evaluated thermal relaxation times of 20 amino acids in water, in picoseconds, against the residence times of codon-anticodon association process in state 1. Amino acid relaxation times were computed using transient non-equilibrium molecular dynamics simulations. Synonymous codons for each amino acid are averaged, with the doublet codons coding for Ser and Leu excluded. Vertebrate Mitochondrial genetic code is used for averaging, as it is the most symmetric of all codes. Amino acids are segregated by color into four groups, coded by SYN, WYN, SRN and WRN codons. Pearson r and p-values are given for both the entire dataset of 20 amino acids and for the case where each of the four groups above are fit individually and combined using Fisher's method. Both methods result in statistically significant correlation. The correlation seen in the plot suggests evolutionary matching of energy supply rates from the codon-anticodon association process to the energy dissipation rates at the amino acid end into the aqueous medium, possibly arising from evolution selecting for efficient energy transfer. This correlation also helps us to set the timescale of codon-anticodon association process to be of the order of picoseconds. Heat relaxation times data from ref. \cite{Hamzi2022}.}
\label{heatrelaxationtimes}
\end{figure}

The first level is simply the entirety of the 64 codons, denoted by NNN in the fig. \ref{codon_org}. In contrast to other states (see below), state 1 residence times do not segregate the 64 codons into distinct groups. This level is included to present correlations of state 1 residence times of codon-anticodon association process with various properties of cognate amino acids. One of the central results of this paper is the surprising finding of the existence of an approximately linear relationship between the energy supply timescales from the codon- anticodon association process and the energy dissipation timescales at the receiving amino acid end.  Fig. \ref{heatrelaxationtimes} shows a plot between the residence times in state 1 of the codon-anticodon association process and the heat dissipation times of amino acids \cite{Hamzi2022}. Heat relaxation times of all 20 proteinogenic amino acids were computed in ref. \cite{Hamzi2022} using transient non-equilibrium molecular dynamics simulations. Approximate linearity between the residence times and the dissipation times of amino acids, grouped into 4 sets coded by SRN, WRN, SYN and WYN codons, is seen. The segregation into four groups is done solely because it brings out the linearity better. Pearson r and p-values for both the full dataset of size 20, and of individually calculated r and p for each of the four groups of SRN, WRN, SYN and WYN codons and combined using Fisher's method are provided in the fig. \ref{codon_org}. The correlation is seen to be significant. We find that segregating the dataset into four groups result in better statistics. An explanation for this observation will be provided later.  For now, let us simply note that, using this result, we can fix the time scale of our theoretical results, which has hitherto remained undetermined.  Thus far, the residence times have been calculated on the scale of $k_0^{-1}$ (see part I), the timescale of hydrogen bond formation between two ribonucleotides within the codon-anticodon system. From the figure, we can tentatively conclude that, since both X and Y axes have dimensions of time, parts of the codon-anticodon association process should happen at the time scale of 1 to 10 picoseconds.

\begin{figure}
\centering
\includegraphics[trim=80pt 20pt 20pt 20pt,width=1.05\textwidth]{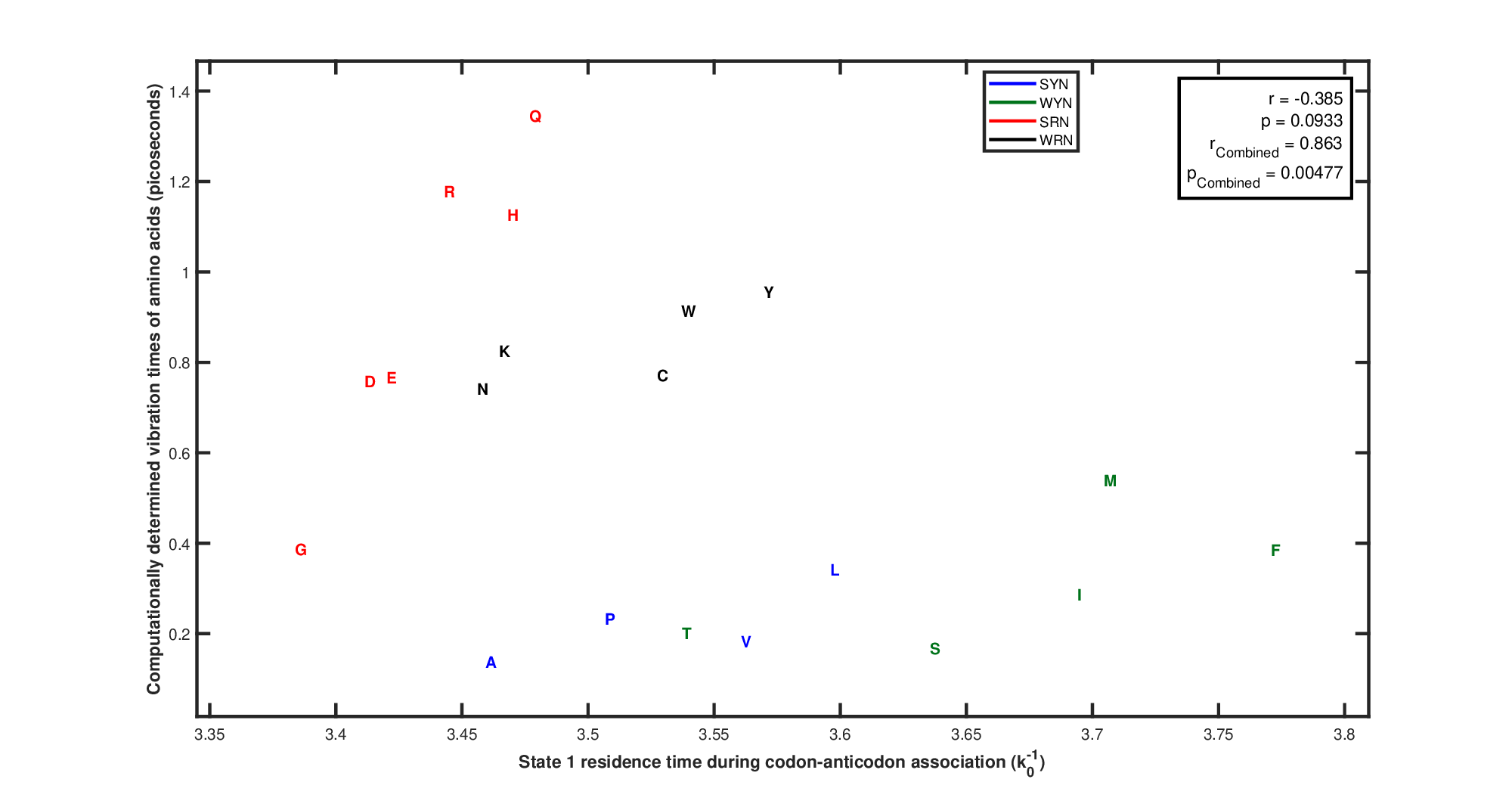}
\caption{Vibrational time periods in picoseconds, of 20 proteinogenic amino acids in zwitterionic form, computationally determined using semi-empirical quantum chemistry methods, is plotted against the residence times of codon-anticodon association process in state 1, calculated in part II of this article. Synonymous codons for each amino acid are averaged. For NRN-coded amino acids, the lowest vibrational mode data is used, whereas, for NYN-coded amino acids, the third-lowest mode is used. Linearity is observed when the 20 amino acids are segregated into four groups, coded by SRN, WRN, SYN and WYN codons, and are quantified in the $r_{combined}$- and $p_{combined}$ values. This p-value is statistically significant. Doublet codons coding for Ser and Leu are excluded. The observed correlation can be used to fix the timescale of the codon-anticodon association process to be of the order of picoseconds. Amino acid vibrational frequency data from table 4 in ref. \cite{Scherbakov2016}.}
\label{compvibtimes}
\end{figure}

\begin{figure}
\centering
\includegraphics[trim=80pt 20pt 20pt 20pt,width=1.05\textwidth]{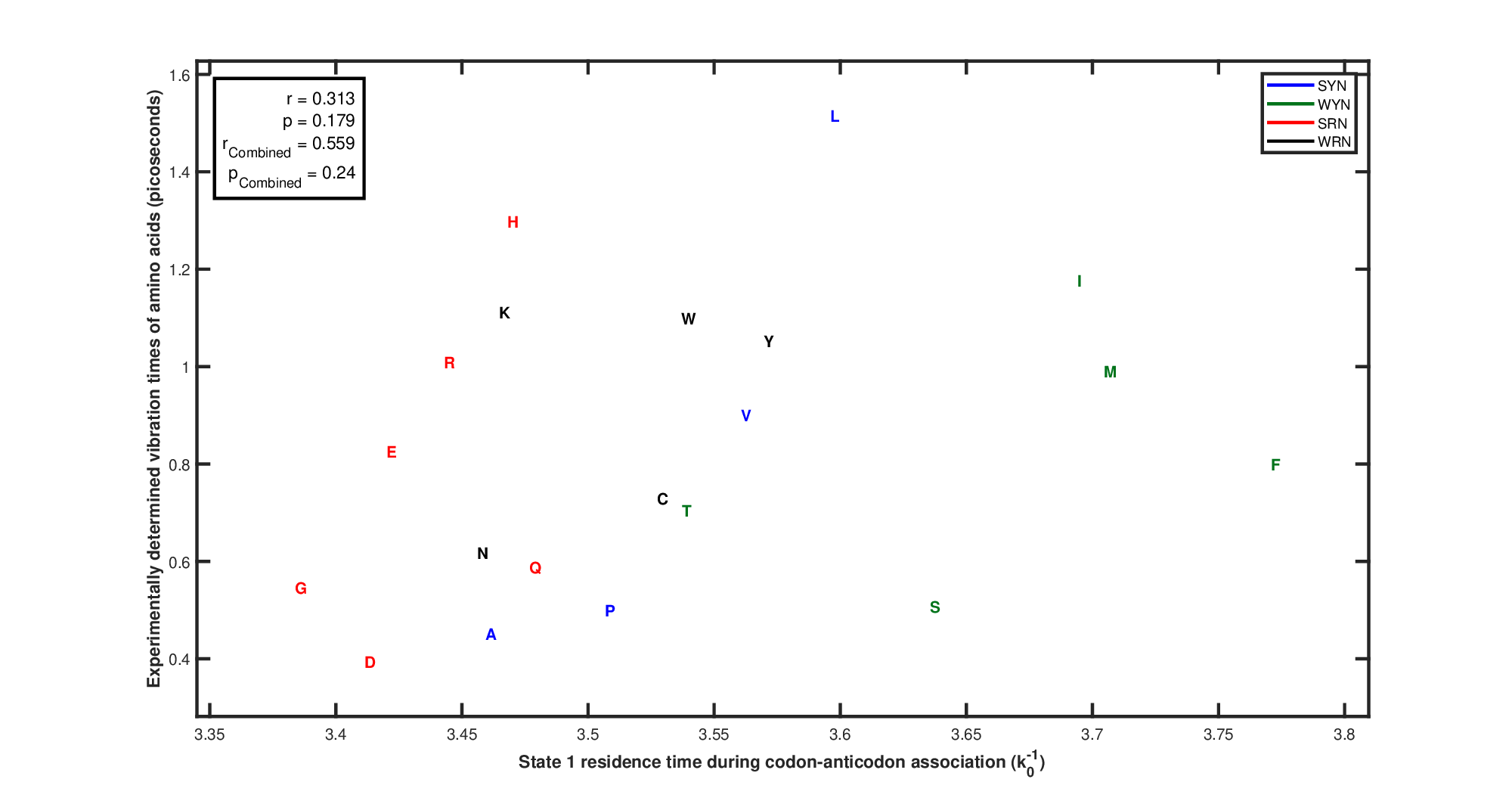}
\caption{Experimentally determined vibrational time periods of 20 proteinogenic amino acids in picoseconds, obtained by THz spectroscopy, versus the residence times of codon-anticodon association process in state 1. Synonymous codons for each amino acid are averaged, excluding doublets of Ser and Leu. The lowest vibrational frequencies of the amino acids are used to calculate the time period. Correlations of SNN-coded amino acids appear better than that of WNN-coded amino acids, although the overall correlation is seen to be statistically insignificant. This partial correlation confirms the time scale of the codon-anticodon association process to be of the order of picoseconds. Amino acid vibrational frequency data from ref. \cite{Yi2016Broadband}.}
\label{expvibtimes}
\end{figure}

We verified this timescale through another correlation that we found between computationally and experimentally determined vibration times of amino acids and the state 1 residence times of the codon-anticodon association process (figs. \ref{compvibtimes} and \ref{expvibtimes}), again arriving at picosecond timescales. Ref. \cite{Scherbakov2016} determined the three lowest vibrational modes of all 20 amino acids using semi-empirical quantum-chemistry programs in both neutral and zwitterionic states. We have used the zwitterionic modes $\nu 1$ (Z-I) for NRN-coded amino acids, and $\nu 3$ (Z-I) for NYN-coded amino acids to calculate the vibrational timescales. Experimental vibrational timescales were calculated from the lowest frequency vibrational modes in ref. \cite{Yi2016Broadband}, which used terahertz spectroscopy to obtain the spectra of all 20 amino acids. We used the frequencies of the lowest vibrational modes of all 20 amino acids for the correlation analysis. Both these datasets show approximate linearities with state 1 residence times, suggesting that energy flow from codon-anticodon system to the vibrational modes of amino acids is somehow involved in the genetic code assignments. While the segregated and combined r and p-values of the correlation of computationally determined vibrational times with state 1 are significant, the correlation with experimental vibrational times are non-significant, but included here because SYN and SRN groups show correlation while the other two groups do not.

\begin{figure}
\centering
\includegraphics[trim=80pt 20pt 20pt 20pt,width=1.05\textwidth]{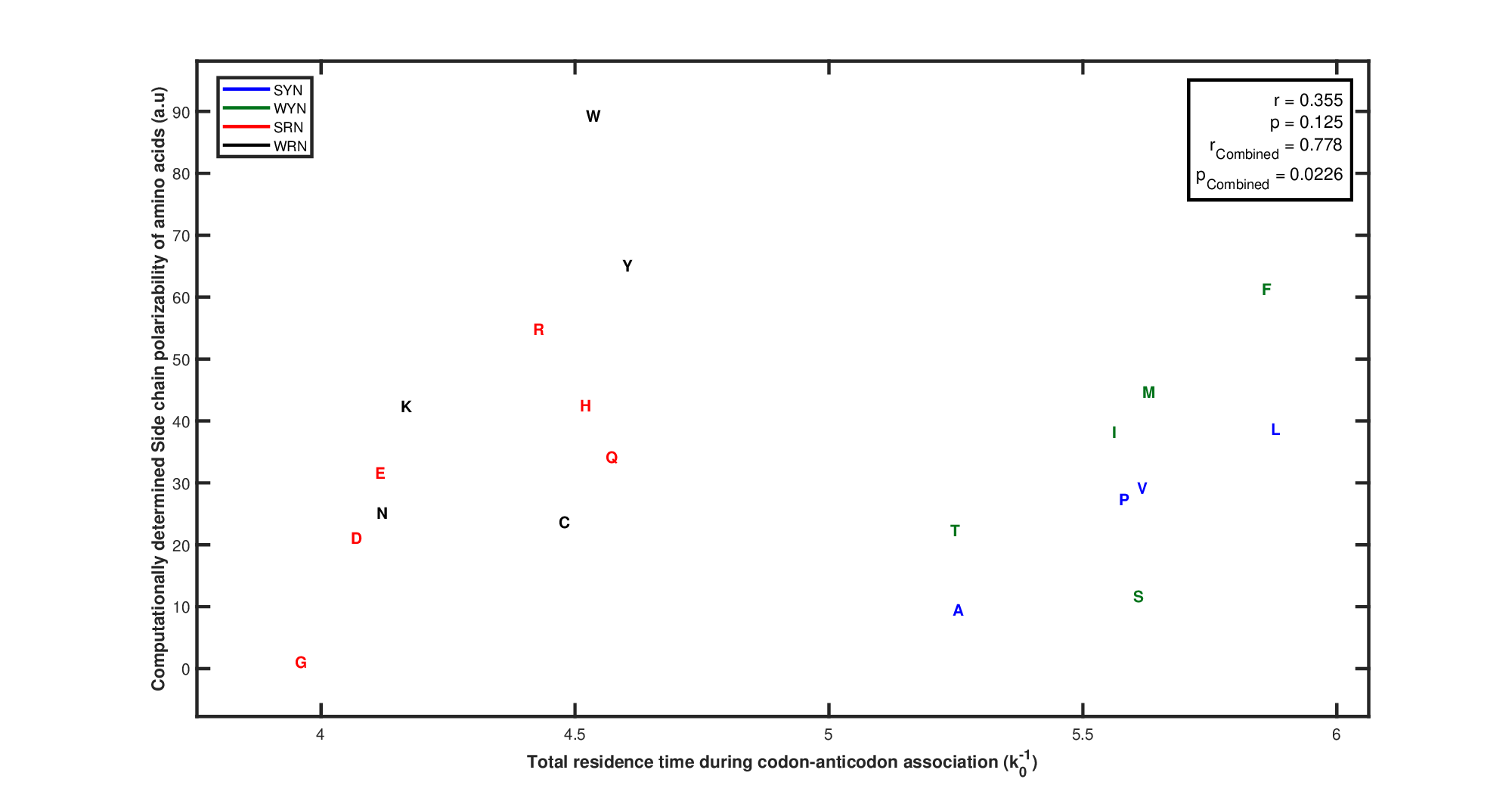}
\caption{Computationally determined side chain polarizabilities (trace of polarizability tensor) of 20 amino acids, calculated using \textit{ab initio} methods, in atomic units (a.u), versus total time taken for codon-anticodon association process. The correlation is seen to be better for NYN codons, which code for mostly hydrophobic amino acids. When segregated into four groups of SRN, WRN, SYN and WYN codons, fit and combined using Fisher's method, we find significant correlation. Amino acid polarizability data from ref. \cite{Krishtal2009}.}
\label{sidechainpol}
\end{figure}

\begin{figure}
\centering
\includegraphics[trim=80pt 20pt 20pt 20pt,width=1.05\textwidth]{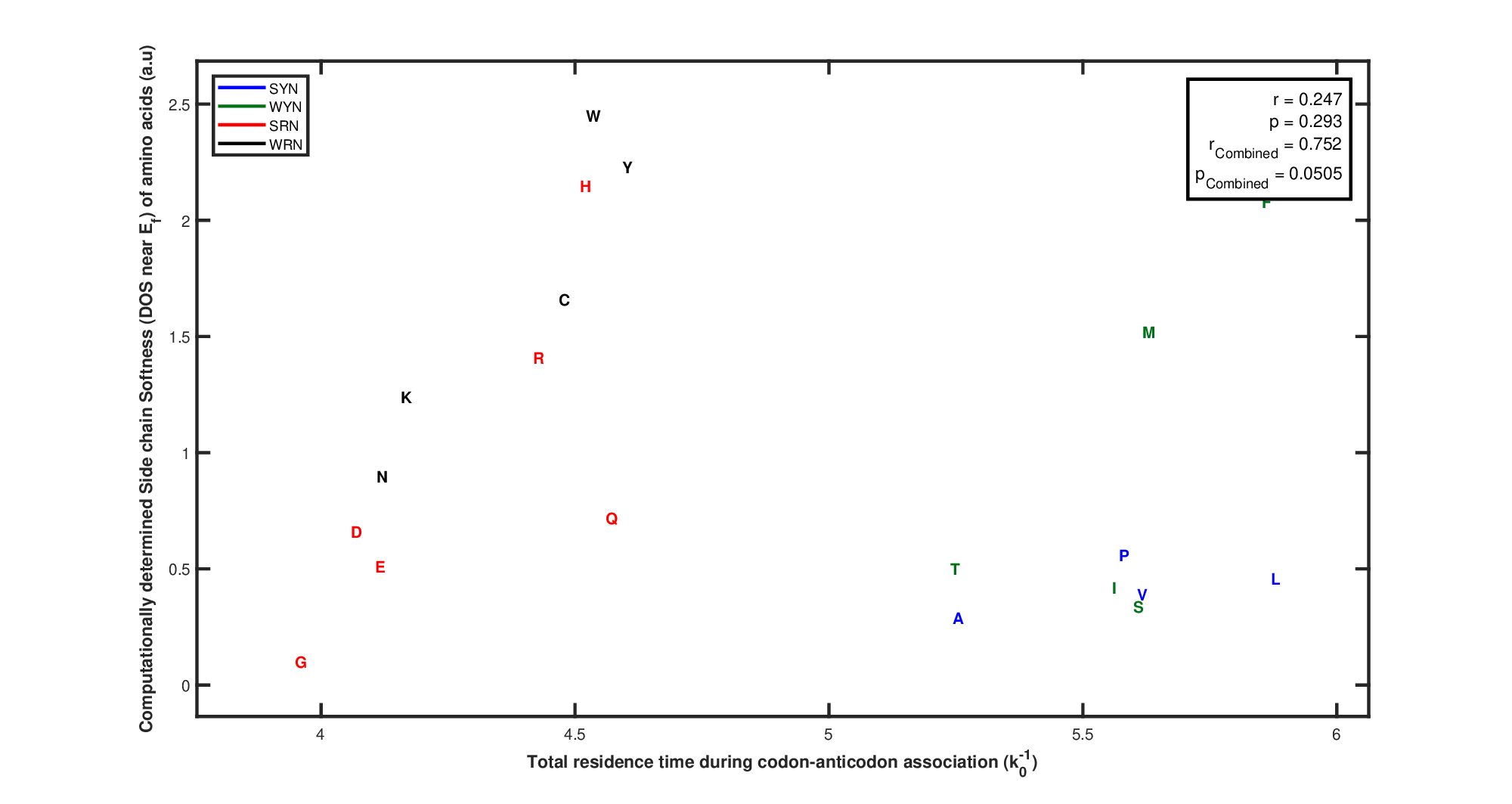}
\caption{Computationally determined side chain softness of 20 amino acids, calculated using \textit{ab initio} methods, in atomic units (a.u), versus total time taken for codon-anticodon association process. Softness measures the reactivity of a molecular system, defined as the derivative of chemical potential w.r.t external potential. The correlation is seen to be better for NRN codons, which code for hydrophilic amino acids. Segregated and combined r and p-values, given in $r_{combined}$ and $p_{combined}$, are almost statistically significant. This correlation suggests that hydrogen bonding ability of hydrophilic amino acids with water molecules is important in setting their energy dissipation rates. Amino acid softness data from ref. \cite{Krishtal2009}.}
\label{sidechainsoftness}
\end{figure}

 Another clue to the genetic code puzzle is found in the correlations seen between amino acids’ side chain polarizabilities (fig.\ref{sidechainpol}) and its related concept of side chain softnesses (fig.\ref{sidechainsoftness}), on one hand, and the total residence times of codon-anticodon association process on the other. Total residence times of the 64 codons are calculated by summing all the seven residence times in the intermediate states. Side chain polarizabilities and softnesses of the 20 amino acids, computed in ref. \cite{Krishtal2009} using \textit{ab initio} methods, are seen to correlate with the total residence times of cognate codon-anticodon association processes. While the residence times of NYN codons, which code mostly for hydrophobic amino acids, correlate with amino acids' side chain polarizabilities better, residence times of NRN codons, coding for hydrophilic amino acids, correlate with the side chain softnesses better. The segregated and combined r and p-values of both the correlations are provided in the figures, and can be seen to be significant. It can be seen that segregating the dataset into four groups of SRN, WRN, SYN and WYN codons results in better statistics, since the four groups are mostly independently linear with the x-axis. Both polarizability and softness are linear response functions. Polarizability quantifies the response of the electronic distribution to an external electric field, whereas softness quantifies the response of electron number (or density) to changes in chemical potential. Given the correlations between the codon-anticodon residence times and amino acid relaxation times seen above, these additional correlations suggest electronic energy (for hydrophobic amino acids) or electron flow (for hydrophilic amino acids) between these two systems, pointing to a non-equilibrium origin for the genetic code.

In the figures referred to above and all the figures that follow, to maintain the symmetry of the genetic code, we avoid averaging over all the six codons that code for the two amino acids, serine (S) and leucine (L) (vertebrate mitochondrial standard code),  and instead average over only the four codons that form a quadruplet group that code for these two amino acids. These averages are denoted with upper case S and L respectively, in the figures. The other two codons that form a doublet corresponding to these two amino acids are averaged over separately and are denoted by the lowercase letters (s and u) in the figures, where necessary.

These correlations suggest the following mechanism for the emergence and maintenance of the genetic code: the perturbation of the electronic degrees of freedom of the ribonucleotides due to the codon-anticodon association process is transmitted, through-space (in the primordial scenario) or more probably, through-bond (in extant systems, through the tRNA molecule), to the cognate amino acid, which dissipates the electronic energy at 1 - 10 picosecond time scales, through vibronic coupling, into the surrounding water molecules. Please note that water molecules have been shown to be present inside the ribosomes and are important for translation \cite{Wallin2010}. The approximately linear relationship between the energy supply timescales from the codon-anticodon bonding process and the energy dissipation timescales at the amino acid end, seen in fig.\ref{heatrelaxationtimes}, suggests that the energy is transmitted from the mRNA at the fastest possible rate. The linearity suggests a matching of energy supply and dissipation timescales, which in turn suggests a form of impedance matching between codon-anticodon system and its cognate amino acid, leading to our hypothesis of fastest energy transfer rates. This matching may therefore underlie the specificity of mapping between the codons and their corresponding amino acids, seen in the genetic code, and results in the correlations seen above. This will be elaborated further in the discussion section. The fast removal of energy stabilizes the codon-anticodon bonding and possibly enables recognition by the ribosome. This mechanism coherently explains the correlations that we found between the residence times of codon-anticodon association process and polarizability/softness (figs.\ref{sidechainpol} and \ref{sidechainsoftness}), vibration times (figs.\ref{compvibtimes} and \ref{expvibtimes}), and heat dissipation times of amino acids (fig.\ref{heatrelaxationtimes}), shown above. Since hydrophobic amino acids, coded by NYN codons, solvate slower than NRN-coded hydrophilic amino acids (see figs.\ref{sidechainpol} and \ref{sidechainsoftness}) at the picosecond time scale due to their hydrophobic nature, their energy dissipation is dictated more by intramolecular degrees of freedom, which might explain their better correlation of residence times with amino acids' side chain polarizability (fig.\ref{sidechainpol}). The solvation timescales are larger for molecules with large polarizabilities, because of the ability of amino acids with large polarizabilities to instantaneously readjust their electronic structure to the changing solvent, thereby decreasing the restoring force for the solvent to return back to its equilibrium during the energy transfer \cite{Bursulaya1995}. Whereas, hydrophilic amino acids, coded by NRN codons, additionally solvate due to their hydrogen bond forming ability with water \cite{Hamzi2022}. This could explain the better correlation of NRN codons with softness of amino acids, which quantifies reactivity \cite{Pearson2005}. Increased solvation times with higher softness values (fig.\ref{sidechainsoftness}) is possibly due to weaker hydrogen bonds of softer amino acids.

\subsubsection*{Second level: NYN-NRN, 0.1 - 1 picoseconds}

\begin{figure}
\centering
\includegraphics[trim=80pt 20pt 20pt 20pt,width=1.05\textwidth]{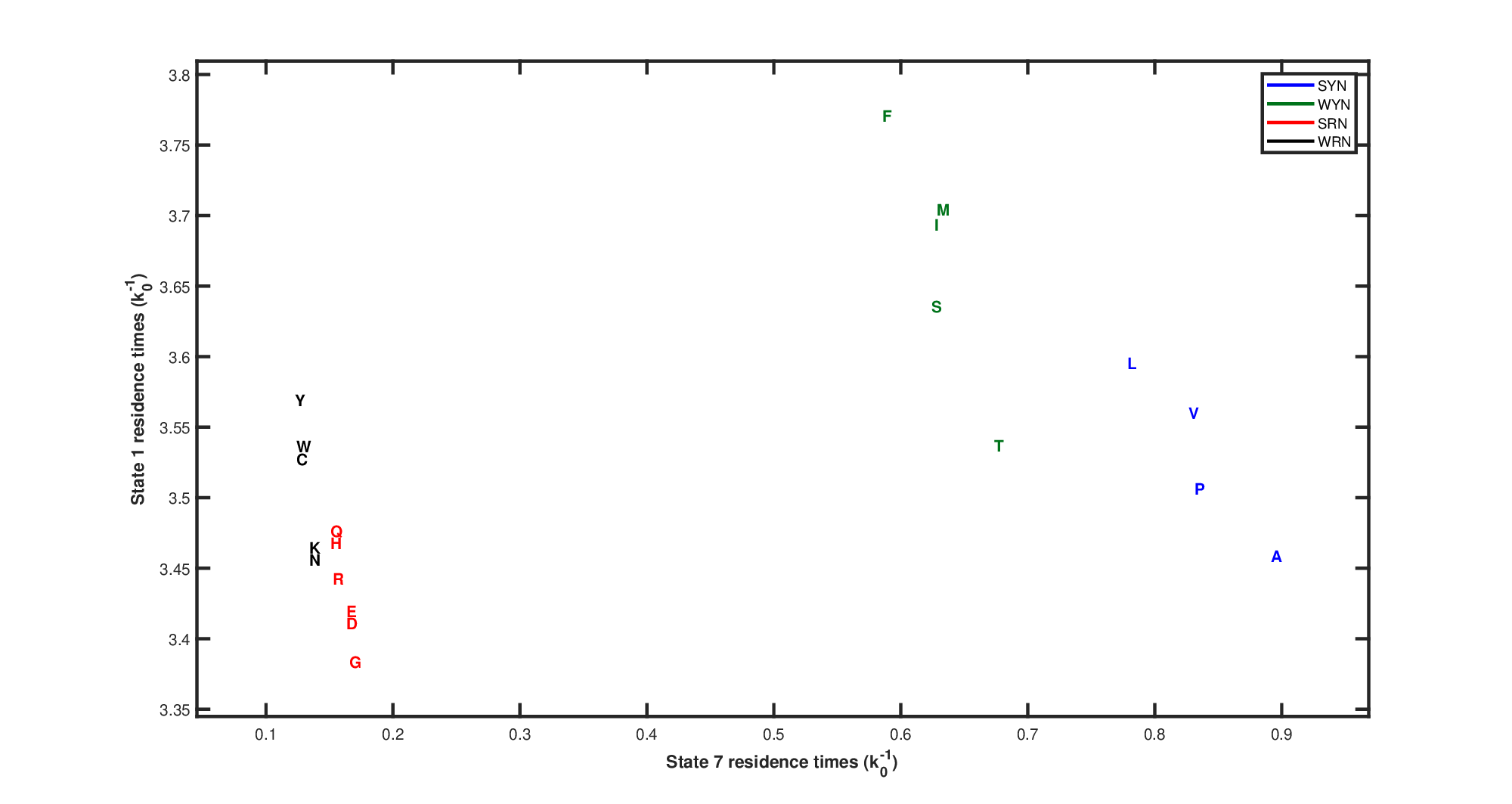}
\caption{Residence times in state 1 of codon-anticodon association process is plotted against the residence times in state 7. This plot demonstrates the separation of timescales between NRN-coded hydrophilic amino acids and NYN-coded hydrophobic amino acids. The timescale of separation between these two clusters in state 7 is about 0.1 - 1 picoseconds. Hydrophilic amino acids appear to solvate faster than hydrophobic amino acids at this timescale.}
\label{state1_state7}
\end{figure}

\begin{figure}
\centering
\includegraphics[trim=80pt 20pt 20pt 20pt,width=1.05\textwidth]{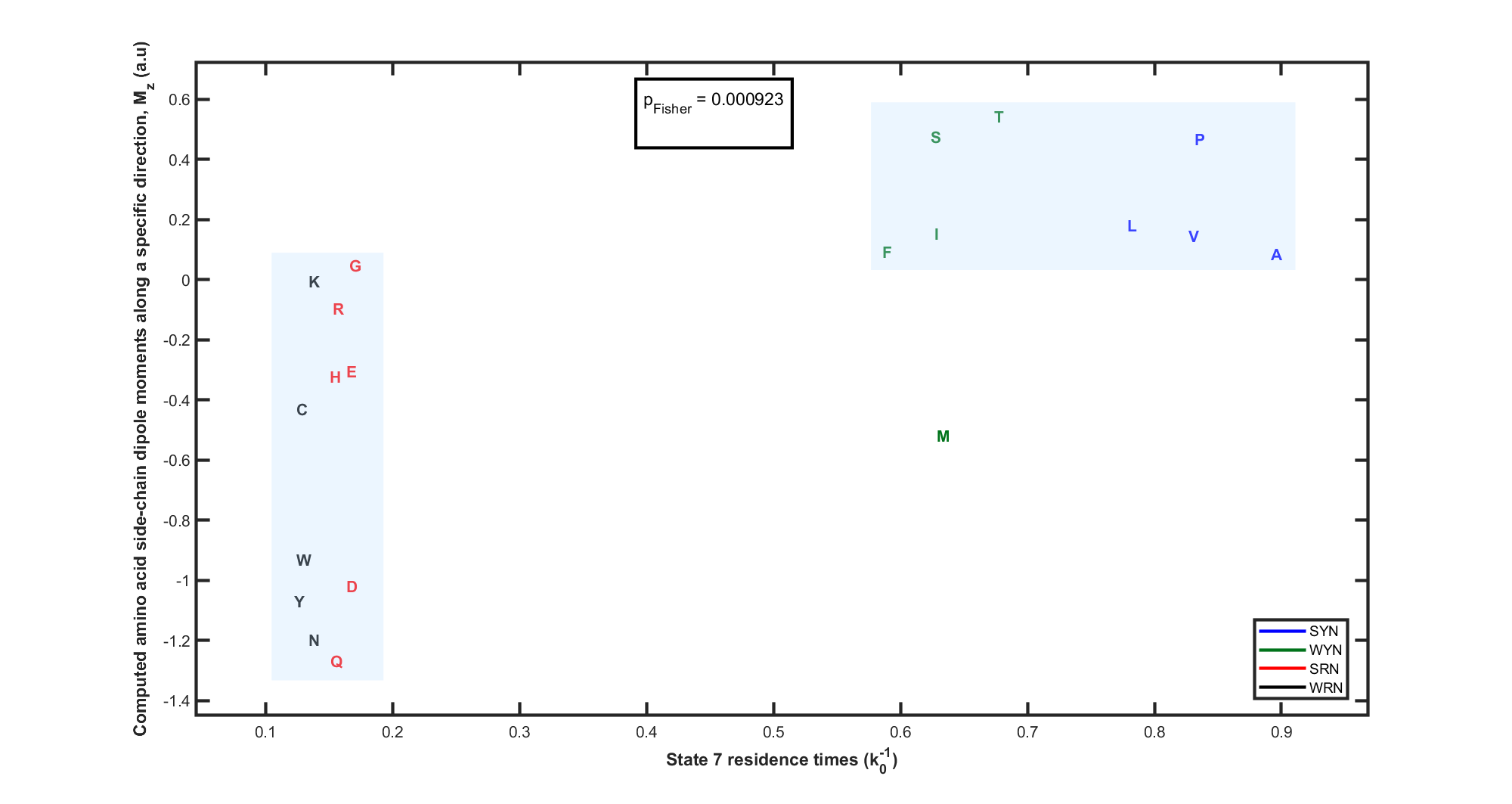}
\caption{Computationally calculated dipole moments of the 20 amino acids along a specific direction, in atomic units, after aligning them along a common orientation, versus the appropriately averaged residence times in state 7 during the corresponding codon-anticodon association process. With the exception of Met (M) and Gly (G), all the NYN-coded hydrophobic amino acids are seen to have positive dipole moments along this specific direction, and relax slower, whereas, all NRN-coded hydrophilic amino acids have negative dipole moment values, and relax faster. The two blue rectangular backgrounds highlight the segregation of NYN- and NRN-coded amino acids. The p-value calculated from two-sided Fisher's exact test is statistically significant for this segregation. The dipole moment direction that correlates with the residence times has the maximum variance among the three directions calculated. The reason for the association of faster NRN codons with amino acids with negative dipole moments could be that the dissipation rates of these amino acids are faster due to the attraction of water hydrogen ions towards the more accessible negative ends of the dipoles. It is therefore conceivable that the interaction of water molecules with the distinct charge distributions around the hydrophilic and hydrophobic amino acids dictate their respective relaxation rates at the 0.1 - 1 picosecond timescales. Amino acid dipole moment data from ref. \cite{Bohrquez2003}.}
\label{dipolemoment_state7}
\end{figure}

\begin{figure}
\centering
\includegraphics[trim=80pt 20pt 20pt 20pt,width=1.05\textwidth]{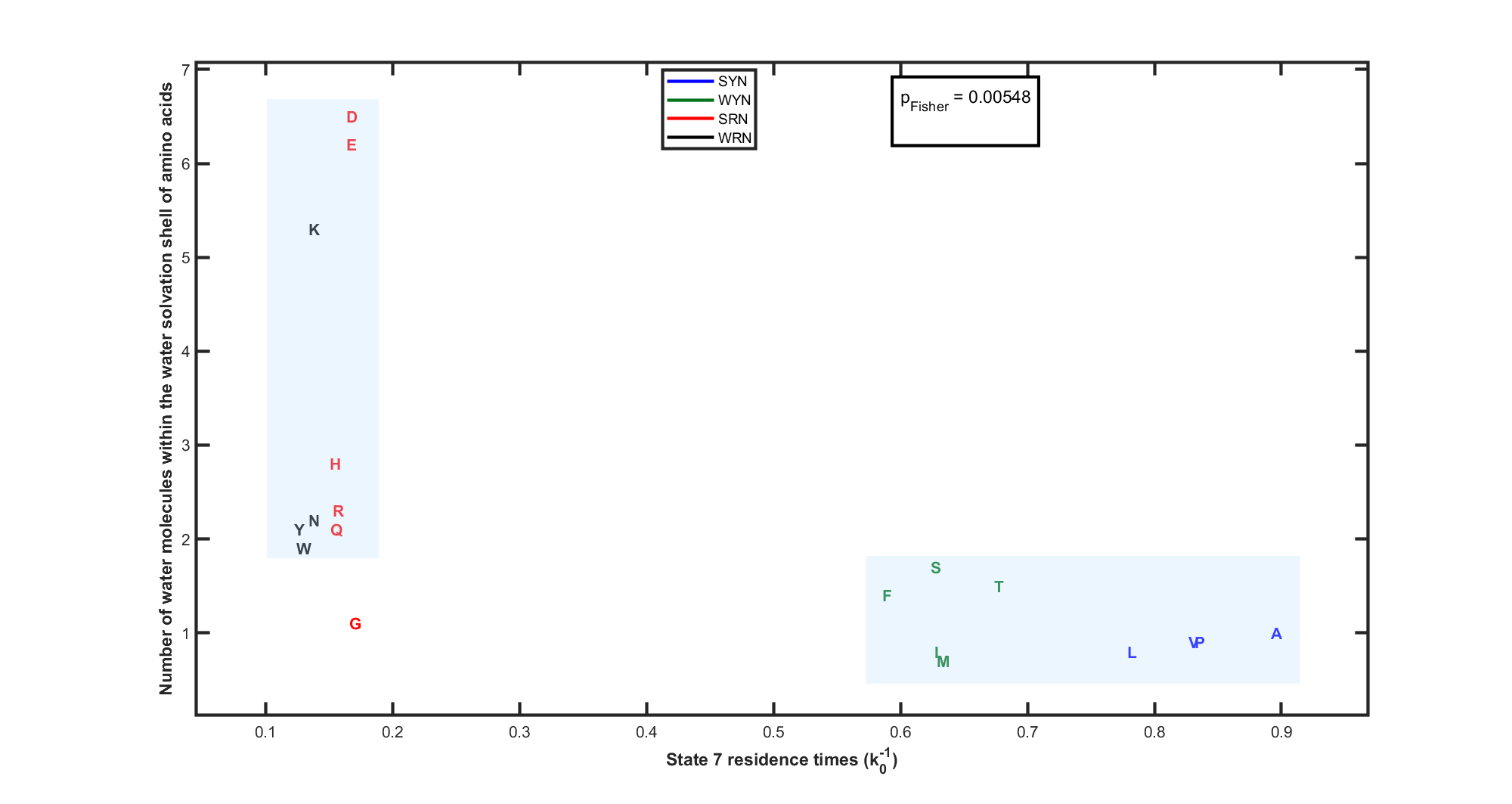}
\caption{Number of water molecules within the solvation shell of amino acids is plotted against the residence times of the corresponding codons in state 7. NRN-coded hydrophilic amino acids are seen to have more water molecules within their solvation shell, thus enabling better dissipation and hence faster relaxation timescales, compared to NYN-coded hydrophobic amino acids with smaller numer of water molecules surrounding them, at these timescales. The two blue rectangular backgrounds highlight the segregation of NYN- and NRN-coded amino acids. The p-value from two-sided Fisher's exact test is statistically significant. This observation agrees with our conjecture that the negative dipole ends of NRN-coded amino acids attract more water molecules, compared to NYN-coded amino acids. The smallest amino acid Gly (G) is an exception. Data for Cysteine (C) is unavailable. Amino acid data from ref. \cite{Hopfinger1977}.}
\label{solvationnumber_state7}
\end{figure}

Moving on to the second level of the hierarchical arrangement of the genetic code (fig. \ref {codon_org}), we see that the 64 codons are split into two sets, NRN and NYN. This splitting is seen in the clustering of residence times in state 7 into two sets, in fig.\ref{state1_state7}, which shows a plot between state 1 and state 7 residence times. The timescale of separation between these two sets can be seen to be about 0.1 - 1 picoseconds. We can say this with some certainty, since we have concluded from the correlations in the first level that the state 1 residence times are in the range of 1 - 10 picoseconds.  On the amino acid front, the biophysical property that best segregates the amino acids that correspond to these two groups of codons is hydrophobicity, as mentioned above. This observation has been made earlier as well \cite{Harrison2022,wolfenden1979water}. A plot between the side chain dipole moments along a specific direction of the twenty amino acids and average state 7 residence times of the corresponding codon families, fig.\ref{dipolemoment_state7}, neatly segregates the two sets, with the exception of Met (M).  These side chain dipole moments were calculated using ``Atoms In Molecule'' method, after aligning all the 20 amino acids in such a way that the common atoms of all the amino acids bonded to their side chains are similarly located in a common coordinate system \cite{Bohrquez2003}. Of the three dipole moment directions calculated in the paper, the direction that has the maximum variance across all amino acids is the one that segregates the NRN and NYN codon sets, possibly to enhance discrimination. It has been shown that both the magnitude and the direction of the ground state dipole moment in an amino acid influences its interaction with nearby water molecules \cite{Hamzi2022}, primarily due to the asymmetric distribution of positive and negative charges in a water molecule, with the negative charges on the oxygen ion buried deeper, and hence is less reactive than the positive charges on the hydrogen ion. The amino acids with positive dipole moment values are associated with NYN codons with larger state 7 residence times, while the ones with negative values, which can interact strongly with the water hydrogen cations, are coded by faster NRN codons \cite{Rey2015}(fig. \ref{dipolemoment_state7}), with the exception of Met (M). The foregoing is substantiated by another correlation found between the number of water molecules within the solvation shell of amino acids \cite{Charton1982,Hopfinger1977} and the residence times in state 7, fig.\ref{solvationnumber_state7}. As can be observed in the plot, more water molecules within the solvation shell of amino acids lead to faster dissipation, and therefore, these amino acids are associated with the faster NRN codons.

All these observations above again suggest that the differences in \textit{sub-picosecond}-scale relaxation pathways of hydrophobic and hydrophilic amino acids, discriminated by the magnitude and orientation of their dipole moments and their interactions with water molecules \cite{Morozova2016}, underlie the segregation of NYN and NRN-coded amino acids. We speculate that the matching of the energy supply rates from state 7 of the association process of NYN (NRN) codons with their corresponding anticodons, on one hand, and the \textit{sub-picosecond}-scale relaxation rates of hydrophobic (hydrophilic)  amino acids, on the other, leads to the observed NYN/NRN codon group - hydrophobic/hydrophilic amino acid group assignments in the genetic code. We know that solvation of amino acids in water happens at multiple timescales (tens of femtoseconds to tens of picoseconds), corresponding to distinct relaxation pathways \cite{Laage2017,Maroncelli1989}. We argue that relaxation rates of amino acids, at distinct timescales, match with the energy supply rates from codon-anticodon association processes, at corresponding timescales, with each timescale specific to a particular hierarchical level of the codon organization, shown in fig. \ref{codon_org}. Moreover, from the above correlation plots, we find that the energy dissipation or solvation rates of amino acids, at these sub-picosecond time scales, is slower for hydrophobic amino acids, possibly due to weaker interactions with the surrounding waters. This corresponds, on the codon front, to slower energy supply rates from state 7 of the NYN codon-anticodon association process, as observed in the correlation plots.

\subsubsection*{Third level: SYN-WYN, 0.1 - 1 picoseconds}

\begin{figure}
\centering
\includegraphics[trim=80pt 20pt 20pt 20pt,width=1.05\textwidth]{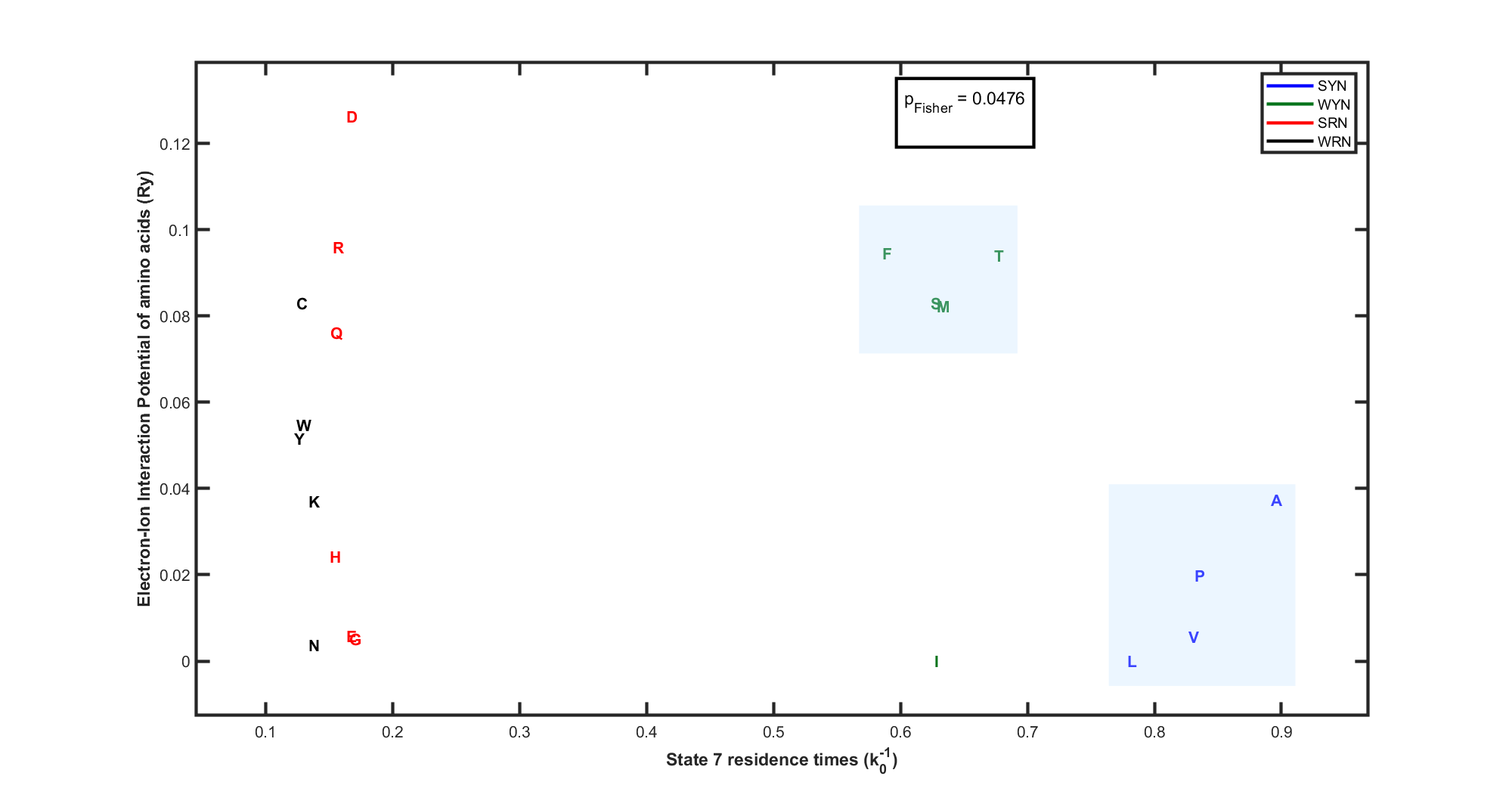}
\caption{Electron-ion interaction potential, quantifying electron-phonon interactions in the amino acids, versus state 7 residence times. This metric reasonably segregates SYN- and WYN-coded amino acids, with the exception of Ile (I). These two groups are highlighted in a light blue background. Higher amounts of intramolecular electron-phonon interaction enables more efficient transfer of excess electronic energy transmitted from codon-anticodon association process into the vibrational modes of the amino acid molecules, possibly decreasing the relaxation time for amino acids coded by WYN, compared to SYN. Thus the amino acids with faster (slower) transduction of electronic energy into the vibrational modes are mapped to the faster (slower) WYN (SYN) codons. The segregation into two groups is statistically significant, from the p-value calculated. Amino acid data from ref. \cite{Pirogova2003}.}
\label{EIIP_7}
\end{figure}

\begin{figure}
\centering
\includegraphics[trim=80pt 20pt 20pt 20pt,width=1.05\textwidth]{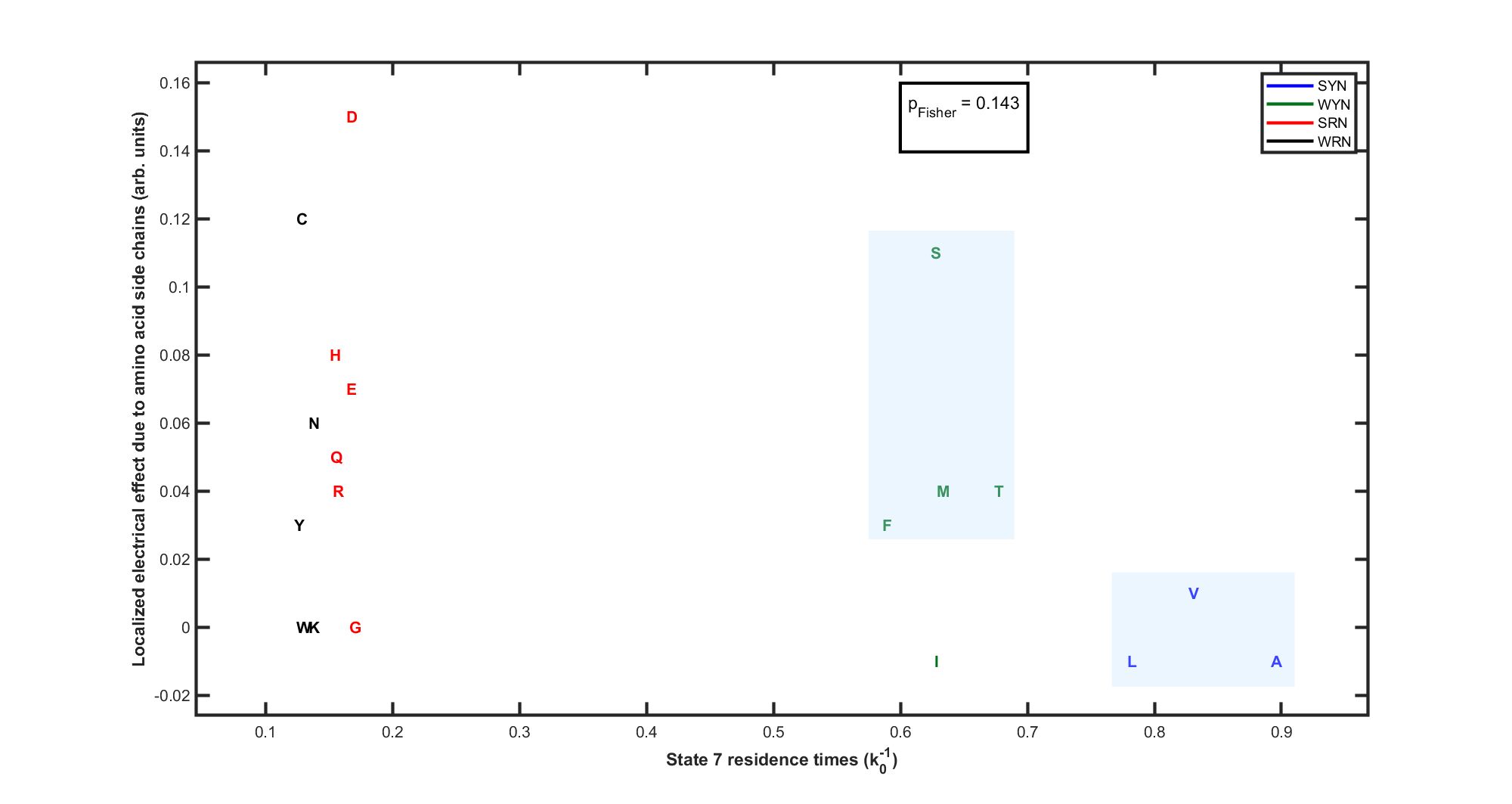}
\caption{Localized electrical effect, which parametrizes intramolecular dipole - dipole, dipole - induced dipole, and induced dipole - induced dipole interactions within an amino acid molecule, versus state 7. This parameter has been shown to be well-correlated with amino acid hydrophobicity scales \cite{Charton1982}. Higher levels of such interactions enable different parts of the amino acid molecule to coordinate in transferring the excess electronic energy into other lower-energy, higher-entropy modes, possibly reducing the relaxation time of the energy for WYN-coded amino acids, which exhibit larger localized electrical effects, compared to SYN-coded amino acids. These faster dissipating amino acids are therefore mapped to faster WYN codons. The segregation of the groups of WYN- and SYN-coded amino acids are highlighted with a light blue background. The p-value calculated using two-sided Fisher's exact test is statistically not significant, due to less number of data points. Ile (I) is an exception. Pro (P) data is unavailable. Amino acid data from ref. \cite{FAUCHRE1988}.}
\label{LEE_7}
\end{figure}

 At the third level, the NYN codons are split into two sets of SYN and WYN codons, as can be seen in figs.\ref{codon_org} and \ref{state1_state7}. The timescale of segregation of SYN and WYN codons is of the same order as that of NYN-NRN segregation, 0.1 - 1 picoseconds, from fig.\ref{state1_state7}. The SYN-WYN splitting, on the amino acid front, appear to be dictated by electron-ion interaction potential \cite{Pirogova2003}, which parametrizes the effect of ionic cores of amino acids on electron motion (mediating electron-phonon interactions) \cite{Veljkovi1972}, and localized electrical effect,  which parametrizes dipole - dipole, dipole - induced dipole, and induced dipole - induced dipole interactions within an amino acid molecule \cite{FAUCHRE1988}. Figures \ref{EIIP_7} and \ref{LEE_7} show that these two properties split the two amino acid sets corresponding to SYN and WYN reasonably well, with the exception of Ile (I). Higher the electron-phonon interaction, faster the energy transfer from electronic modes to vibrational modes of amino acids, and hence faster the relaxation times. Therefore, the amino acids with higher values of electron-ion interaction potential and localized electrical effect are assigned to smaller residence time WYN-codons. It has been shown that the relaxation lifetimes of correlated electrons in polycyclic aromatic molecules from XUV ionization, due to electron-phonon interactions, is of the order of tens of femtoseconds \cite{Herv2020}, about an order of magnitude smaller than the timescale range suggested by the model, above. The question of how hydration of amino acids would shift this experimental relaxation time range is difficult to answer, without explicitly evaluating the change in the frequency of all relevant vibrational modes of every one of the 20 amino acids due to hydration. Please note that the amino acids' physicochemical properties that correlate with the residence times at this timescale are intramolecular in nature, understandable because we are dealing with hydrophobic amino acids.

\subsubsection*{Fourth level: SRN-WRN, 0.01 - 0.1 picoseconds}

\begin{figure}
\centering
\includegraphics[trim=80pt 20pt 20pt 20pt,width=1.05\textwidth]{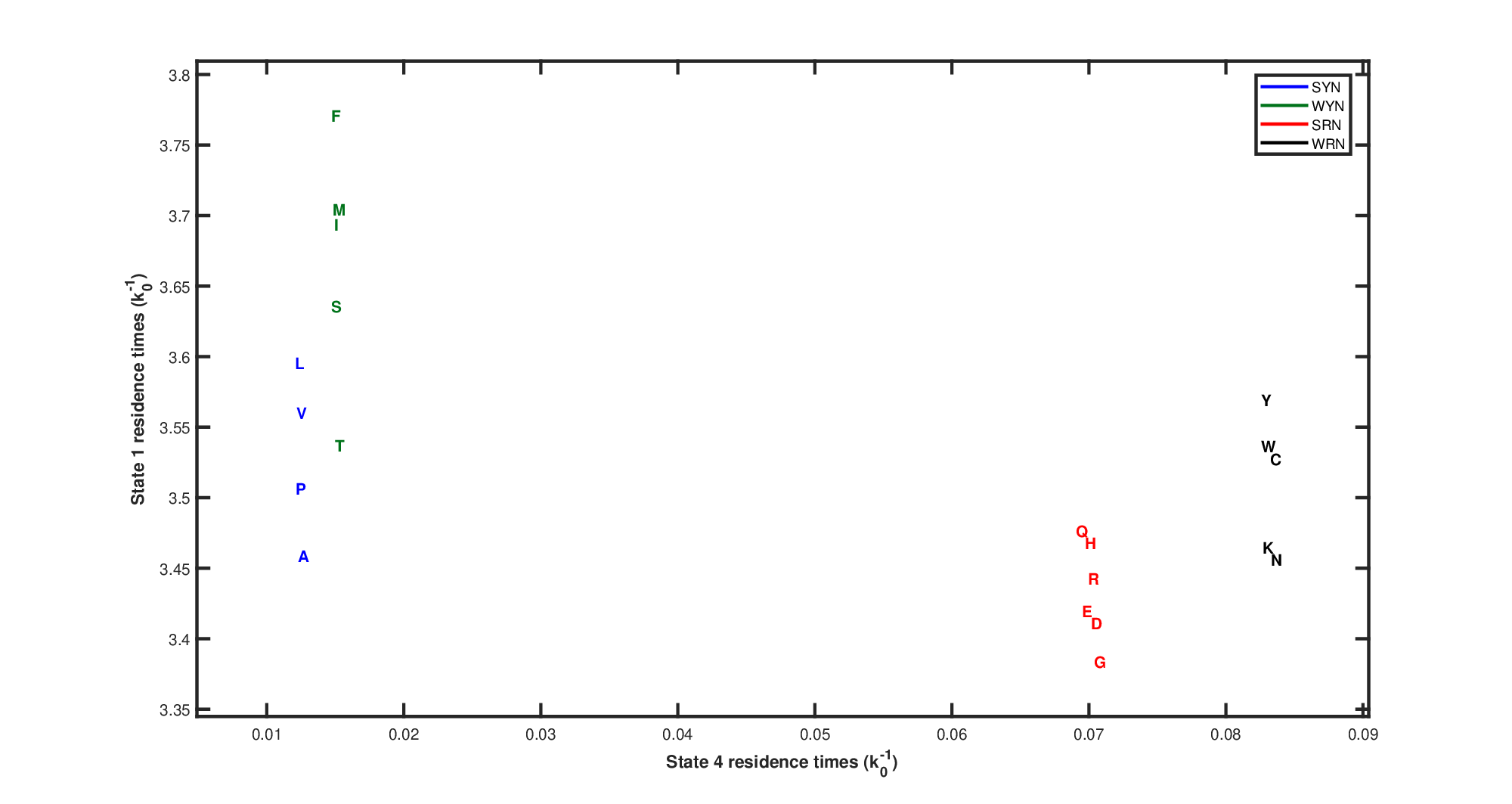}
\caption{Residence times in state 1 of codon-anticodon association process plotted against the residence times in state 4. The timescale of separation between SRN and WRN codons is seen to be in the range of 0.01 - 0.1 picoseconds. Relaxation time differences at this timescale between SRN- and WRN-coded amino acids appear to be dictated by the number of hydrogen bond acceptors in the amino acids. Interfacial thermal conductance and hydropathy scales derived from partitioning of amino acids between water and organic solvents also better segregate these two groups. Note that at this timescale, in contrast with the previous 0.1 - 1 ps timescale, hydrophobic amino acids-coding NYN codons are faster. }
\label{state1_state4}
\end{figure}

\begin{figure}
\centering
\includegraphics[trim=80pt 20pt 20pt 20pt,width=1.05\textwidth]{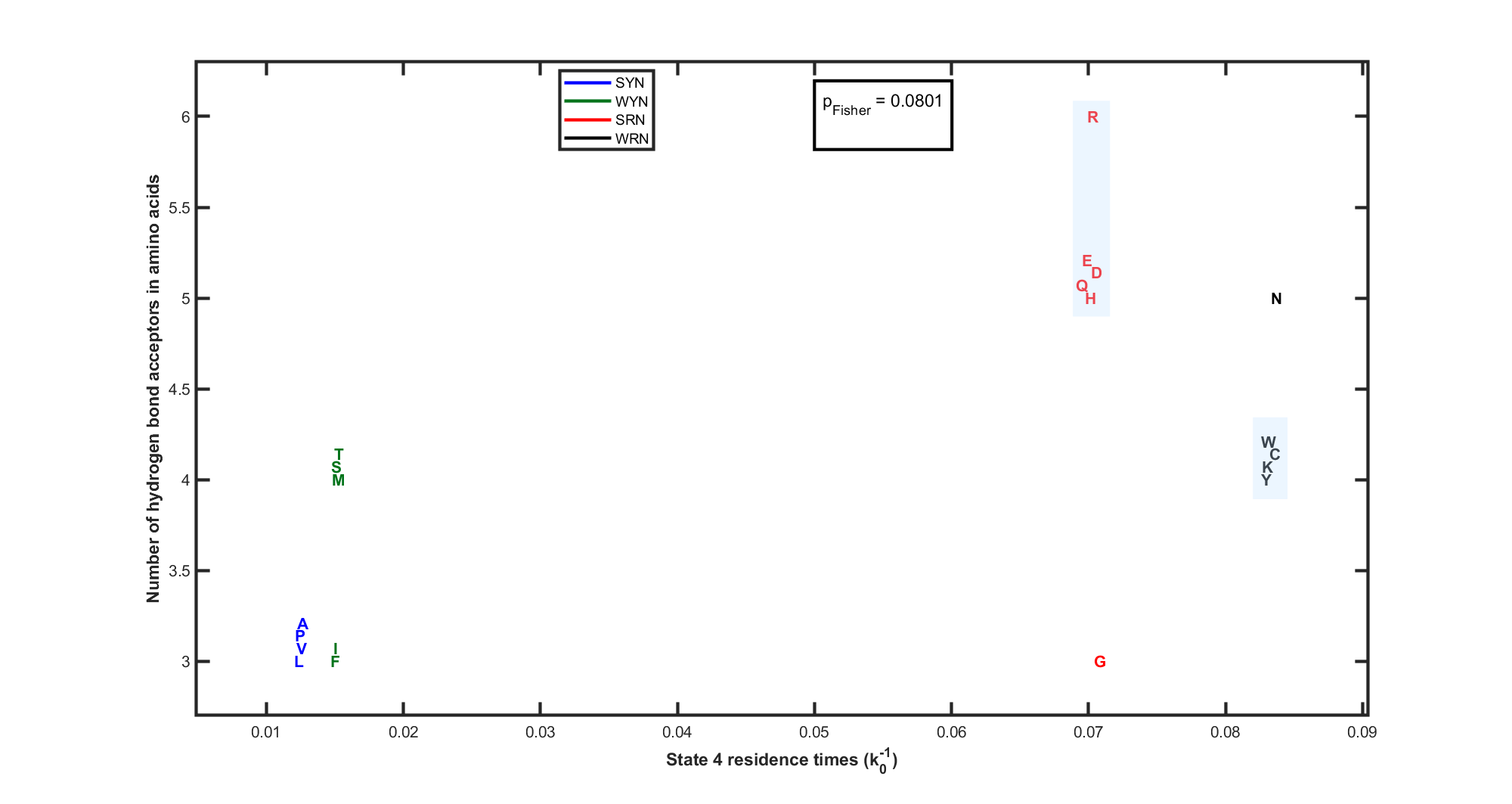}
\caption{Number of hydrogen bond acceptors, such as carbonyl and carboxylate groups in the amino acids, versus residence times in state 4 during the codon-anticodon association process. SRN-coded amino acids is seen to have higher number of hydrogen bond acceptors, compared to WRN. These two groups' segregation is illustrated by shading the groups in light blue. The p-value of segregation, calculated using two-sided Fisher's exact test, is not significant, because of two outlier amino acids and the small number of data in the clusters. Since these groups can better attract the positive hydrogen ion in water, water molecules are more densely packed around hydrogen bond accceptors than around hydrogen bond donors. This might result in better dissipation and hence faster relaxation at these timescales, as seen in the graph. The amino acid labels are slightly displaced vertically for visual clarity. Data from ref. \cite{Yang2010}.}
\label{Hbondacceptors_4}
\end{figure}

In the fourth level, the group of NRN codons, coding mostly for hydrophilic amino acids, splits into SRN and WRN codons, with the difference in time between these two groups, as seen in fig.\ref{state1_state4}, is in the 0.01 - 0.1 picoseconds range. The amino acid properties that appear to segregate these two groups are interfacial thermal conductance, number of hydrogen bond acceptors, and certain hydropathy scales. Please note that these properties are intermolecular in nature, in contrast with the previous level segregation of hydrophobic amino acids, which was governed by intramolecular properties. Fig. \ref{Hbondacceptors_4} shows the segregation of SRN and WRN codon sets along both X and Y axes. State 4 residence times, along X-axis, segregate these two sets clearly, whereas, the number of hydrogen bond acceptors in an amino acid \cite{Yang2010}, such as carbonyl and carboxylate groups, which donate electrons to form hydrogen bonds with nearby water molecules, segregates these two sets along Y-axis, with the exception of Asn (N). More hydrogen bond acceptors in an SRN-coded amino acid would allow it to form more hydrogen bonds with water molecules, possibly enabling better energy transfer. More over, the electronic charge distribution at these acceptor ends would attract the hydrogen ions of water, making the water molecules denser near the surface, compared to hydrogen bond donors (fig. 11 in \cite{Rey2015}), further improving solvation. These interactions may lead to faster solvation times for SRN-coded amino acids, compared to WRN-coded amino acids. Hydrogen bond donor numbers do not substantially vary across the 20 amino acids and they do not correlate with the residence times. While hydrogen bond donors retard the water dynamics at the picosecond timescale, similar to hydrophobic solutes, hydrogen bond acceptors have been shown to alter the water dynamics dramatically by either slowing down or accelerating the water dynamics. Acceptors that form strong hydrogen bonds, such as Gln (Q), His (H), Asp (D) and Glu (E), strongly slow down water dynamics at the picosecond timescale \cite{Sterpone2010}. How this picosecond-scale slowdown alters the dissipation rates at sub-picosecond scales in not clear from the literature. Gly (G) and Asn (N) are exceptions.

\begin{figure}
\centering
\includegraphics[trim=80pt 20pt 20pt 20pt,width=1.05\textwidth]{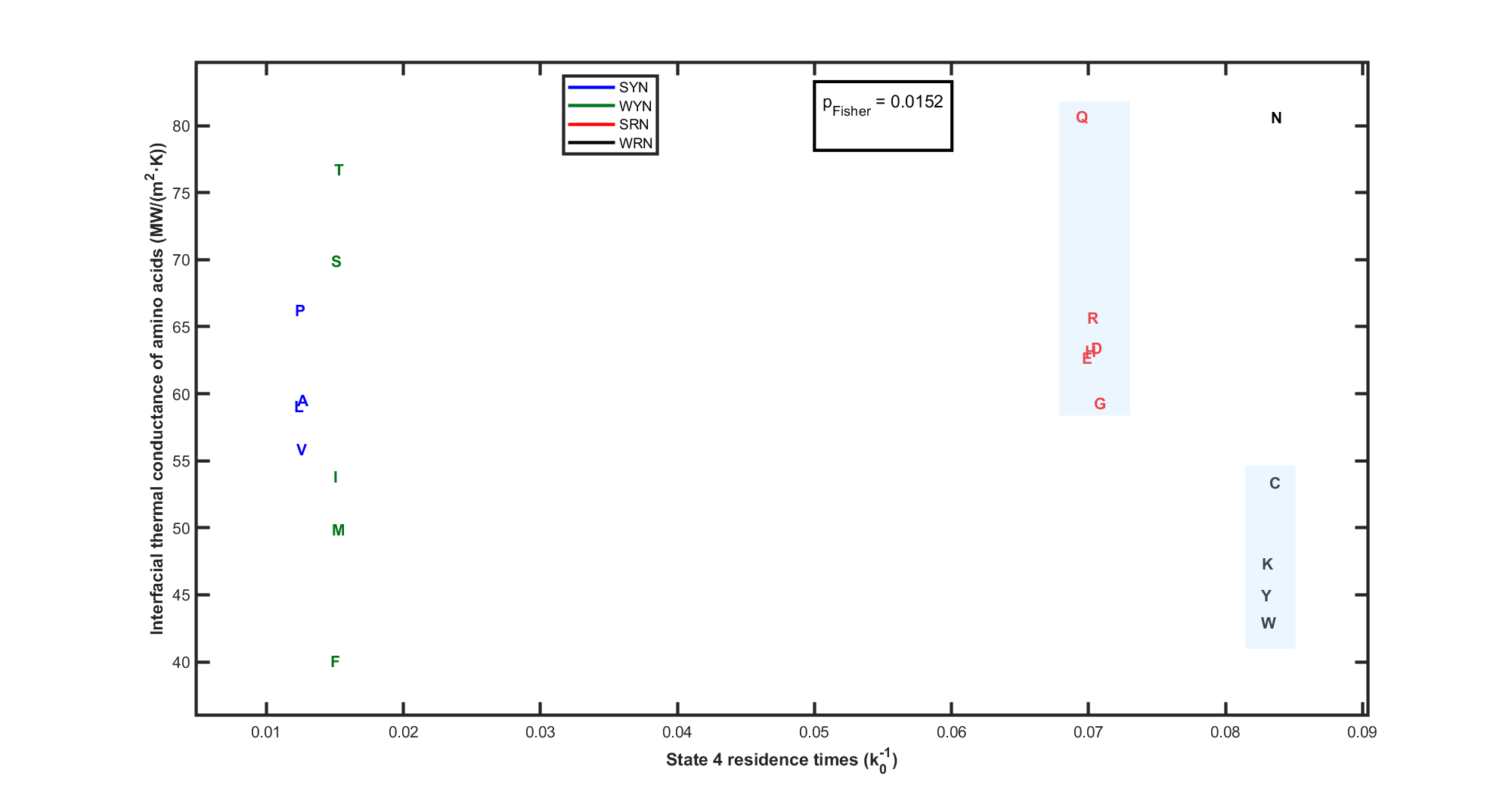}
\caption{Computationally derived interfacial thermal conductance ($MW/m^2.K$), measuring the ability of amino acid surfaces to transfer heat into the surrounding water solvent, versus state 4 residence times. Amino acids having higher interfacial conductance and hence faster relaxation rates at these timescales are paired with faster SRN codons. Again, Asn (N) is an exception. The p-value for segregation of the two groups, highlighted in light blue, is statistically significant. Data from ref.  \cite{Hamzi2022}.}
\label{themalconductance_4}
\end{figure}

\begin{figure}
\centering
\includegraphics[trim=80pt 20pt 20pt 20pt,width=1.05\textwidth]{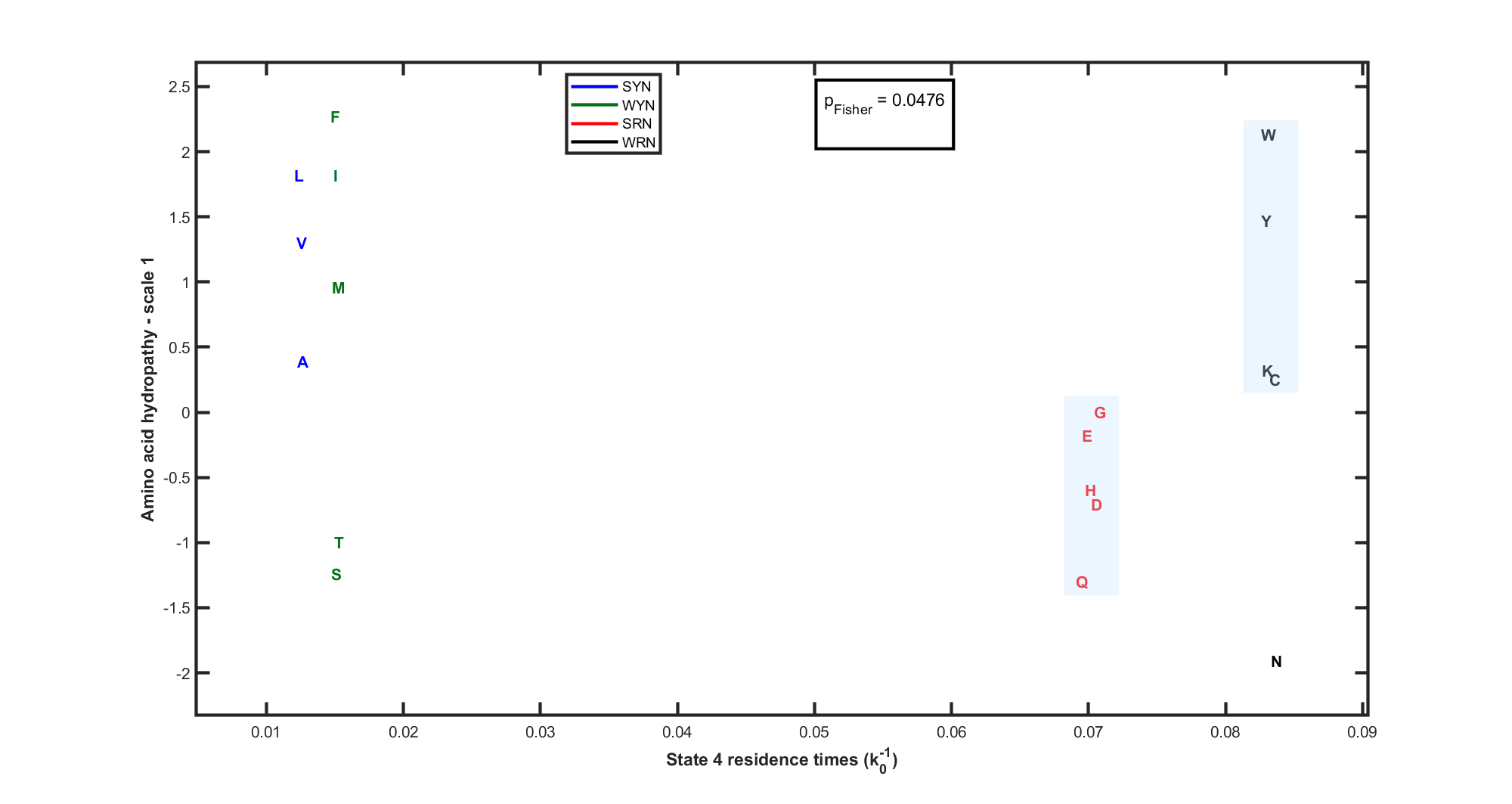}
\caption{Computed hydropathy of amino acids versus state 4 residence times during codon-anticodon association process. Hydropathy is theoretically calculated from transfer free energies of amino acid side chains from water to an organic solvent, octonol, under the structure-additivity assumption. Positive hydropathy values indicate hydrophobic tendency. More hydrophilic amino acids are coded by faster SRN codons, whereas, less hydrophilic ones are coded by slower WRN codons, both groups highlighted in light blue background. Notice that Asn (N) is consistently an exception. The p-value for segregation is statistically significant. Data from ref. \cite{Roseman1988}.}
\label{hydropathy1_4}
\end{figure}

\begin{figure}
\centering
\includegraphics[trim=80pt 20pt 20pt 20pt,width=1.05\textwidth]{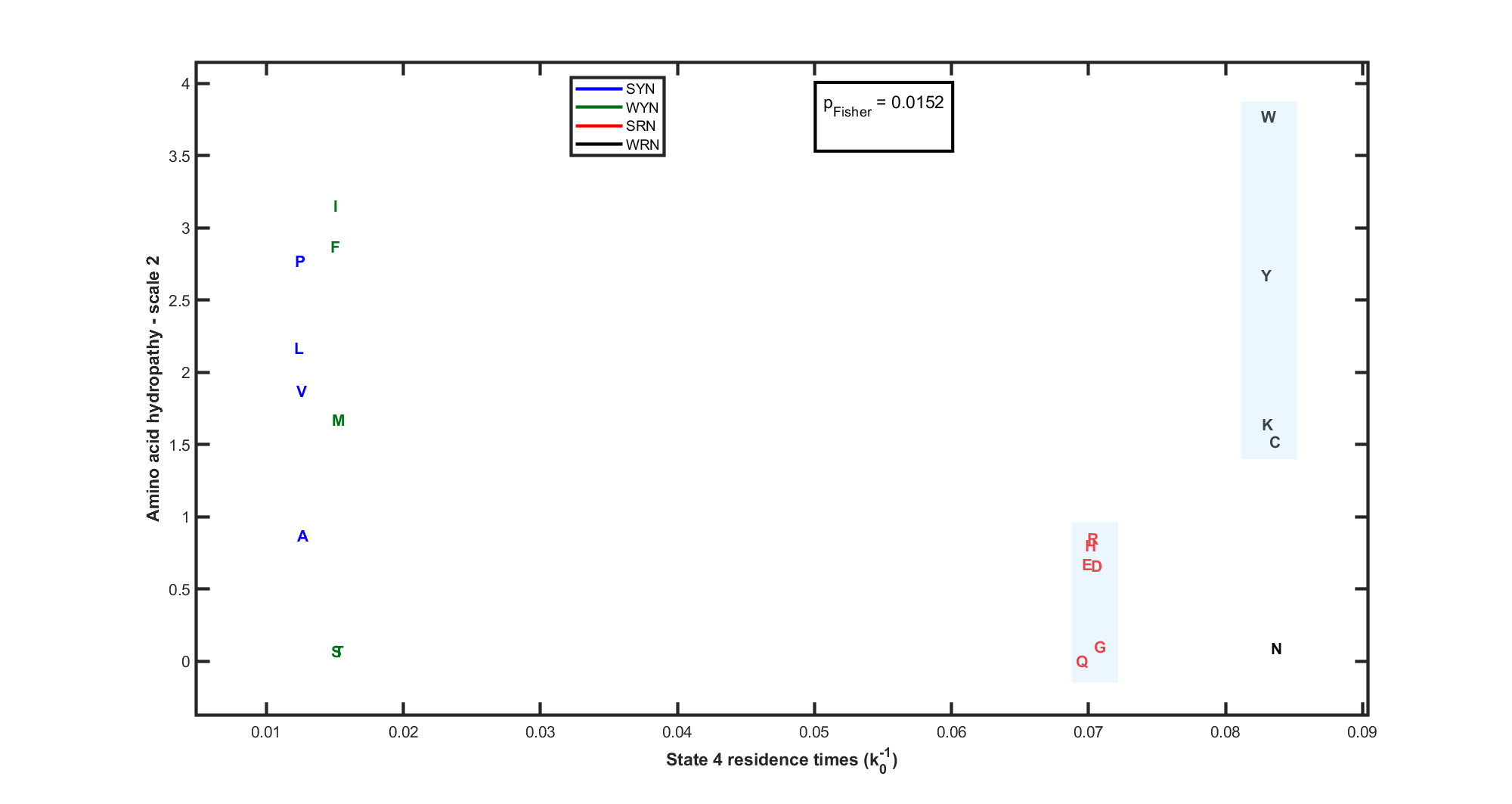}
\caption{Another hydropathy scale shows similar segregation, with significant p-value, as in fig.\ref{hydropathy1_4} of SRN- and WRN-coded amino acids with respect to state 4 residence times, with the exception of Asn (N). Data from ref. \cite{Bhaskaran1988}.}
\label{hydropathy2_4}
\end{figure}

The foregoing is corraborated by the correlations we identified between the state 4 residence times and interfacial thermal conductance and two hydropathy scales (figs.\ref{themalconductance_4}, \ref{hydropathy1_4} and \ref{hydropathy2_4}). Amino acids' interfacial thermal conductance \cite{Hamzi2022}, which measures the ability of a molecular surface to transport thermal energy from the solute to solvent, segregates SRN and WRN-coded amino acids, again with the exception of Asn (N). SRN-coded amino acids have higher conductance, have shorter relaxation times at this timescale, and therefore are assigned to codons with smaller residence times. Similarly, a hydropathy scale \cite{Roseman1988}, theoretically derived from transfer free energies of amino acid side chains from water to an organic solvent, octonol, also segregates the two groups, with more hydrophilic amino acids assigned to faster SRN codons. Another hydropathic \cite{Bhaskaran1988} scale reproduces the above segregation. In all these correlations, Asn (N) is found to be an outlier.

At the timescale of 10 - 100 femtoseconds, librational modes (hindered rotation) dominate the relaxation pathways of amino acids \cite{Rey2015,Yagasaki2009Ultrafast}. Another relaxation pathway is the resonant energy transfer into the OH-stretch vibrational modes of water, which is shown to occur at timescales faster than 100 fs \cite{Woutersen1999}. It is possible that more effective energy transfer into these modes by SRN-coded amino acids with more intermolecular hydrogen bonds compared to WRN-coded amino acids result in the correlations seen above \cite{Laage2017}.

\subsubsection*{Mezzanine level 5M: Quadruplets and doublets}

This level is introduced as a prelude to the next level and classifies all 64 codons into two sets, quadruplets, in which four codons code for an amino acid, and doublets, in which two codons code for an amino acid. The four SYN-, WYN-, SRN- and WRN-coded amino acid sets each split into quadruplet-coded and doublet-coded sets, with the exception of SYN, which has no doublet-coded amino acids, and WRN, which has no quadruplet-coded amino acids.

\begin{figure}
\centering
\includegraphics[trim=80pt 20pt 20pt 20pt,width=1.05\textwidth]{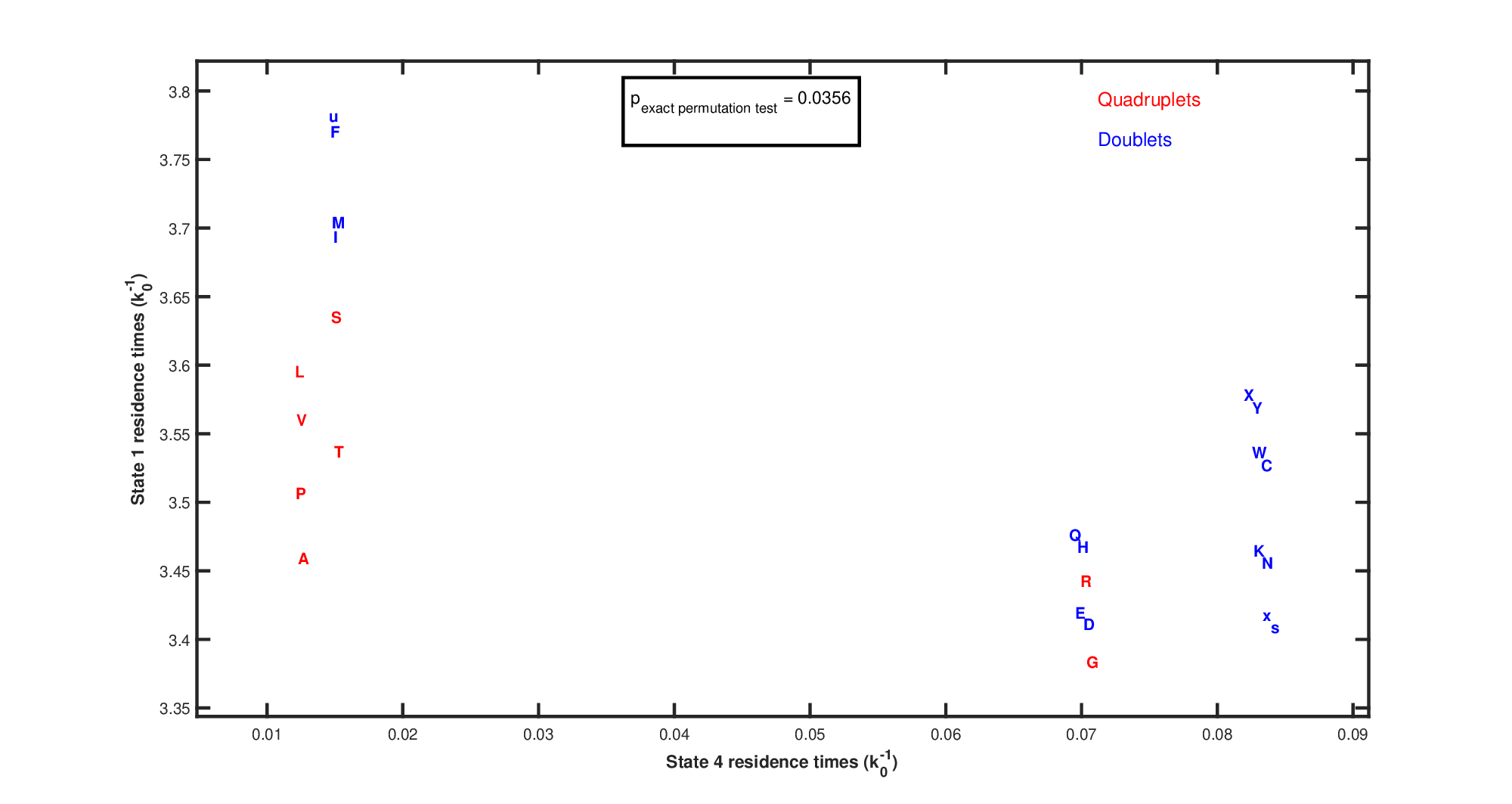}
\caption{State 1 versus state 4 residence times during the codon-anticodon association process, with quadruplet-coded amino acids and doublet-coded amino acids denoted by distinct colors. With the exception of Arg (R), all other amino acids coded by quadruplet codons have smaller state 1 residence times, compared to amino acids coded by doublet codons. State 4 is chosen for the abscissa because this state clusters the four groups SYN, WYN, SRN and WRN cleanly. This statistically significant segregation suggests the presence of a temporal resolution limit in the genetic code deciphering mechanism in the ribosome, below which NNR and NNY codons cannot be discriminated. The amino acids that are coded by six codons (a quadruplet and a doublet) in many alternate genetic codes, Ser (S), Arg (R) and Leu (L), are seen to be present either near the interface between the quadruplet and doublet clusters or amongst the doublets.}
\label{state_14_QD}
\end{figure}

\begin{figure}
\centering
\includegraphics[trim=80pt 20pt 20pt 20pt,width=1.05\textwidth]{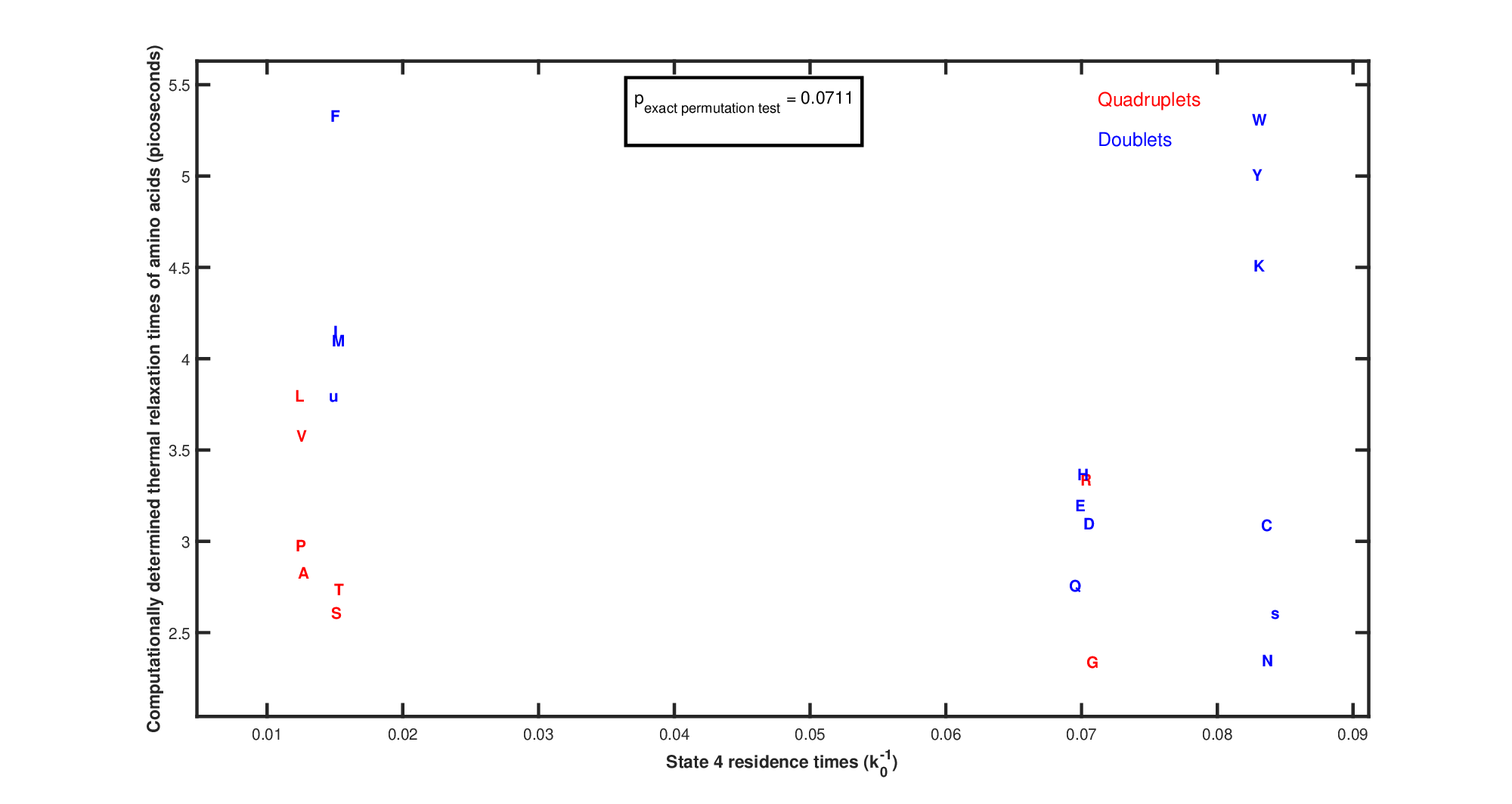}
\caption{The quadruplet-doublet segregation holds at the amino acid front as well, with computationally determined amino acid heat relaxation times at the picosecond time scale exhibiting similar segregation of quadruplets and doublets as in fig. \ref{state_14_QD}. This is not surprising, since we have already shown that the picosecond heat relaxation times correlate with state 1 residence times in fig. \ref{heatrelaxationtimes}. Quadruplet-coded amino acids dissipate heat faster compared to doublet-coded amino acids, and are therefore associated with faster codons. Leu (L) and Arg (R), frequently coded by six codons, are seen at the interface between the two groupings or as an outlier. Due to the small number of permutations possible, the statistical significance is low. Data from ref. \cite{Hamzi2022}.}
\label{heatrel4_QD}
\end{figure}

\begin{figure}
\centering
\includegraphics[trim=80pt 20pt 20pt 20pt,width=1.05\textwidth]{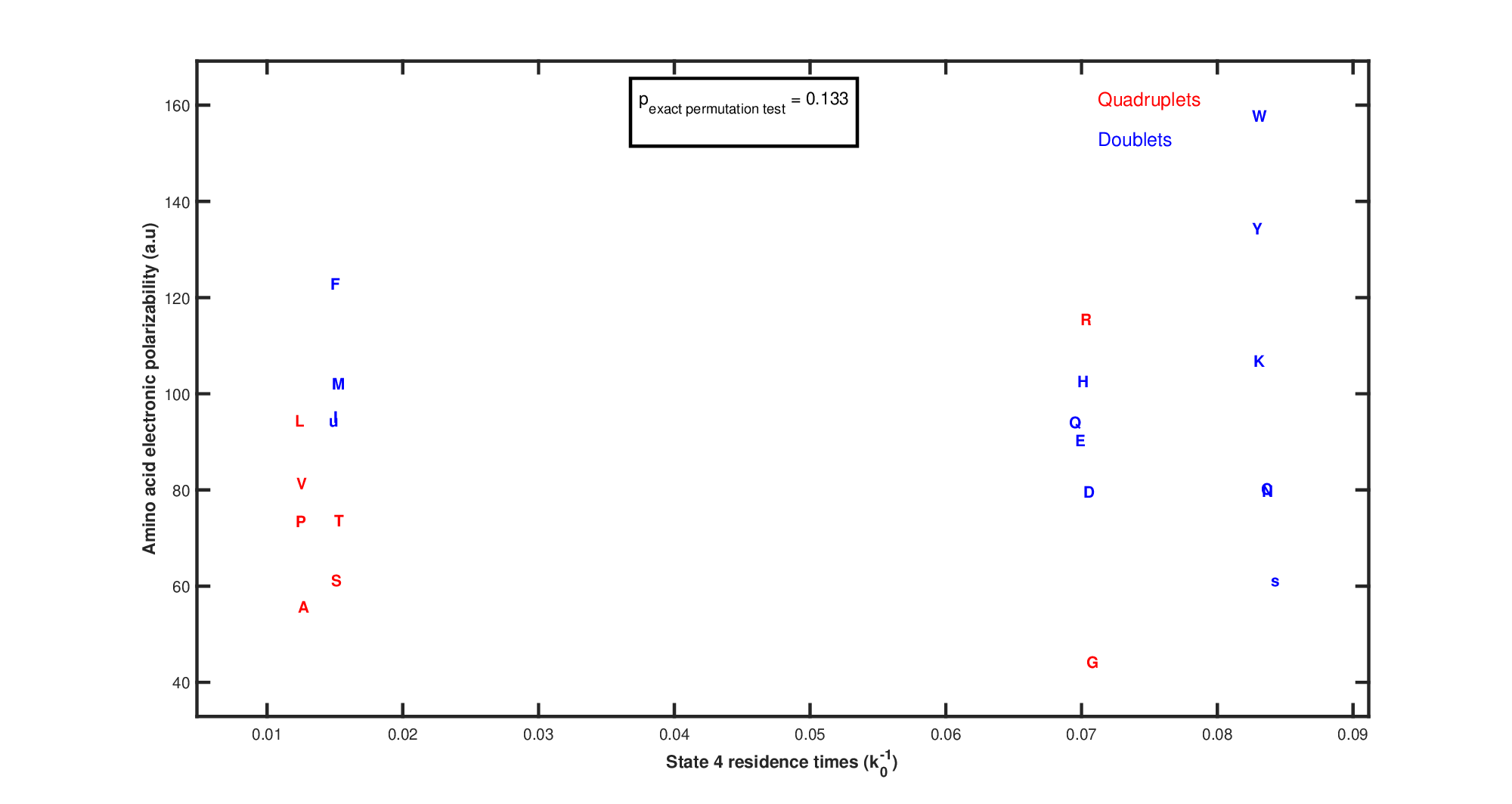}
\caption{As with heat relaxation times, experimentally determined electronic polarizability of amino acids also segregate quadruplet-coded amino acids from doublet-doded amino acids, with amino acids having low polarizability assigned to faster, quadruplet codon groups. As argued in the text, highly polarizable amino acids adjust their electronic densities more effectively to the solvent electronic states, and hence reduce the restoring force, leading to slower charge oscillations, which are vibronically coupled to thermal relaxation. This supports our proposal of the presence of a temporal resolution limit, below which discrimination between NNR and NNY is not possible. Again, Leu (L) and Arg (R) are seen to be closer to doublet cluster. Again, due to the low number of permutation possibilities, the p-value, calculated using exact permutation test, is insignificant. Data from ref. \cite{Millefiori2008}.}
\label{pol4_QD}
\end{figure}

From fig. \ref{state_14_QD}, we see that the quadruplet-coded and the doublet-coded amino acids can be segregated by drawing a straight line through the graph, with the exception of Arg (R). The quadruplet-coded amino acids are coded by faster codons, and doublet-coded ones, by slower codons. Such segregation suggests that there exists a temporal resolution limit in the genetic code decoding mechanism, below which the difference in the kinetics of association of NNY and NNR codons with their corresponding anticodons cannot be resolved. Below this resolution limit, codons are four-fold degenerate, and above it, two-fold degenerate. Curiously, the amino acids that are coded by both four-fold degenerate and two-fold degenerate codons in many alternate genetic codes, Ser (S), Arg (R) and Leu (L), are seen to be present either near the interface between the quadruplets and doublets, or as an outlier. Since the absolute residence times of synonymous codon groups were unconstrained during optimization to fit the genetic code, and only relative residence time vector distances between these groups were optimized, in part-II, the observation of such segregation between doublets and quadruplets is unexpected, and is therefore an emergent property. On the amino acid end, heat relaxation times \cite{Hamzi2022}, which was shown to correlate with state 1 residence times earlier, and electronic polarizabilities \cite{Hamzi2022} segregate the quadruplet and doublet groups well, with the exception of Arg (R), as seen in figs. \ref{heatrel4_QD} and \ref{pol4_QD}. Amino acids having smaller heat relaxation times and smaller polarizabilites are seen to be coded by four-fold degenerate codons. Again, Leu (L) and Arg (R), frequently coded by six codons, are seen either near or within the doublet region in these plots.

\subsubsection*{Fifth level: NNY-NNR, 0.0001 - 0.01 picoseconds}

In the fifth and the final level of the hierarchy, each of the eight sets of doublet-coded amino acids or stop codons above are further classified into amino acids coded by codons containing either a pyrimidine (Y) or a purine (R ) in the third, wobble position. The timescale associated with this segregation is about 0.0001 - 0.01 picoseconds. This timescale assignment is arrived at from fig. \ref{state_14} and from state 6 residence times. At this level, we include the doublet codons coding for Leu (u), Ser (s) and stop codons (X) in the plots. The plot \ref{state_14} shows, interestingly, that the canonical stop codon UAR occupies a special location, where it occupies the slowest column in the state 4 residence time (X axis), and, in that column, the slowest position vertically, along the state 1 residence time (Y axis). The amino acids' physicochemical property that appear to determine the segregation of amino acids according to their cognate codons’ third position nucleotide type is a steric parameter called STERIMOL side chain length, measured in the direction in which the side chain is attached to the glycine backbone, as calculated by the STERIMOL computer program \cite{FAUCHRE1988,Verloop1976}. From figs. \ref{sterimol5} and \ref{sterimol6}, one can see that amino acids with a side chain length of around 5 units are coded by codons with a pyrimidine in the third position, whereas, amino acids with longer side chains are coded by codons with a purine in the third position. Amino acids with side chain lengths shorter than about 4.5 units are all coded by quadruplet codons. This side-chain length dependence on nucleotide type at the third codon position has been noted earlier in \cite{Harrison2022}(fig.4). Exceptions to the above observation are Ser (s), Leu (L,u) and Arg (R), all frequently coded by both quadruplets and doublets in alternate genetic codes.

Fig. \ref{sterimol5}, which plots STERIMOL side chain length of amino acids against state 5 residence times of codon-anticodon association process, shows interesting correlations. All quadruplet-coded amino acids with side chain length below about 4.5 units show an approximately linear relationship with the state 5 residence times, with the exception of Thr (T). NNY- and NNR-coded amino acids show an even more interesting pattern, where, the difference between the side chain lengths of amino acids coded by codons with the same first and second nucleotides linearly decreases with increasing state 5 residence times. Note that the timescale of state 5 is too large to distinguish the time difference between the amino acids coded by codons with the same first and second nucleotides. To observe the difference in the timescales of NNR and NNY-coded amino acids, we use state 6 residence times, the shortest of the residence times of all the states. The time difference between NNY- and NNR-coded amino acids, such as (N,K), (D,E), (C,W), (I,M) and (H,Q) is of the order of 0.0001 - 0.01 picoseconds, as seen from the fig. \ref{sterimol6}. One can observe in the figure that NNR-coded amino acids with larger side chain lengths in the above pairs are paired with codons with larger residence times in state 6.

Since the sub-femtosecond timescales of state 6 correspond to energies in the electronvolts range, we explored the relationship between the state 6 residence times of codon-anticodon association process and the inverse of HOMO-LUMO gaps of all 20 amino acids \cite{Krishtal2009}, which are of the order of electronvolts. Fig. \ref{homo_lumo_6} plots softness of amino acids, which is defined as the inverse of HOMO-LUMO gap at 0 K \cite{Pearson2005}, converted to the units of femtoseconds, and the state 6 residence times. Except for the six amino acids R, W, S, P, L and Q, we see an approximately linear relationship between the two quantities in the figure. Please note the matching of the sub-femtosecond timescales along X and Y axes.

The six amino acids R, W, S, P, L, and Q (including the stop codons), which are outliers in the fig. \ref{homo_lumo_6}, are either fully or partially coded by codons with a YNR motif. Please recall from part II of this series that this sequence motif results in a high kinetic barrier for the middle nucleotide N, due to the asymmetrically cooperative inhibitory kinetic influences of the pyrimidine on the left and the purine on the right on the middle base pair,  resulting in slow transitions from and to the state 6 (state 101). This kinetic inhibition slows down the codon-anticodon association process at these timescales and isolates the codons corresponding to these six amino acids from other codon groups along the X-axis in the fig. \ref{homo_lumo_6}.


When we attempt to interpret the above results, we face a limitation imposed by our modeling choice: our classical Markov chain model is constructed to describe hydrogen-bond formation and dissociation between codon and anticodon nucleotides on picosecond timescales. At ultrafast timescales (approximately 0.1–10 fs), however, nuclear motion is minimal and the dynamics are dominated by electronic degrees of freedom that are not explicitly represented in the model. In this regime, we are effectively extrapolating our classical model beyond its domain of validity, and the sub-femtosecond residence times should not be regarded as quantitatively reliable physical observables. We therefore include here the caveat that the interpretation that we provide below is more speculative, and is provided solely for hypothesis generation purposes. The correlations that we find in figs.\ref{sterimol5}, \ref{sterimol6} and \ref{homo_lumo_6} need to be independently validated by more appropriate quantum mechanical models or by experiments that are capable of resolving electronic dynamics. More importantly, the energetics of the interaction at this timescale between codon-anticodon system and its cognate amino acid is not favorable: while the codon-anticodon binding releases energies of the order of millielectronvolts, the HOMO-LUMO gaps are of the order of electronvolts. Even when we take into account the energies of O-H/N-H vibrational quanta, which are of the order of an electronvolt, we have to combine multiple such quanta to match the HOMO-LUMO gaps of amino acids, making such interactions very improbable.

Keeping in mind the above caveat, we speculate that the matching of timescales and the resulting correlation between the residence times in state 6 of codon-anticodon association process and the electronic oscillation time-periods of cognate amino acids (fig. \ref{homo_lumo_6}), measured by the inverse of the HOMO-LUMO gap, again suggests some form of energy transfer from the codon-anticodon system to the amino acid system, although the quantum of energy and the real timescales of transfer are not quantifiable within the model. The timescale matching allows for efficient electronic energy transfer via tRNA, akin to impedance matching in electrical systems and electronic circuits, albeit in the quantum regime. The correlations of state 6 residence times with STERIMOL side chain length fits with this interpretation, since amino acids with larger side chains generally would have smaller HOMO-LUMO gaps, and hence slower electronic oscillations, which are therefore assigned to slower codons in fig. \ref{sterimol6}.  Since dissipation of energy into the environment of the amino acid thermodynamically is usually not possible at these ultrafast timescales, the environment instead serves possibly as an agent of decoherence, leading to information loss due to the loss of coherence in the electronic coupling between the codon-anticodon system and the cognate amino acid, and hence to entropy production, locally at the amino acid end \cite{oppenheim2002}. If evolution optimized for efficient energy transfer, a convergence of timescales of energy supply, absorption and decoherence would result \cite{Alterman2024,Lloyd2011,Shabani2014,Cao2009,Skourtis2002}, as in photosynthesis, which is a signature of environment-assisted quantum transport (ENAQT) \cite{Rebentrost2009}. We have reported on the convergence of the supply and absorption timescales in fig. \ref{homo_lumo_6}, while the timescales of decoherence of all the 20 amino acids in aqueous medium have not been quantified in the literature and hence is not available. Although we cannot determine the absolute decoherence timescales, we can roughly estimate the trend of decoherence timescales by correlating the propensity of amino acids to interact with water (system-bath interactions), that is, their hydropathy \cite{Roseman1988}, with state 6 residence times. Fig. \ref{hydropathy_6} shows that, with the exception of the six amino acids listed above, more hydrophobic amino acids map to slower cognate codons, suggesting that slower codons code for amino acids with weaker system-bath interactions, and therefore slower decoherence times.

We speculate that the loss of correlation between the HOMO-LUMO gap inverses and hydropathies of the six amino acids (R, W, S, P, L, Q and the stop codons), with state 6 residence times in figs. \ref{homo_lumo_6} and \ref{hydropathy_6} as due to opening up of dissipative channels (e.g. vibronic coupling) in the amino acids at the slower end of this timescale range (a few femtoseconds), resulting in the broadening of the ``resonance'' peak in energy transfer, and hence loss of specificity of mapping. This can be tested by quantifying the error rates or misincorporations of amino acids in the polypeptide chain corresponding to these YNR codons during translation and comparing them to non-YNR codon error rates. It is interesting to note that these same six amino acids are frequently re-assigned to different codon groups in alternate genetic codes (\url{https://www.ncbi.nlm.nih.gov/Taxonomy/taxonomyhome.html/index.cgi?chapter=cgencodes}), with approximately 70 percent of the re-assignments arising from these six amino acids (and stop codons) alone.



\begin{figure}
\centering
\includegraphics[trim=80pt 20pt 20pt 20pt,width=1.05\textwidth]{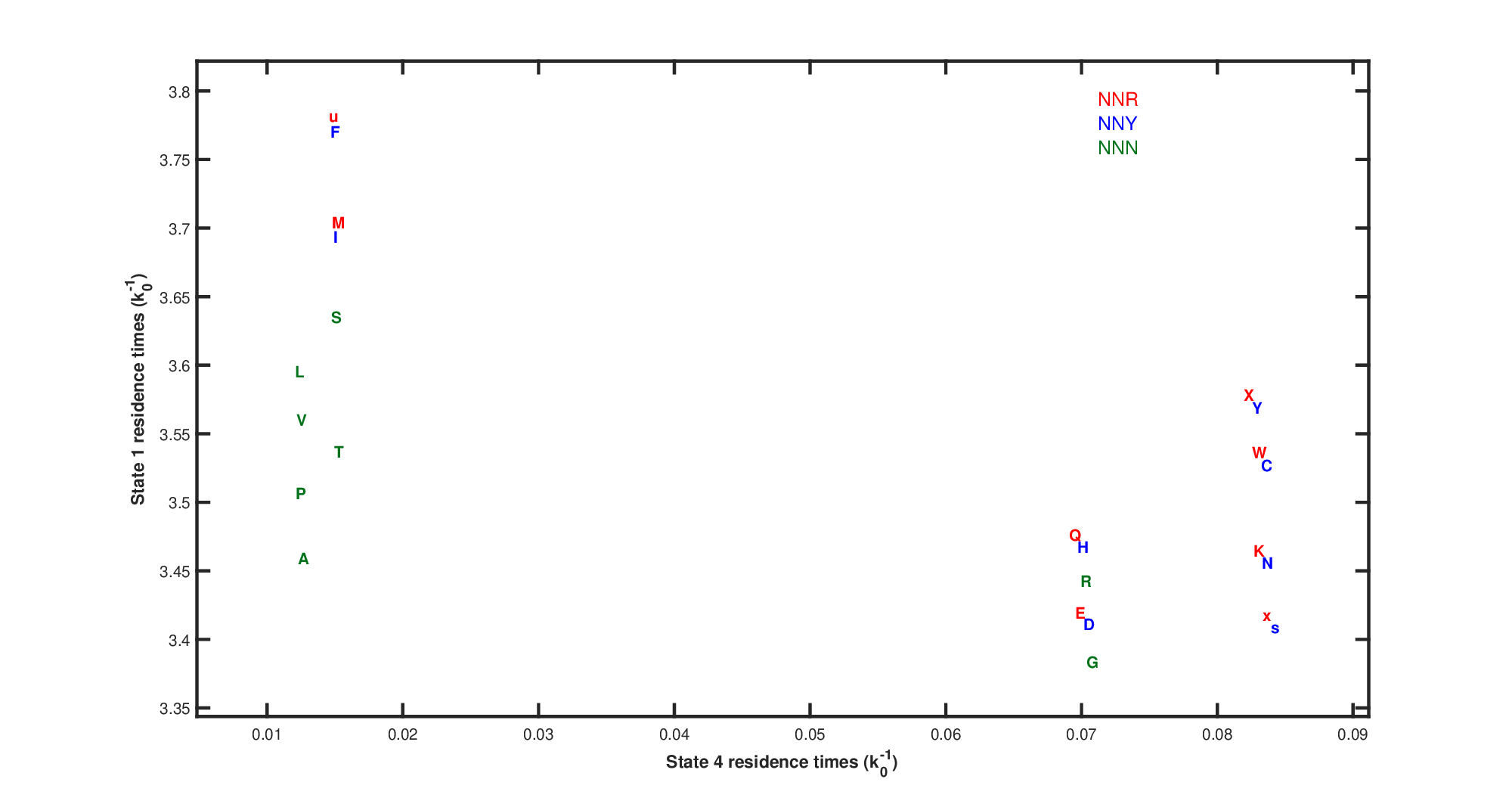}
\caption{State 1 residence times versus state 4 residence times during codon-anticodon association process, now highlighting the small time scale difference between NNY- and NNR-coded amino acids, by color-coding them differently. The time scale of the processes distinguishing NNY from NNR codons is in the range of 0.0001 - 0.01 picoseconds, as we will see in fig. \ref{sterimol6}. The plot includes the doublet codons coding for Leu (u) and Ser (s), and stop codons (X). Note that the canonical stop codon doublet, UAR, is in the slowest column, and in it, at the slowest position. NNR codons are seen to be slower than NNY codons. This shows that it would be erroneous to generalize by associating the presence of purines (R) in the codon sequence to faster codons and of pyrimidines (Y) to slower codons, as observed in the NYN-NRN splitting at 0.1 - 1 ps time scale, where the presence of purines lead to faster codons. Whether purines and pyrimidines in the codon sequence leads to faster or slower association process depends on the position of them within the codon sequence, and on the specific time scale in the multi-timescale association process. }
\label{state_14}
\end{figure}

\begin{figure}
\centering
\includegraphics[trim=80pt 20pt 20pt 20pt,width=1.05\textwidth]{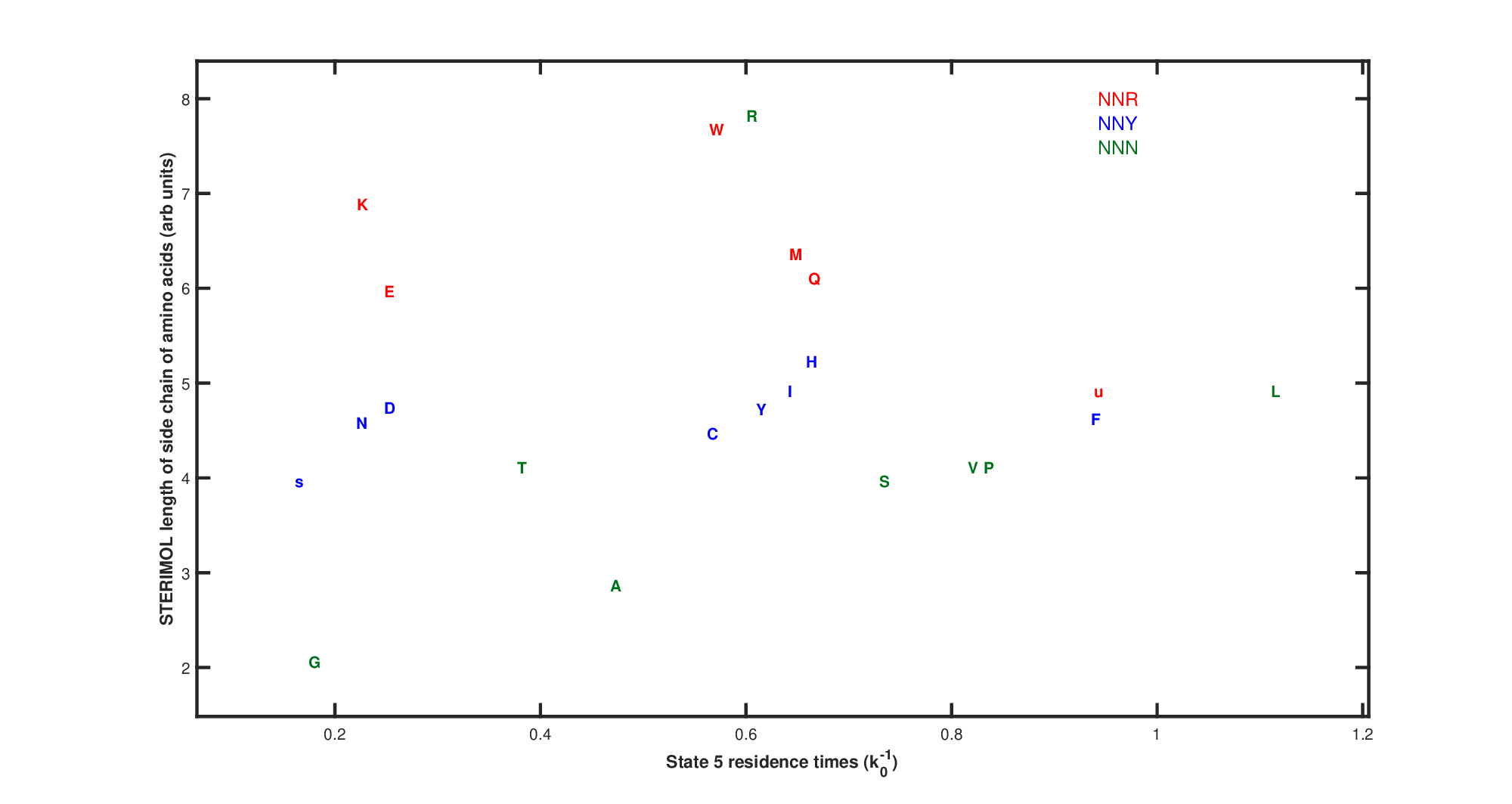}
\caption{STERIMOL side chain length of the 20 amino acids, evaluated using the STERIMOL program, in arbitrary units, versus state 5 residence times in the codon-anticodon association process. The NNR- and NNY-coded amino acids are only segregated along the ordinate, and not along the abscissa. With a few exceptions (Phe (F), Thr (T) and Arg (R)), one can observe regularities in the plot: the residence times of quadruplet-coded amino acids are linearly related to the side chain length, whereas, the doublet NNY- and NNR-coded amino acids show an interesting pattern, with the two groups appearing to linearly converge towards a common length. Since these regularities are very difficult to capture in a model, no statistics is done. The NNR stop codons corresponding to the NNY-coded Ser (s) and Tyr (Y) are not shown. The biophysical origin of this regularity cannot be deciphered within the current top-down approach. Dedicated bottom-up approaches are required to understand this relationship. Data from \cite{FAUCHRE1988}.}
\label{sterimol5}
\end{figure}

\begin{figure}
\centering
\includegraphics[trim=80pt 20pt 20pt 20pt,width=1.05\textwidth]{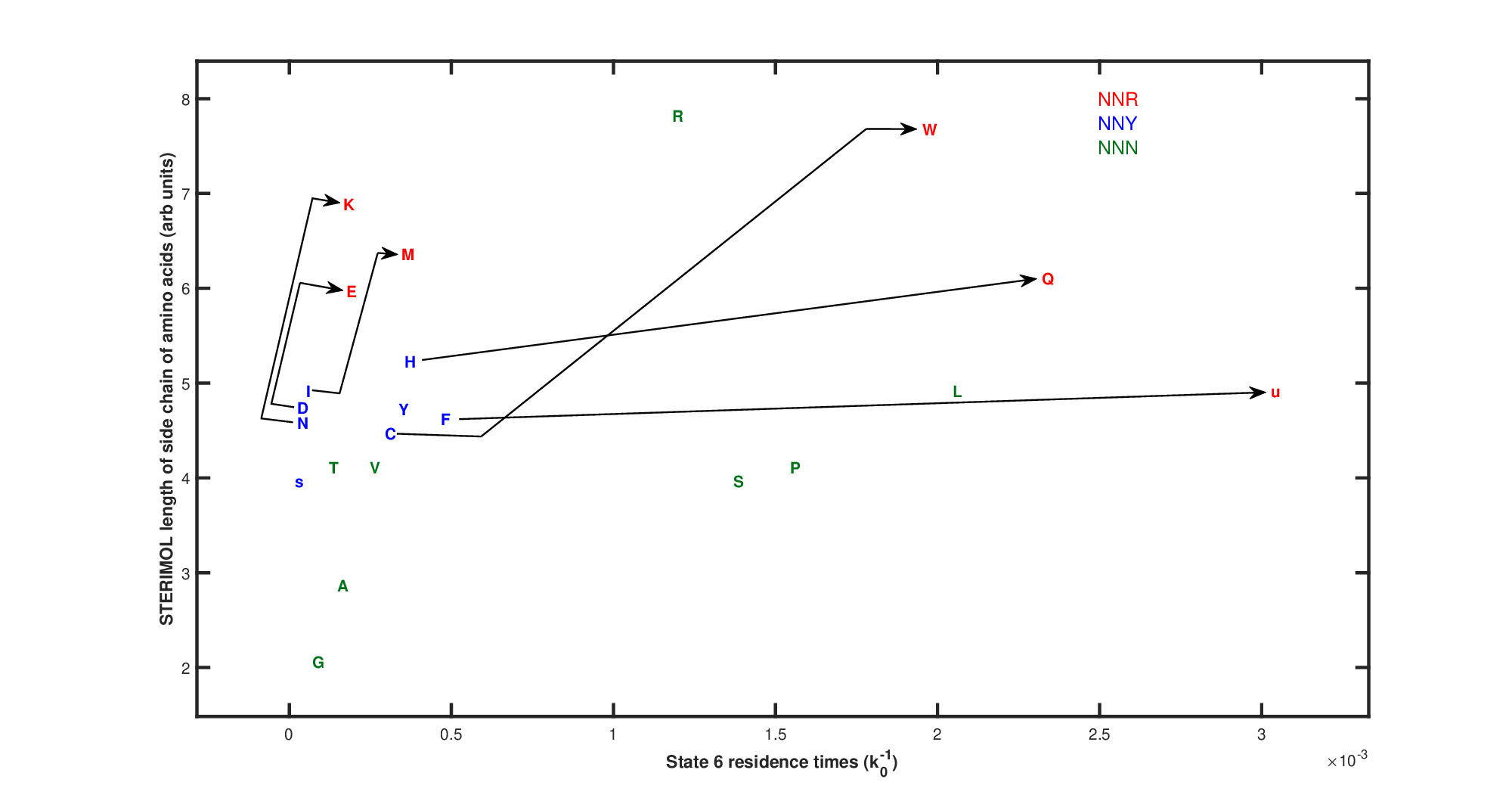}
\caption{STERIMOL side chain length of the 20 amino acids versus state 6 residence times. The time scale separating the NNY- and NNR-coded amino acids is in the range of 0.1 - 10 femtoseconds. Arrows connect all NNY-NNR pairs with the same first two nucleotides NN. This plot shows that NNR-coded amino acids have longer side chain lengths and larger state 6 residence times than NNY-coded amino acids, given the same first two nucleotides NN. Data from \cite{FAUCHRE1988}. }
\label{sterimol6}
\end{figure}

\begin{figure}
\centering
\includegraphics[trim=80pt 20pt 20pt 20pt,width=1.05\textwidth]{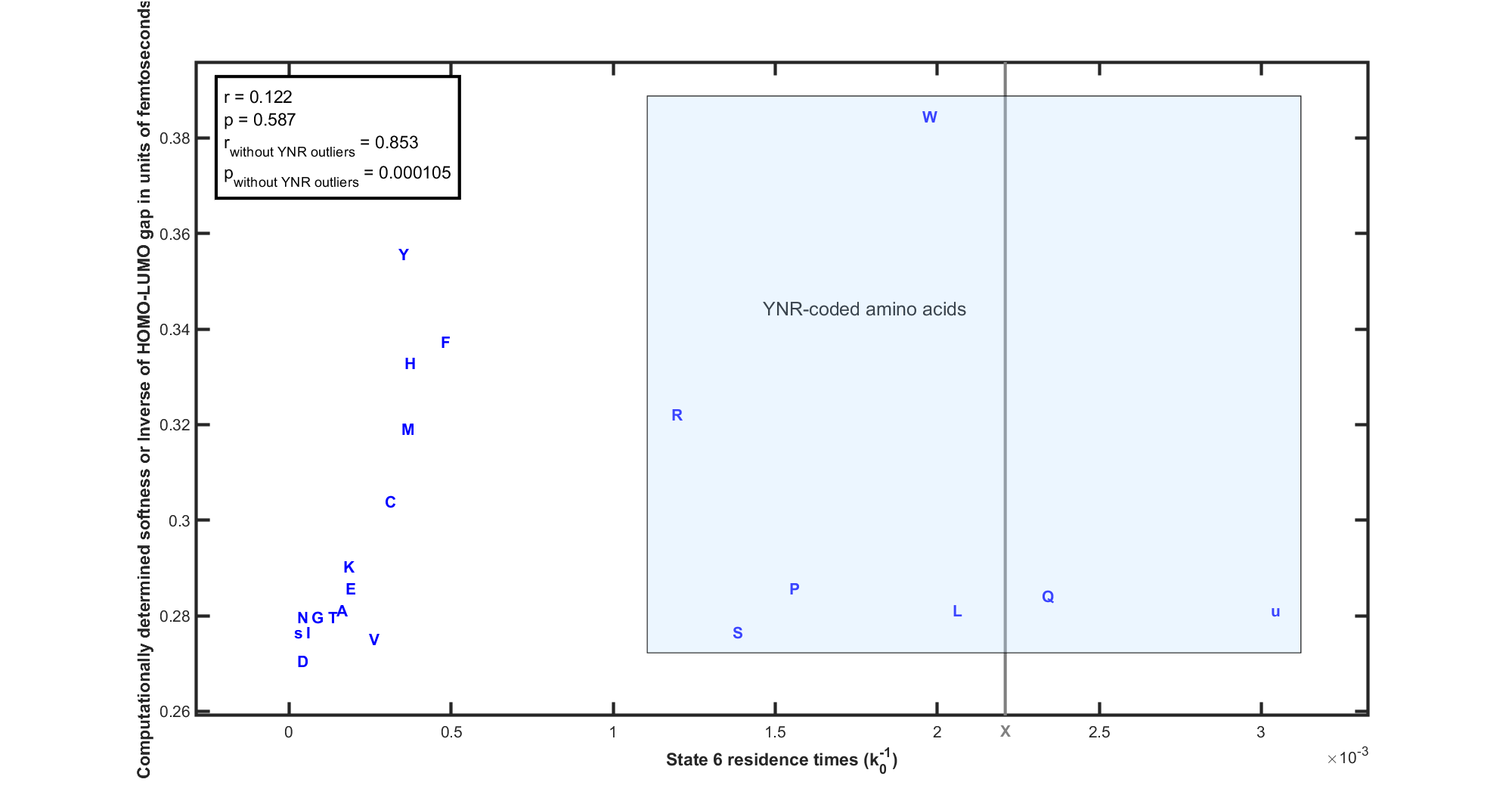}
\caption{Inverse of HOMO-LUMO gaps or softnesses of 20 amino acids in femtosecond units versus state 6 residence times of the corresponding cognate codon groups. Note the matching of timescales (0.1 femtoseconds) of both the axes for the amino acids that show correlation. Except for six amino acids coded by YNR codons, significant correlation between the electronic oscillation timescales of amino acids and the state 6 residence times can be observed. Two correlations are reported, one with YNR coded amino acids included, and one after excluding them. The latter correlation is statistically significant. UAR stop codons are denoted by an X along the X-axis. Note that the location of the stop codon along the X-axis, denoted by a gray line, is close to that of Trp (W) and Gln (Q), which commonly exchange places with Stop (X) in alternate genetic codes. This is due to the optimization procedure explained in part II of this series. Leu (u), which includes the start codon UUG, is at the right-most end. We speculate that this partial correlation arises due to energy transfer between the codons and the cognate amino acids at these timescales. The outlier amino acids R, W, S, L, and Q, excluding Pro (P), are frequently reassigned to different codon groups in alternate genetic codes. Data from \cite{Krishtal2009}.}
\label{homo_lumo_6}
\end{figure}

\begin{figure}
\centering
\includegraphics[trim=80pt 20pt 20pt 20pt,width=1.05\textwidth]{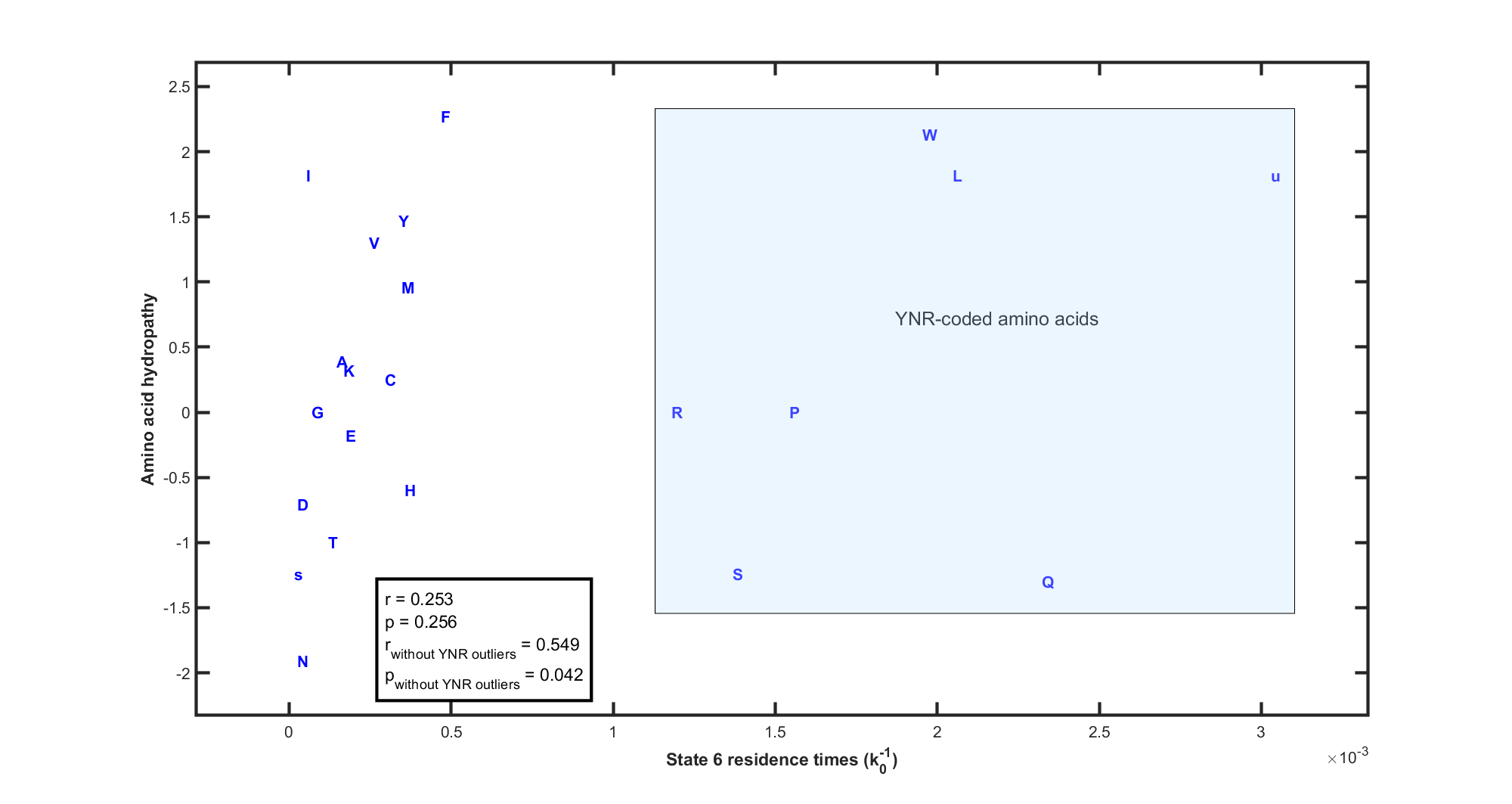}
\caption{The decoherence rates of electronic interaction between codon-anticodon association process and its cognate amino acid through tRNA depends on system-bath interactions between the amino acid and its environment, quantified here by hydropathy. The plot shows the variation of hydropathy of amino acids with state 6 residence times. Again, two correlations are reported, one including YNR-coded amino acids, and another without them, with the latter being statistically significant. For amino acids with small residence times, one can observe a linear relationship, suggesting that the system-bath interactions are weaker for amino acids coded by slower codons. This may lead to longer decoherence times, possibly resulting in the convergence of timescales of energy supply, dissipation and decoherence, a signature of environment-assisted quantum transport. Data from \cite{Roseman1988}.}
\label{hydropathy_6}
\end{figure}

\subsubsection*{Correlation with evolutionary order of amino acid recruitment}

\begin{figure}
\centering
\includegraphics[trim=80pt 20pt 20pt 20pt,width=1.05\textwidth]{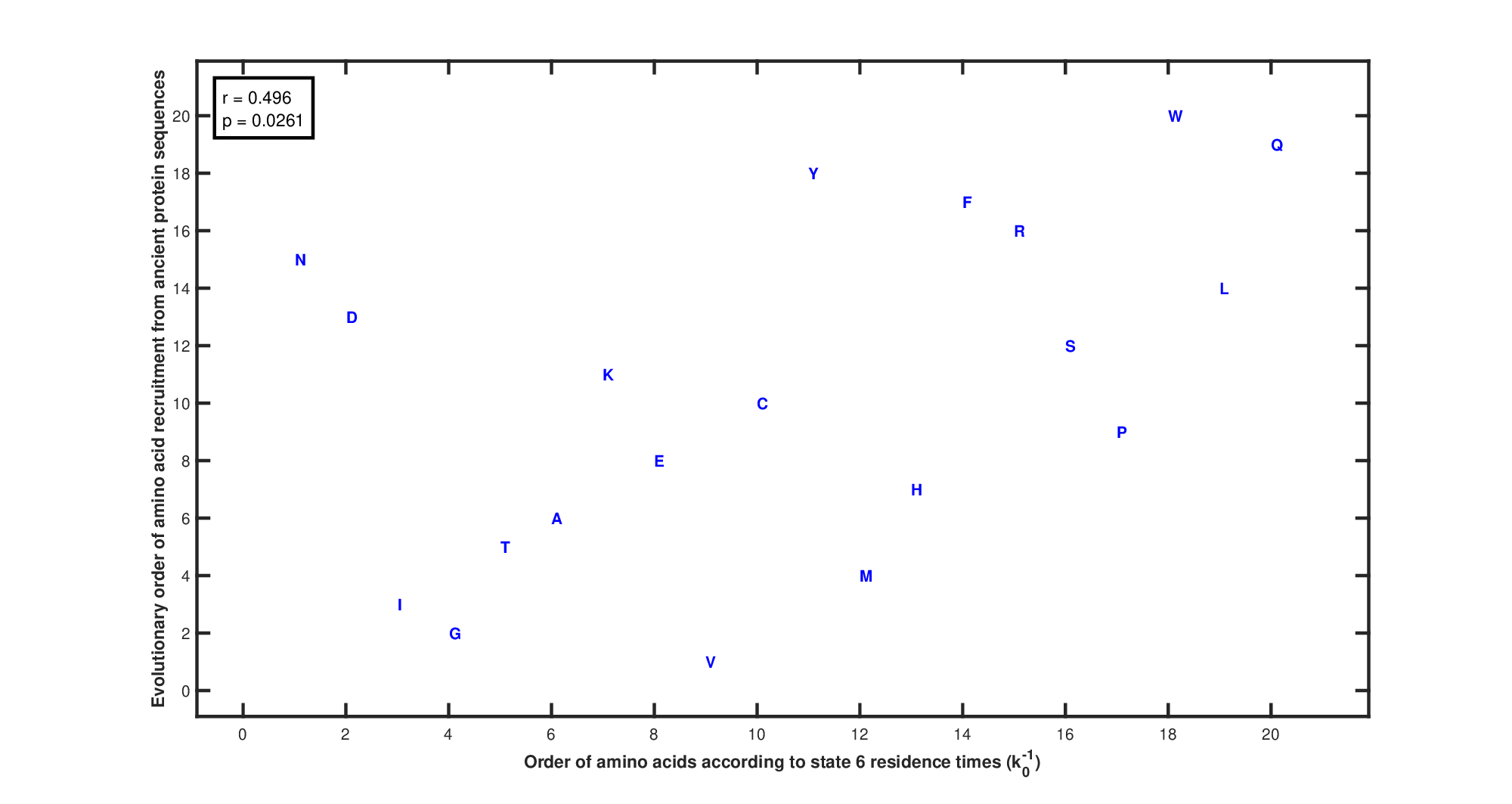}
\caption{Evolutionary order of amino acids derived from the proteins of LUCA (\cite{Wehbi2024}) versus the ordering of amino acids according to the state 6 residence time of codon-anticodon association process, from the smallest to the largest times. With the exceptions of Asn (N) and Asp (D), the state 6 residence time ordering appears to roughly follow the evolutionary ordering determined from phylogenetic analysis. The correlation is statistically significant. }
\label{evolutionaryorder6}
\end{figure}

The importance of each timescale mentioned above for energetically stabilizing the interacting codon-anticodon system and for dissipating the energy generated without damage to the mRNA can be gauged by calculating the power transferred from the codon-anticodon system to the amino acid system at each timescale. As can be seen above, the sub-femtosecond timescales carry most power away from the codon-anticodon system, due to the large energy scales (electronvolts) and very small timescales involved (sub-femtoseconds). Therefore, under the assumption that evolution's purpose was to transfer as much energy away from the actively-transcribed mRNA (in prokaryotes) as quickly as possible for its stability, evolution would have recruited amino acids in the ascending order of their shortest timescales, i.e., the inverse of their HOMO-LUMO energy gaps. Another reason for the importance of the fastest sub-femtosecond timescales in the evolutionary ordering of amino acid recruitment lies in the specificity of mapping of the genetic code. At these ultrafast timescales, environmental degrees of freedom are mostly quiescent and the phase space into which energy dissipation and decoherence happens is much smaller, leading to enhanced specificity of molecular recognition events, with the assignment of the fastest amino acids being more specific. A case has already been made for a correlation between amino acids' HOMO-LUMO gaps and their evolutionary order of appearance (fig.1 of \cite{Granold2018})  on the basis of the evolutionarily late amino acids' redox activity, with small-gap, redox-active, amino acids recruited later. Motivated by the correlation between inverse HOMO-LUMO gaps and state 6 residence times, observed in fig.\ref{homo_lumo_6}, we correlate the state 6 residence times of codon groups corresponding to the 20 amino acids with the consensus evolutionary ordering or amino acid recruitment, obtained from phylogenetic analysis \cite{Wehbi2024}. Fig. \ref{evolutionaryorder6} shows that there is a correlation (with the exception of asn (N) and asp (D)) between the order of state 6 residence times, which is the shortest of the residence times of all the seven states,  and the evolutionary order of appearance of amino acids in LUCA. The correlations of the evolutionary order of amino acids with other states with larger residence times are minimal. This correlation leads us to speculate that the evolutionary rationale for the sequential recruitment of amino acids is for the stabilization of the mRNA undergoing translation, through efficient energy transfer from the codon-anticodon system to the amino acids, with the highly efficient amino acids recruited earlier.

\section*{Discussion}

The physical origin of the genetic code, by which we mean the physical process that mediates the interactions between codons and amino acids that underlie the genetic code mapping, remains elusive, despite considerable attention having been paid to the problem. Much of the effort has concentrated on finding structural, short range, specific interactions between codons and their cognate amino acids, limiting the search to thermodynamic equilibrium interactions. For a process that lies at the core of life, with the latter helpfully defined as agents that efficiently and rapidly convert free energy available in the environment to entropy without destroying themselves, investigating the non-equilibrium origin of the genetic code is obviously more appropriate. In non-equilibrium, kinetics play a crucial role in determining the dynamics of a system, which living systems have taken advantage of, by creating catalysts that significantly accelerate energy transformation processes that would take much longer times if performed abiotically. This has enabled living systems to compete with abiotic processes successfully in degrading free energy, after utilizing part of the energy for self-organization, thereby gaining foothold even in apparently hostile environments.

By creating a model of heteropolymer replication that explicitly addressed the \textit{kinetics} of replication (a non-equilibrium process), instead of thermodynamics, we have been able to connect sequence information with replication kinetics, and explain some fundamental properties of DNA and RNA, apart from the emergence of prebiotic information predating transcription and translation \cite{subramanian2018evolutionary,subramanian2020evolutionary}. This model, when appropriately parametrized, allows us to calculate the rates of codon-anticodon association process as well. This is an example of a top-down approach \cite{Szathmry1999Origin}, as opposed to a bottom-up approach, in understanding the structural characteristics of biomolecules, where, we have utilized evolutionary reasoning to identify or rationalize molecular properties that are beneficial for self-replication. We have shown, in part I and II, that this model can reproduce the genetic code organizational structure, by which we mean the dependence of genetic code degeneracy on the class of nucleotides (strong/weak and purine/pyrimidine) present in the three positions of the codon. We validated our model's results by correlating them with experimental data derived from ribosomal footprinting experiments that measure the residence times of ribosomes at every codon of mRNA.

In this part of the series, we have shown that the very same \textit{sequence-dependent} zipping/unzipping kinetics, that dictates the rate of  DNA/RNA self-replication, thereby enabling evolutionary competition for resources between different sequences and kick-starting prebiotic evolution\cite{subramanian2020evolutionary}, also possibly underlie the origin of the genetic code. Specifically, we argue that the multiple timescales at which codon-anticodon association process happens are correlated with similar timescales of relaxation processes happening at the cognate amino acid's end. This matching of energy supply timescales from the codon-anticodon association process with the energy dissipation timescales from the codon's cognate amino acid to the surrounding aquatic medium, suggests maximization of the efficiency of an electronic energy transfer process operating between the two systems. At the timescales and lengthscales involved, the energy transfer must be through near-field interactions (within or without the tRNA). Moreover, the presence of at least four timescale ranges (1-10, 0.1-1, 0.01-0.1, 0.0001-0.01 picoseconds) over which the supply-dissipation matching happens between codons and cognate amino acids, suggests a lock-and-key-like mechanism which brings about the specificity of the genetic code, with each timescale acting as a lever in the lock, replacing the frequently invoked spatial complementarity with temporal complementarity.

A crucial question that remains to be addressed is the apparent inconsistency between the timescales of ribosome profiling experiments, which were shown to correlate with the total codon-anticodon residence times in part II of this series, and the timescales for the residence times we arrive at in this part of the series. The timescale of ribosome movement along the mRNA track is of the order of $10^{-1}$ seconds per amino acid \cite{Sharma2019}, whereas the codon-anticodon association timescales are of the order of $10^{-12}$ seconds, from the correlations shown above. The correlation between ribosome residence times at various codons on mRNA and the codon-anticodon association process residence times, observed in figs. 5 and 6 of part II of this series, then appears improbable, due to this large timescale separation. This requires an explanation for the mechanism behind the ability of ribosomes to translate picoseconds-scale codon-anticodon dynamics to seconds-scale ribosomal dynamics, which would result in the correlation seen in figs. 5 and 6.

We suggest that the amplification (or gating) mechanism(s) that connect fast timescale dynamics to slow timescale dynamics in the following biological processes may also underlie the multi-timescale ribosome dynamics mentioned above: (a) magnetoreception of cryptochrome proteins in migratory birds, where quantum dynamics of photo-induced radical pairs in the proteins, when influenced by geomagnetic field, guides the physiological behavior of birds \cite{Maeda2012,Ikeya2021}, (b) the effect of quantum tunneling of hydrogen ion in enzymes on the kinetic behavior of the enzyme at milliseconds timescale, detected through Kinetic Isotope Effects \cite{Meyer2008,Klinman2013,Zaragoza2023}, (c) the microsecond proton tranfer between amino acids in a peptide triggered by picosecond electron transfer \cite{MacAleese2016} (d) the classical coupling of fast micro- to millisecond timescale dynamics of various protein domains to much slower chemical cycles, connected to enzyme catalysis \cite{Haran2023,jager1997chromophore}, (e) pico- to nanosecond local atomic fluctuations in enzymes coupling to slower lid motions and thus directly affecting enzyme catalysis \cite{Henzler-Wildman2007,Ojeda-May2021,Frost2022,Buchenberg2015,Struts2011}, and (f) nanosecond timescale interactions of protein surfaces with water hierarchically coupling to longer time scale protein backbone motions, necessary for catalytic function \cite{Bellissent-Funel2016,aryal2015hydrophobic}.

A common thread running through all these experimental observations above, where slow, global dynamics of a system are shown to be constrained by its fast, local dynamics, bridging timescales and length scales orders of magnitude apart, is the model of hierarchically constrained dynamics \cite{Palmer1984,Huberman1985}, proposed initially for relaxation in glassy systems. Within this model, the dynamics of a system is categorized into multiple, hierarchical, spatiotemporal levels, with the faster, lower level of the hierarchy constraining the dynamics of the slower, upper level right above it, resulting in a power-law or a stretched-exponential relaxation behavior. This dynamical picture arises from rugged energy landscapes with multiple local minima at multiple scales of spatiotemporal resolution. This unifying picture has been successfully used theoretically and experimentally to understand proteins as hierarchically constrained systems \cite{Frauenfelder2009,ansari1985protein,Arnlund2014}. Such a hierarchically constraining arrangement allows proteins to efficiently transduce energy generated by fast (sub-picosecond), local chemical reactions into slow (microseconds to seconds), global functional motions, or into the solvent by dissipating through global dynamical modes, and therefore must have been evolutionarily selected \cite{Arnlund2014}. The evolutionary selection for efficiency of energy dissipation is reflected in the fractal nature of protein structures and dynamics \cite{Lewis1985,Reuveni2010}, ubiquitously found in nature where rapid dissipation or efficient transport of energy or nutrients across the interface between two interacting subsystems is crucial.

Our results in this article extend this multi-timescale, hierarchical, dissipative dynamics of proteins down to single amino acids. These results are consistent with the experimentally obtained broad relaxation spectra of aqueous solutions of amino acids \cite{Nandi2000}. Both the codon-anticodon association and the amino acid dissipation dynamics are hierarchical, extending over four to five decades of timescale, and importantly, the timescales of both the systems appear matched, as seen from the correlations shown in the Results section. This strongly suggests that the amino acids are evolutionarily selected for maximally efficient energy transfer and dissipation, from the codon-anticodon system and into the water molecules surrounding the cognate amino acid. This matched timescale hierarchy possibly ensures that bottlenecks to the energy flow at any timescale can be circumvented by utilizing the flow through other timescales, such that the mRNA molecule is structurally stable and available for continuous energy transport, ensuring its persistence \cite{wong2023roles}. This is similar to broadband impedance matching in electrical and electronic systems, which ensures efficient, unhindered energy flow across the frequency scale. Whether the amino acids are hierarchically constrained such that the fast processes at sub-femtosecond timescales influence slower dissipation dynamics at picosecond timescales is not known; however, one can easily imagine a scenario where the energy transferred into an amino acid at sub-femtosecond timescales electronically reorganizes its immediate aqueous environment at sub-femtoseond timescales, affecting longer timescale dissipation.

In support for the above ``efficient energy transfer'' hypothesis from mRNA-tRNA interactions to their cognate amino acids, we point to experiments that show that, as long as an mRNA is properly translated, its degradation is delayed by the celluar machinery and the mRNA is stabilized \cite{Shoemaker2012Translation,Isken2007}.  Ribosomal binding to mRNA has also been shown to have a protective effect on mRNA \cite{vargas2020regulation}. Whereas, stalled \cite{Hayes2003,Doma2006}, damaged, slow \cite{Bol2016,Presnyak2015Codon}, or untranslated mRNAs are readily marked for degradation. This suggests that translation plays a protective role for the stability of mRNA. Within the model we develop here, this translation-assisted stability arises from efficient energy dissipation, whereas accumulation of excessive energy within the actively-transcribed mRNA molecule marks it for decay. This is possibly the rationale for the invention of translation by evolution in the first place, when the evolutionarily-early mRNA translation products were useless.

 What is the evolutionary advantage of removing energy from mRNA, and what is the source of this free energy being removed?  Why would evolution invent such an elaborate dissipative mechanism to maintain the stability of mRNA? It is important to remember that nearly all biomolecules, such as DNA, RNA, cytoskeleton and cell membrane, exist in an \textit{energetically metastable} state \textit{in vivo}, which can be hydrolyzed, however slowly, into their constituent monomers, without external energy input. The factor that differs between these biomolecules is their kinetic stability, that is the energy barrier height dictating the rate of spontaneous decay of these molecules into their constituent monomers. Evolution has designed ribozymes and proteins to modulate these kinetic barriers, thereby modulating the persistence of these different molecular species, for the purpose of persistence of the whole selective unit\cite{wong2023roles}. Since living systems are persistent systems that are stabilized through energy flow, exhibiting dynamic kinetic stability \cite{Pross2004,Pross2005}, sub-systems such as RNA, DNA and proteins too must have been evolutionarily selected to facilitate energy flow generated from self-replication, and derive kinetic stability from it. Therefore, one can imagine that as long as mRNAs carry energy from cellular processes outward into peptides through translation, they are stabilized. When translation stops and energy builds up within mRNA, it decays, along with its energy source, due to their metastability, assisted by various enzymes. This provides a plausible answer for Eugene Koonin's question: ``what would be the selective force behind the evolution of the extremely complex translation system before there were functional proteins? '' \cite{Koonin2009}.

Our proposed identification of energy transfer as the physical mechanism behind the origin of the genetic code immediately explains the frequently observed relationship between amino acids' hydrophobicity/hydrophilicity and their genetic code assignments. The interactions of amino acids with surrounding water molecules entirely determine the relaxation rates of the amino acids (even within the ribosome), right after they receive the energy transmitted from the codon-anticodon association process, and therefore, directly impact the genetic code assignments. This proposed process of efficient removal of energy released during the codon-anticodon association process, thereby stabilizing and promoting the association process, possibly was the evolutionary rationale for the induction of individual amino acids, as catalysts or cofactors, in the prebiotic regime. Conversely, this implies that non-cognate amino acids on tRNA or deacylated tRNA cannot efficiently ``cool down'' the codon-anticodon association process and should result in faster dissociation of the codon-anticodon complex than otherwise.

Please note that the correlation plots involving the state 4 (and to some extent state 7) only segregate the codon groups of SYN, WYN, SRN and WRN cleanly, as seen in the fig. \ref{state1_state4}. These states do not discriminate between the codons within each of the four groups, leading to the presence of degeneracy in the genetic code at the timescales corresponding to these states. However, State 1 discriminates between codons within each of these four codon groups, but has no discriminatory capability across these four groups. To effectively discriminate between each of the 24 synonymous groups of codons, one must take into account information from both state 1 and state 4 (apart from other states). In other words, one must be capable of bridging the timescales of states 1 and 4, 1-10 picoseconds and 10-100 femtoseconds respectively, for effective discrimination. In the next section, we argue that aminoacyl tRNA synthetases were invented by evolution precisely for this purpose: to integrate information from all timescales for effective discrimination between the 24 synonymous codon groups, thereby improving the specificity of mapping between codon groups and their cognate amino acids.

The model's predictions can be tested at multiple levels: Apart from testing for the presence of sequence-independent and sequence-dependent asymmetric cooperativities in RNA/DNA association kinetics through dedicated unzipping/zipping kinetics experiments, working out the kinetics of codon-anticodon association kinetics, at sub-picosecond timescales, for all the 64 codon-anticodon pairs, would provide direct evidence for or against the model's predictions. Although the correlation plots between theoretical codon-anticodon association time predictions and experimental ribosomal residence time evaluations, shown in part II, are somewhat convincing, the lack of direct evidence for the timescales involved is a drawback that needs to be rectified. Evidence for electronic energy transfer between the codon-anticodon and the amino acid systems may be obtained by detecting the electroluminescence or the evanescent electromagnetic waves generated during the single-molecule codon-anticodon association process, using non-reciprocal optical elements to avoid disturbing the system, both in the presence and absence of amino acids \cite{Imada2016,Yang2019}. The experimental machinery that has been brought to bear on proving the quantum nature of energy transfer in photosynthesis may also be employed here. In general, experiments with very high spatial resolution for tracking single molecule interactions, and with sub-picosecond temporal resolution for tracking the kinetics, are required to satisfactorily corroborate or refute the predictions of the model.

\subsubsection*{Future directions}
It should be clear that we have simply scratched the surface in understanding the phenomenon of information transfer between two central subsystems of life (nucleic acids and amino acids), that lies at the core of life itself. Dedicated bottom-up investigations are required to unravel the precise physical causative mechanism behind every single plot above, that exhibit correlations between the theoretical residence times of codon-anticodon binding process and the various physicochemical properties of amino acids. The precise roles that the amino acid physicochemical properties, such as polarizability, softness, molecular vibrations, dipole orientations, hydrophobicity, electron-ion interactions, intramolecular dipole-dipole interactions, side chain lengths and HOMO-LUMO gaps play in receiving the energy transferred from the codon-anticodon association process and in dissipating the received energy into the surrounding aquatic medium need to be elucidated. It has to be noted that the physicochemical properties that we mentioned above are probably not the most appropriate variables to describe the energy transfer process, but were used simply because of the availability of data. This implies that dynamically appropriate physicochemical (structural, vibrational, electronic and optical) variables have to be defined and measured for accurate description of the energy transfer process. 

More generally, the origin of biological information may be imagined and modeled, not as complementarities of \textit{forms}, as in the lock-and-key model of enzyme-substrate interactions, but as evolutionarily-tuned \textit{processes} that transform energy from one form to another, from one molecule to another, through molecular structures whose preservation or persistence depends on the energy transfer itself. The persistence of a molecule A (here, mRNA) is ensured by the rapid transmission of excess free energy to another molecule B (here, amino acids) through timescale matching, which dissipates that free energy into the surrounding aqueous medium, thereby preventing the dissociation of molecule A. Information in the genetic code lies in the matching of the timescales of energy supply, which ensures the persistence of mRNA, introducing a temporal dimension to information storage and transfer. In molecular evolution, a slightly more tautological \textit{survival of the persistent} therefore precedes \textit{survival of the fittest} as a selection force.


\subsubsection*{Conclusions}
Employing a kinetics-based model that we have constructed earlier to explain some fundamental observations regarding DNA/RNA, such as their strand directionality, anti-parallel strand orientation and heteromolecular base-pairing, we have theoretically calculated the time taken for the association of codons with their corresponding anticodons, and showed that these residence times cluster appropriately, after fitting, to reproduce the structure of the genetic code (parts I and II). In this third part, we showed that certain physicochemical properties of amino acids, and where available, their relaxation times, correlate with the residence times in different states of the codon-anticodon association process. These correlations between the amino acids' physicochemical properties and the codon-anticodon residence times suggest that the genetic code emerged through matching of the rate of energy supply from the codon-anticodon association process with the rate of dissipation of the supplied energy to the aquatic medium at their cognate amino acids' end, at multiple timescales. This probably happens through electronic energy transfer, where, the energy generated during the association of codons with anticodons is relayed to the vibronic modes of cognate amino acids, thereby dissipating the generated energy into the surrounding water molecules and stabilizing the codon-anticodon association process. This explains the frequently found relationship between the genetic code and the hydrophobicities and hydrophilicities of the amino acids. If experimentally proven, the proposed mechanism underlying the genetic code emergence and maintenance would solve one of the central problems of biology. Moreover, by associating genetic information, embedded in the DNA/RNA sequences, with the rates of energy transfer,  the model explicates and clarifies what we mean by biological information, and how this information gets expressed and translated. The role of sequences in dictating both the DNA/RNA replication kinetics and in dictating the codon-anticodon zipping kinetics that underlies genetic code goes to show that both DNA/RNA replication and translation have a common kinetic determinant, effected by the DNA/RNA's sequence, a result of the non-equilibrium origins of life itself.

\clearpage
\begin{center}
    {\Large\bfseries IV: Emergence of two classes of aminoacyl tRNA synthetases} \\[0.5cm]
\end{center}

\section*{Introduction}
Aminoacyl-tRNA synthetases (aaRSs) are ancient enzymes that catalyze the acylation of tRNAs with their cognate amino acids\cite{Gomez2020,Woese2000}. A fundamental challenge in explaining the origin of aaRSs is the apparent circularity of the translation system: synthetases are essential for protein synthesis, yet they are themselves proteins whose production depends on the very machinery they help establish. Resolving this bootstrapping problem is important for understanding the emergence of translation. Another challenge is the presence of two classes of aaRSs with little homology observed between their coding sequences, molecular structure or evolutionary pathways \cite{ibba2000aminoacyl,Eriani1990Partition,burbaum1991structural,carter2017coding}. Why do two classes exist? The existence of Class I synthetases appears to presuppose the existence of Class II synthetases, which are required for the translation of Class I proteins, and vice versa, introducing a second layer of circularity in the evolution of the translation system.

There has been multiple attempts to understand the assignments of class-I and II aaRSs in terms of amino acid properties and the genetic code, and to resolve the above paradoxes. Regularities in the assignments have frequently been observed. The division of aaRSs into two classes has been found to be strongly correlated with the chemical and biological properties of their cognate amino acids \cite{Cavalcanti2004}. It is well-known that Class I enzymes preferentially activate large, hydrophobic amino acids, whereas, Class II enzymes generally activate smaller, polar, and hydrophilic amino acids \cite{Cavalcanti2004,rodin1996presence,Carter2015}.  The partition of the codon table between Class I (which aminoacylates the 2'-OH of the tRNA via the minor groove) and Class II (which aminoacylates the 3'-OH via the major groove) has been shown to be mathematically reproduced through a highly deterministic cascade of successive binary decisions (e.g., symmetry breaking at the first, second, and third bases) \cite{Delarue2007}. Such regularities suggest that the aaRS class assignment is not random, but emerges from specific physicochemical interactions between the amino acids, tRNA and the aaRS enzymes.

Despite such advances, there exists a glaring gap in our understanding of aaRS evolution and the evolution and functioning of the translation system in general that is caused by our inability to demonstrate a mapping between codons/anticodons and amino acids \textit{in the absence of aaRS}. Once this centrality accorded to the aaRS enzymes as being responsible for such a mapping is diluted, and a direct mapping between codons/anticodons and amino acids is found based on physicochemical principles without requiring the presence of aaRS enzymes, the circularities or the bootstrapping problems mentioned above are automatically resolved. In part III of this series, we have speculated, with some evidence, that such a mapping is present between the codon-anticodon system and cognate amino acids, on the basis of the existence of electronic energy transfer between the two systems. The specificity of interactions between the two systems arises, in our view, from the condition of impedance matching, wherein, the energy supply rates at the codon-anticodon end and the energy dissipation rates at the cognate amino acid end are matched, at multiple timescales, leading to fastest possible energy transfer rates. Importantly, such an explanation does not require the presence of aaRS proteins to instantiate the mapping. If specific intermolecular attraction between codon-anticodon system and its cognate amino acid + water system is taken into account on the basis of efficient electronic energy transfer and dissipation between the two systems \cite{McClare1974Resonance,Andrews2008,Perez-Martin2025,Lechelon2022,Sherkunov2009,Sasihithlu2024,frohlich1968long,frohlich1970long,Sthr2019,Barik2022}, aaRS enzymes become redundant and instead require justification for their evolution and existence. The aaRS enzymes probably evolved to improve the specificity in the interactions between the codon-anticodon system and the amino acid + water system, after the establishment of the translational apparatus, which initially required no synthetases, due to the pre-existence of direct attractive interactions between the two systems. It has to be remembered that the existence of attractive interactions between the two systems presupposes the existence of non-equilibrium, which most in-vitro experiments do not satisfy, and therefore, the phenomenon is out of their scope. The foregoing resolves the problem of circularity in the evolution of aaRS and the translational apparatus.

The question is now refined to asking how exactly does the aaRS enzymes enhance specificity of interactions between codon-anticodon system and the cognate amino acid + water system, and why are there two classes of these enzymes. Earlier, in part III (Title: Kinetic basis of the genetic code: III - Non-equilibrium Origin), we pointed to the existence of degeneracies in the residence times, the time taken by the codon-anticodon system in various states during the association process, specifically in state 4, which only distinguishes between SYN, SRN, WYN and WRN codons, at the timescales of 10-100 femtoseconds. Moreover, State 1, with its characteristic timescale of 1-10 picoseconds, helps discriminate codons within each of the four groups above, but not across the four groups. For the removal of such degeneracies, we require machines that are capable of integrating information coming in from events happening across multiple timescales and producing a specific output. Enzymes, as we argued in part III, are capable of transducing events happening at short timescales into action at longer timescales, thereby bridging multiple timescales. In the Results section below, we argue that these degeneracies are removed and the specificity of the attractive interactions are enhanced if there exist multiple enzymes, that are capable of integrating information from multiple timescales, and each of which maps synonymous codons to their cognate amino acids. It has to be emphasized that the lack of specificity in the mapping between synonymous codons and their cognate amino acids, due to degeneracies in state 1 and 4, does not imply a complete lack of specificity; interactions between the two systems at timescales other than that of states 1 and 4, such as state 6 (sub-femtoseconds) would have been specific enough to bootstrap translation early on, however inaccurately, without the help of aaRSs.

\section*{Methods}
As in part III of this article, for this section, we use, unaltered, the theoretically calculated residence times data obtained by tuning the asymmetric cooperativity model's four parameters in part II (Title: Kinetic basis of the genetic code: II - Quantitative reproduction of the codon table structure and its experimental validation) to reproduce the genetic code organizational structure, by which we mean the dependence of codon degeneracies on the types of nucleotides present in the three positions of a codon (strong, weak, purine or pyrimidine). The term ``residence times'' implies the average time spent by the codon-anticodon system in each of the seven states during its sojourn from completely dissociated state to fully bonded state (please see part II), where all three hydrogen bonds between the codon and the anticodon are present. No fitting or tuning of parameters is involved in generating the plots below, and the residence times are used as evaluated in part II. Specifically, no information about synthetases or the class I/II division entered our calculations in part II of this series. The experimentally determined ribosomal residence times too are used as they were calculated in part II, without any alteration. 

\section*{Results}

The primary result of this section is shown in fig. \ref{state_14}. It is a plot between state 1 versus state 4 residence times, the average times taken by the codon-anticodon association process to stay in states 1 and 4 during its sojourn from fully unbonded state to fully bonded state, evaluated using a Markov chain model described in part II of this series. The amino acids corresponding to each of the 64 codons, including stop codons denoted by X or x, are shown, now color-coded according to the Aminoacyl-tRNA synthetase classes, with amino acids in blue catalyzed by class II enzymes, whereas, amino acids in red, by class I enzymes. The stop codons are shown in black.

It can be noticed that nearly all codons colored in red, denoted by their cognate amino acid letters, catalyzed by class I enzymes, have larger state 1 residence times, compared to the amino acids catalyzed by class II enzymes, within each of the four groups. The exceptions are Phe (F) and His (H). Such a splitting of class I- and class II-catalyzed amino acids is unexpected, since no information about synthetases entered into our calculations in part II, where we were only concerned with reproducing the genetic code organizational structure by tuning the four asymmetric cooperativity parameters. This result is therefore emergent, suggesting that the evolution of synthetase class I/II classification has its roots in the evolution of the genetic code itself. This figure strongly suggests that evolution has attempted to remove the degeneracies imposed by state 4, which only discriminated between SYN, WYN, SRN and WRN (with the latter two split into two groups each, SRS and SRW, and, WRS and WRW), using state 1 residence times. Although the specificity of energy transfer interactions between the codon-anticodon association process and the amino acid dissipation process is not completely lost due to the above-mentioned degeneracies, due to the presence of other states, such as state 6, which also discriminates between codons, evolution's attempt to improve upon this specificity of the interactions between the codon-anticodon system and the amino acid + water system, by removing the degeneracies imposed by state 4 residence times, is clearly evident in the fig. \ref{state_14}. Evolution probably invented two classes of synthetases to break the symmetry inherent in the state 4 residence times, where, the two classes split each of the four groups of codons (SYN, WYN, SRN and WRN) into two halves, with class I assigned to codons with larger state 1 residence times, and the evolutionarily earlier class II assigned to codons with smaller state 1 residence times. 

\begin{figure}
\includegraphics[trim=80pt 20pt 20pt 20pt,width=1.05\textwidth]{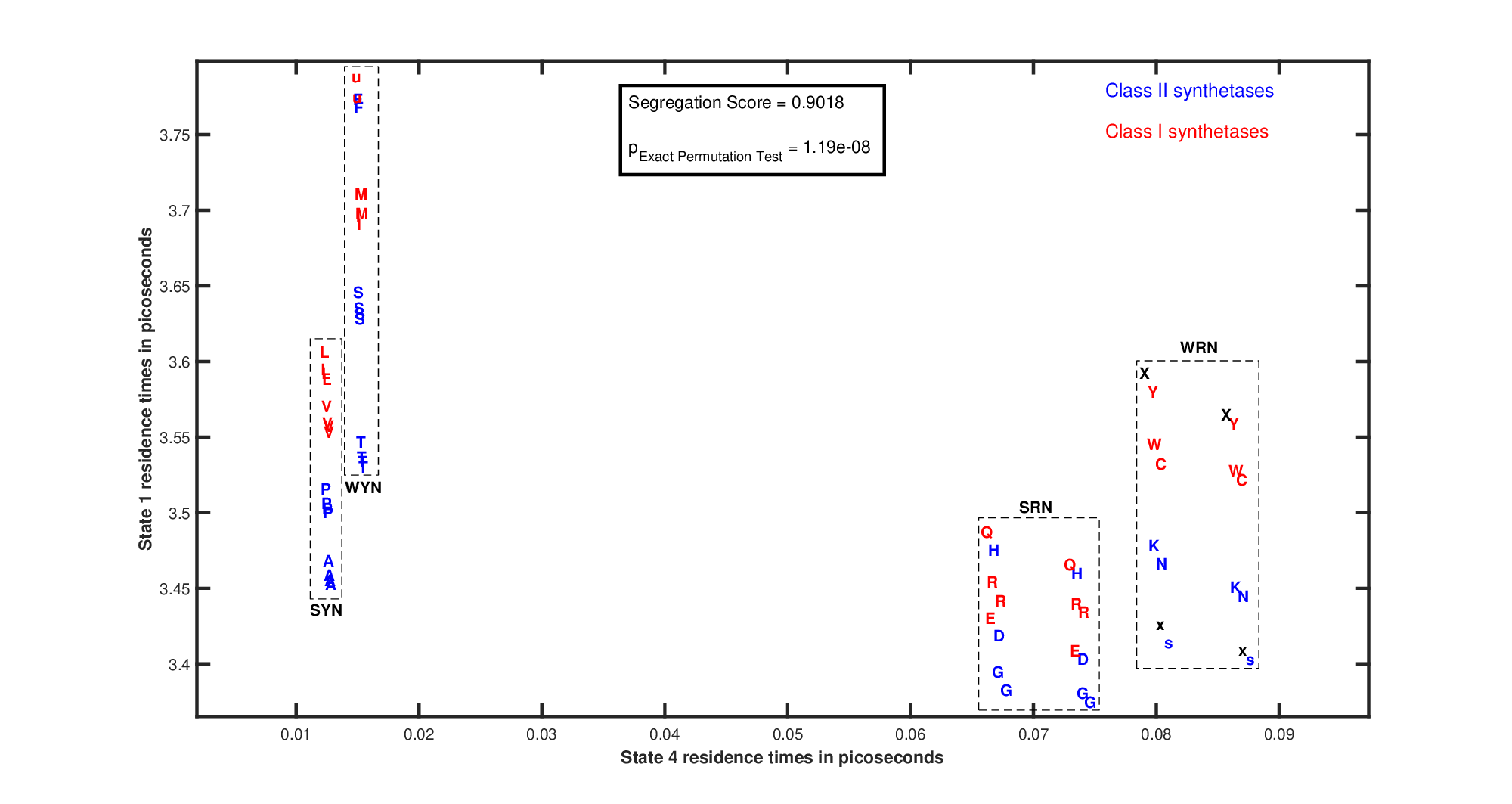}
\caption{The figure shows distribution of codons, denoted by their single letter cognate amino acid notation, according to their state 1 and state 4 residence times. The plot cleanly segregates class-I catalyzed amino acids and class-II catalyzed amino acids, with the exception of Phe (F) and His (H), along the state 1 residence time direction. The degeneracy inherent in state 4 residence times along X-axis, which cluster into four groups of SYN, WYN, SRN and WRN codons, is broken into two sets, creating eight groups. This is an emergent result, since no information about synthetases entered our calculations for generating the residence times in part II. The evolutionary reason for such a breaking of symmetry in state 4 is possibly to improve the specificity of interactions between codon-anticodon system and the amino acid + water system. }
\label{state_14}
\end{figure}

We asked if there is any mapping that takes a codon catalyzed by class II and transforms it into a codon catalyzed by class I enzymes. The mapping should take a codon within any one of the four groups SYN, WYN, SRN and WRN, and transforms it into another codon \textit{within the same} group, thereby breaking the symmetry within each group. Therefore, the transformation must preserve the sequence characteristic of each group, keeping the input codon before transformation and the output codon obtained after the transformation within the same group, while classifying the two codons into distinct synthetase classes. We found a transformation or mapping that does this correctly for seventeen of the twenty amino acids, excepting Phe (F), His (H) and Glu (E). The transformation is simply changing the first two nucleotide letters of a codon such that the nucleotides at these two positions are different from that of the input codon, while their character, i.e., whether they are strong/weak for the first position and purine/pyrimidine for the second position, is preserved. This transformation ensures that the codon stays within the group after the transformation process. For example, the codon GUN is transformed into CCN, which preserves the first position's strong character and the second position's pyrimidine character, thereby keeping both the codons in the group SYN. In terms of amino acids, this mapping example transforms Valine, which is in class II, to Proline, in class I. A surprising feature of this transformation is that, within the vertebrate mitochondrial code under consideration throughout this series due to its symmetry \cite{Delarue2007}, the canonical stop codon UAR maps to AGR, which also codes for stop.

The class I/II classification seen in fig. \ref{state_14} shows that, within each of the four groups SYN, WYN, SRN and WRN, codons and their cognate amino acids whose tRNA aminoacylations are catalyzed  by class I enzymes have slower state 1 residence times. This suggests that the biophysical mechanism at the tRNA end that underlies the classification into class I or II operates at picosecond timescales. We hypothesize that the distinct picosecond relaxation times resulting from the amino acids acylated initially to either 2'-OH or 3'-OH of the ribose ring of the Adenosine nucleotide at position 76 in the acceptor stem of tRNA is the causative factor behind the classification scheme. The distinct biochemical properties of these two chemical groups \cite{Merino2005}, due to their distinct local stereochemical environments, should result in distinct picosecond dissipation dynamics of amino acids attached to these two hydroxyl groups, which can then be used for discrimination by the two classes of synthetases at the picosecond timescale. Before the advent of aaRSs, aminoacylation of tRNA probably happened with some amino acid specificity, due to the specific attractive interactions arising from impedance-matched electronic energy transfer, as mentioned in part III. However, the acylation was not specific with respect to 2'-OH and 3'-OH of the adenosine nucleotide, resulting in some loss of specificity, due to their distinct picosecond dissipation dynamics. This perhaps led to the evolutionary invention of two synthetase classes, specific for each of the two hydroxyl groups. The foregoing explains the enigmatic presence of amino acid attachment to the 2'-OH end by class I synthetases, which undergoes intramolecular transesterification reaction and migrates to the 3'-OH group eventually, nullifying the distinct effects of class I and class II synthetases on aminoacylated tRNAs used for translation within the ribosome. This circuitous route of aminoacylation of tRNA does not make sense from the conventional biochemical point of view, since the end result of a 3'-OH aminoacylated tRNA is the same for both classes of aaRSs. Within our picture, the difference arises from the distinct picosecond dissipation dynamics of amino acids attached to the two hydroxyl groups, which helps in discrimination, and hence is evolutionarily preserved. Moreover, our explanation does not require that the two aaRS classes need to be evolutionarily related. Due to the possibility of evolution of translation predating the evolution of aaRS enzymes, the enzymes that catalyze 2'-OH aminoacylation and 3'-OH aminoacylation can evolve independently of each other. Therefore, no homology is seen between the two classes, in their sequences, structures or evolutionary pathways.

\begin{figure}
\includegraphics[trim=80pt 20pt 20pt 20pt,width=1.05\textwidth]{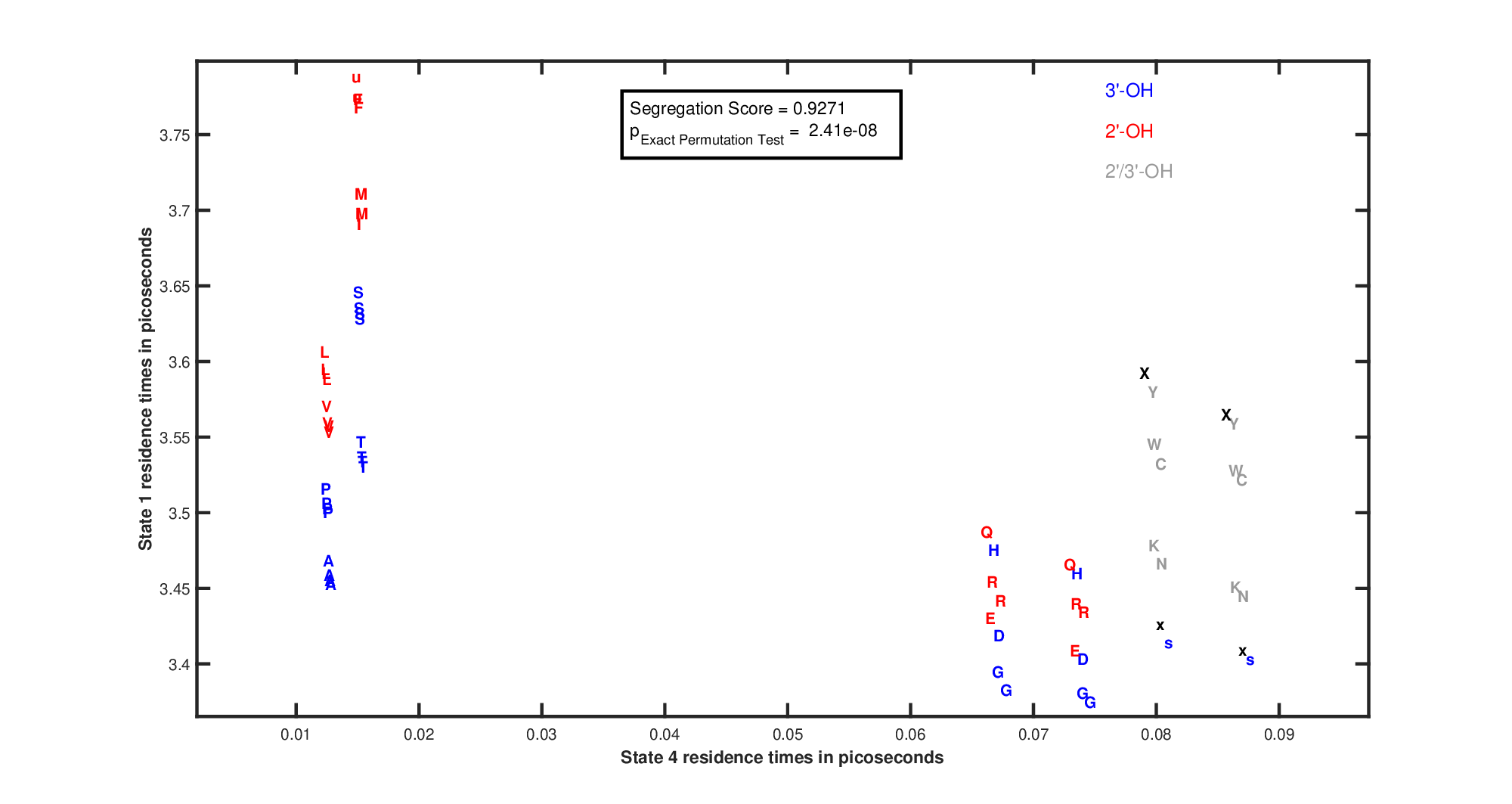}
\caption{Distribution of codons, denoted by their single letter cognate amino acid notation, according to their state 1 and state 4 residence times, now classified according to whether the amino acids are acylated at the 3'-OH end (blue), 2'-OH end (red) or at both ends (gray) of the terminal adenosine of their cognate tRNA. Phe (F), aminoacylated at the 2'-OH end, is not an outlier here. All amino acids coded by WRN codons, which have the largest state 4 residence times, viz. Asn (N), Cys (C), Tyr (Y), Trp (W) and Lys (K), are prone to be aminoacylated at both 2'-OH and 3'-OH ends, and are colored gray. Ser (s) is an exception.}
\label{state14_23OH}
\end{figure}

We have noted above that the mapping we discovered that transforms class I amino acids into class II have three exceptions: Phe (F), whose codons transform into Thr (T) according to the rules mentioned above, both of which are in class II, His (H), which transforms into Gly (G), both class II, and Glu (E) paired with Arg (R), both class I. Of these three, Phe (F) tRNAs are charged at the 2'-OH end, a signature class I behavior. This led us to a search for 2'-OH / 3'-OH aminoacylation specificities of each of the twenty amino acids. A systematic analysis of the ability of each of the aaRS to aminoacylate its cognate tRNA at the 2'-OH / 3'-OH of the terminal adenosine has been performed, by preparing modified tRNAs terminating in  either 2'- or 3'-deoxyadenosine, and using \textit{E.Coli} and Yeast aaRSs \cite{Hecht1976,Sprinzl1975}. The amino acids Asn (N), Cys (C) and Tyr (Y) were initially aminoacylated at both the 2'-OH and 3'-OH ends, while Asp (D), Gln (Q) and Glu (E) were deemed uncertain, combining the data for \textit{E.Coli} and yeast tRNA. Moreover, the above experiment found that Trp (W) aminoacylated at the 2' position in \textit{E.Coli} and 3' position in yeast. We already know that the class rule is violated by Lys (K); in some organisms, its synthetase is class I, while in others it is class II \cite{Woese2000,Ibba1997}. Putting it all together, the tRNAs of amino acids Asn (N), Cys (C), Tyr (Y), Trp (W) and Lys (K) most probably allowed and still allows for aminoacylation at both 2'-OH and 3'-OH ends of the terminal adenosine. The fig. \ref{state14_23OH} shows codons, denoted by their cognate amino acid letters, now colored according to whether they are aminoacylated at the 2'-OH end (red), 3'-OH end (blue) or at both the ends (gray). Stop codons are denoted by a black X or x. An interesting pattern emerges: the tRNAs of nearly all amino acids encoded by WRN codons, with largest state 4 residence times ($\approx 0.1$ picoseconds) of all codons, are prone to getting initially aminoacylated at both 2'-OH and 3'-OH ends. The exception is Ser (s), coded by AGY codons. Moreover, Phe (F) is now classified appropriately, in the 2'-OH group. It appears that the 2'-OH / 3'-OH specificity is lost when the state 4 residence times are large, mirroring the situation we encountered in part III, where large state 6 residence times led to a loss of correlation between state 6 residence times and inverse HOMO-LUMO gaps.

\section*{Experimental Support}
\begin{figure}
\includegraphics[trim=80pt 20pt 20pt 20pt,width=1.05\textwidth]{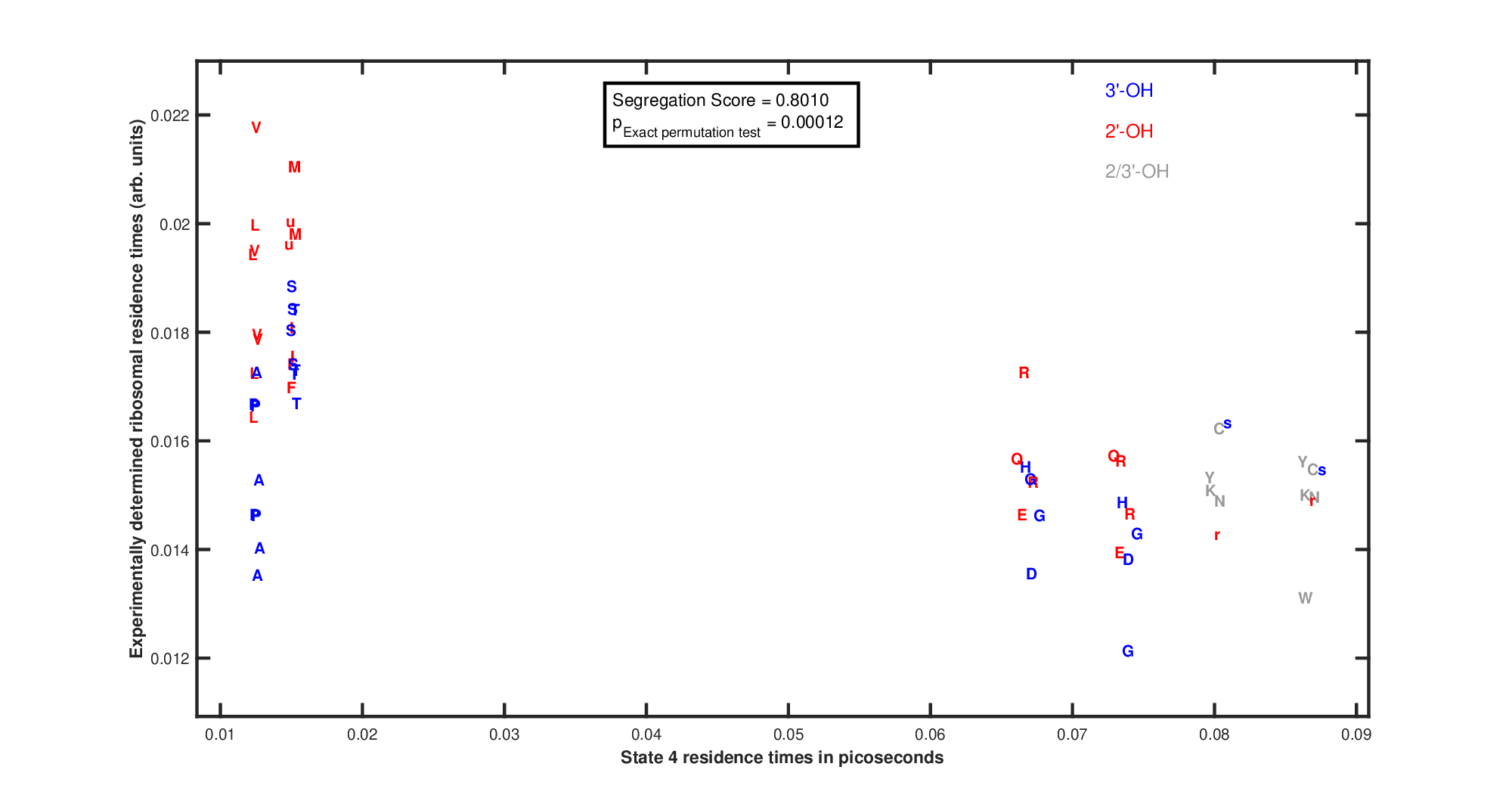}
\caption{A plot of experimentally determined ribosomal residence times of 61 codons, mostly from yeast, converted to their cognate amino acid symbols, against the theoretically calculated state 4 residence times, which groups together 64 codons into four groups SYN, WYN, SRN and WRN. This plot should be compared with fig. \ref{state14_23OH} for similarities and differences. 2'-OH-aminoacylated amino acids generally have higher experimental ribosomal residence times, with some exceptions, partly agreeing with the theoretical data. The amino acids in the WRN group are prone to be aminoacylated at both 2'-OH and 3'-OH ends and therefore are marked in gray.}
\label{rrt_23OH}
\end{figure}

\begin{figure}[!htbp]
\begin{center}
\includegraphics[width=0.9\textwidth]{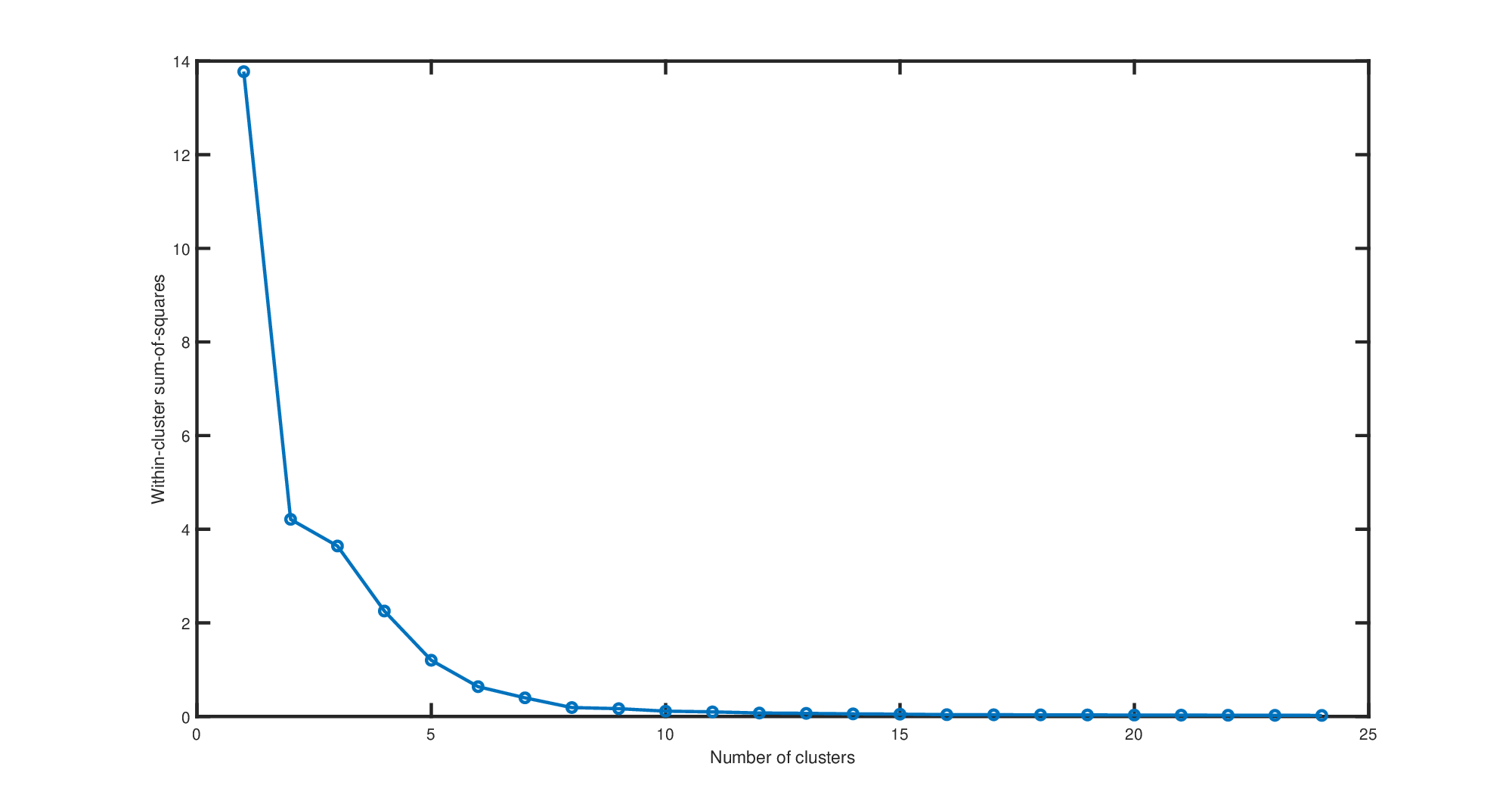}
\caption{Scree plot corresponding to the clustering of residence time vectors corresponding to 64 codons into 24 groups. The optimal number of clusters appears to be around 12. }
\label{screeplot}
\end{center}
\end{figure}

\begin{figure}[!htbp] 
\begin{center}
\includegraphics[width=0.9\textwidth]{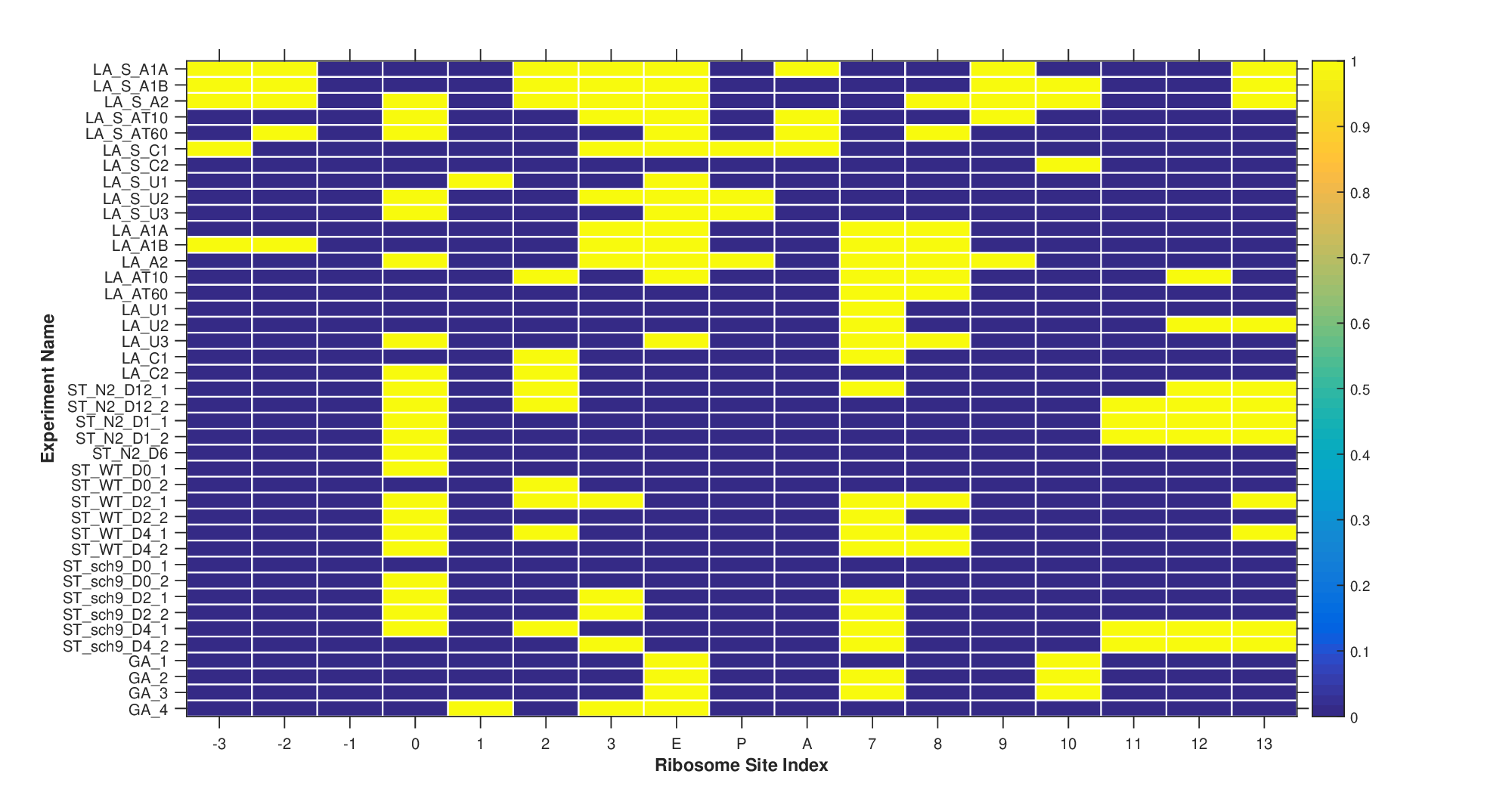}
\caption{Significance of correlation between theoretical mRNA-tRNA association times and experimental ribosome residence times. The correlations themselves are shown in fig. \ref{correlationmatrix}. Cells where p value is less than 0.05 and the correlation is higher than 0.3 or lower than -0.3 are highlighted in yellow.} 
\label{correlationmatrixpvalue}
\end{center}
\end{figure}

In part II, we have shown a correlation between theoretically obtained residence times of codon-anticodon association process and experimentally evaluated ribosomal residence times on 61 codons, obtained from ribosomal footprinting experiments. Here, in fig. \ref{rrt_23OH}, we plot the same ribosomal residence times data along y-axis, but against state 4 residence times, instead of total residence times as done in part II. We color the amino acids according to whether they are aminoacylated at the 2'-OH end (red), 3'-OH end (blue) or at both ends (gray) of the terminal adenosine of tRNA. The ribosomal residence times of all 61 codons is plotted in fig. \ref{rrt_23OH}. The propensity for 2'-OH-acylated amino acids to have larger ribosomal residence times can be seen from the fig. \ref{rrt_23OH}, with the exceptions of Ile (I) and Phe (F). His (H) is an exception in both theoretical calculations and experimental data. The experimental data in fig.  \ref{rrt_23OH} should be compared with the theoretical data in fig. \ref{state14_23OH}. The order of amino acids within the WRN column in fig. \ref{rrt_23OH}, colored mostly in gray, only moderately match with the theoretical data in fig. \ref{state14_23OH}.     

\section*{Discussion}
The evolution of aaRSs, probably dating back to Last Universal Common Ancestor (LUCA) itself, is an enduring mystery, due to the circularity they impose on the translation process: translation requires aaRSs, while construction of aaRSs require translation. There are ample instances of such circularities in biology, since evolution frequently erases its traces. This erasure is most probably caused by the highly effective masking of baseline processes, such as translation without aaRSs, due to the invention of aaRSs which makes translation much more probable and effective, thereby overwhelming the small probability of translation without aaRSs. Moreover, as the LUCA adapted to the presence of aaRSs, the presence of the baseline processes themselves became a liability, due to the LUCA's inability to control such processes, which probably pushed it to adapt away from such processes, leaving it to rely on aaRSs exclusively. To muddy the waters more, evolution evolved aaRSs to rely more on base-pairing dynamics near the tRNA acceptor stem itself for amino acid specificity, and less on codon-anticodon dynamics at the anticodon loop, leading to the emergence of a ``second code'' \cite{Mller1992,schimmel1993operational,rodin1996presence}. Despite such erasure attempts, fortunately, the emergence of aaRSs has left its traces in the genetic code itself, specifically in its degeneracies, thereby providing us a window into its evolutionary rationale.

We speculate that aaRSs evolved to improve the specificity of aminoacylation, well above the specificities arising out of \textit{non-equilibrium} physicochemical interactions between codon-anticodon system and the amino acid + water system. In Part III, we tentatively identified this specific physicochemical interaction as impedance-matched electronic energy transfer between the two systems, leading to maximal power transfer, or in other words, transfer of energy in the shortest possible time. We arrived at this explanation from the observed matching of the residence times in various states of the codon-anticodon association process and the dissipation times at the amino acid end, at multiple timescales. This matching however is incomplete, due to the presence of degeneracies in the residence times of certain states, such as state 4 and state 1, with the former discriminating codon-anticodon residence times only across four groups SYN, WYN, SRN and WRN, while the latter state 1 discriminating only within each of the four groups, and not across them. The dynamics of these two states happen at distinct timescales, 10-100 femtoseconds for state 4 and 1-10 picoseconds for state 1. These degeneracies were unresolvable without the existence of a machine that integrated information across multiple timescales and outputted a decision. Enzymes are precisely suited for this purpose, since they are designed to operate across multiple timescales, as described in part III.

The aaRS enzymes probably evolved to distinguish the distinct picosecond dissipation dynamics of amino acids when they are aminoacylated at the 2'-OH or the 3'-OH ends of the terminal adenosine of the tRNA, which were then mapped to the distinct picosecond association dynamics of the codon-anticodon association process in state 1, through impedance matching. This initially led to the independent evolution of two distinct classes of aaRSs, now without needing each other's presence due to the existence of the baseline specificities of aminoacylation from impedance-matched electronic energy transfer. This may explain the lack of homology between the two aaRS classes. The lack of specificity of aminoacylation at 2'-OH and 3'-OH ends of amino acids belonging to the WRN group is possibly due to the their large residence times in state 4, which opens up and mixes dissipative channels from other timescales, apart from that of state 4, leading to loss of specificity. In conclusion, we have shown that the evolution of two classes of aaRSs was needed to remove the degeneracies in the residence times of certain states during the codon-anticodon association process, thereby significantly increasing the aminoacylation specificity. We have also argued that, due to the presence of baseline specificities arising out of non-equilibrium energy transfer interactions between the codon-anticodon and the amino acid + water systems, translation didn't need aaRSs to evolve, thus breaking the apparent circularity in its evolution.

\bibliographystyle{unsrt} 
\bibliography{library_recovered}

\subsection*{Author Contributions}
HS conceived, researched and wrote the article. 

\subsection*{Declaration of Interests}
The author declares no competing interests.

\subsection*{Funding}
Support for this work was provided by the Science \& Engineering Research Board (SERB), Department of Science and Technology (DST), India, through a Core Research Grant with file no. CRG/2020/003555 and a MATRICS grant with file no. MTR/2022/000086.

\subsection*{Acknowledgements}
I thank Parthasarathi Sahu, Sashikanta Barik and Koushik Ghosh for their critical reading of the manuscript. I thank Partharasathi Sahu for his help with the illustrations. 

\newpage
\section*{Supplementary Material}

\subsubsection*{Residence times table for all 64 codons} 
The table below provides the seven-element residence time vector for each of the 64 codons and their total residence times. 

\begin{table}[htbp]
    \centering
		\tiny
    \begin{tabular}{|l|l|l|l|l|l|l|l|l|}
    \hline
        CODON & STATE 1 & STATE 2 & STATE 3 & STATE 4 & STATE 5 & STATE 6 & STATE 7 & TOTAL \\ \hline
        TTT & 3.773504842 & 0.021857148 & 0.524664186 & 0.014653417 & 0.937202851 & 0.000468432 & 0.587295106 & 5.859645981 \\ \hline
        TTC & 3.768099828 & 0.02191339 & 0.521232557 & 0.014703908 & 0.933931331 & 0.000466969 & 0.583489611 & 5.843837594 \\ \hline
        TTA & 3.788629562 & 0.066528489 & 0.526831088 & 0.014521197 & 0.941346813 & 0.003034384 & 0.589724099 & 5.930615633 \\ \hline
        TTG & 3.775279363 & 0.066843181 & 0.522263342 & 0.014587873 & 0.935898402 & 0.003023206 & 0.584645019 & 5.902540386 \\ \hline
        CTT & 3.592837308 & 0.019133685 & 0.331404409 & 0.012059613 & 1.110792901 & 0.000552015 & 0.77922502 & 5.846004951 \\ \hline
        CTC & 3.588814253 & 0.019194473 & 0.329599304 & 0.012114957 & 1.107232615 & 0.000550432 & 0.775045054 & 5.832551087 \\ \hline
        CTA & 3.606515702 & 0.061668251 & 0.3327291 & 0.011921544 & 1.11555587 & 0.003549261 & 0.782348084 & 5.914287812 \\ \hline
        CTG & 3.595309396 & 0.06198909 & 0.330230126 & 0.011997499 & 1.109495136 & 0.003536967 & 0.776532121 & 5.889090334 \\ \hline
        ATT & 3.696683133 & 0.021875374 & 0.560604634 & 0.01487883 & 0.642294597 & 5.24298E-05 & 0.628424583 & 5.564813582 \\ \hline
        ATC & 3.691239634 & 0.02193237 & 0.55715476 & 0.014930247 & 0.638920867 & 5.21699E-05 & 0.624599767 & 5.548829815 \\ \hline
        ATA & 3.711009058 & 0.065582688 & 0.562731399 & 0.014743823 & 0.644750154 & 0.000346947 & 0.630810021 & 5.629974088 \\ \hline
        ATG & 3.698041706 & 0.065897573 & 0.558167181 & 0.014811601 & 0.640089154 & 0.000344903 & 0.625735278 & 5.603087396 \\ \hline
        GTT & 3.557583383 & 0.019176178 & 0.351336174 & 0.012201665 & 0.816652749 & 6.5983E-05 & 0.82751766 & 5.584533792 \\ \hline
        GTC & 3.553537671 & 0.019237639 & 0.349532221 & 0.012257773 & 0.812975066 & 6.57E-05 & 0.823342407 & 5.570948477 \\ \hline
        GTA & 3.570872954 & 0.061263741 & 0.352623196 & 0.012061652 & 0.819665451 & 0.00043037 & 0.83055206 & 5.647469423 \\ \hline
        GTG & 3.559849226 & 0.061585387 & 0.35014562 & 0.012138587 & 0.814410033 & 0.000428058 & 0.824788484 & 5.623345395 \\ \hline
        TCT & 3.632126451 & 0.014700569 & 0.558041281 & 0.014731291 & 0.729121921 & 0.000363304 & 0.624982483 & 5.5740673 \\ \hline
        TCC & 3.628842019 & 0.014729 & 0.554562242 & 0.014778833 & 0.727442398 & 0.000362581 & 0.621122197 & 5.56183927 \\ \hline
        TCA & 3.646165188 & 0.05788461 & 0.560240108 & 0.01460232 & 0.732221815 & 0.002378753 & 0.627447934 & 5.640940728 \\ \hline
        TCG & 3.635503366 & 0.058147607 & 0.555608133 & 0.014665653 & 0.728913132 & 0.002372941 & 0.622294862 & 5.617505694 \\ \hline
        CCT & 3.503086076 & 0.013307755 & 0.35365746 & 0.012126741 & 0.830662361 & 0.000412314 & 0.832492782 & 5.545745488 \\ \hline
        CCC & 3.50049243 & 0.013338295 & 0.35182097 & 0.012179113 & 0.828834284 & 0.00041153 & 0.828234787 & 5.535311408 \\ \hline
        CCA & 3.516113119 & 0.054952443 & 0.355014664 & 0.011990733 & 0.834108393 & 0.002680465 & 0.835694331 & 5.610554149 \\ \hline
        CCG & 3.506674949 & 0.055217118 & 0.352467202 & 0.012063437 & 0.830470076 & 0.002674083 & 0.829759073 & 5.589325937 \\ \hline
        ACT & 3.533910844 & 0.014578285 & 0.601534873 & 0.014992798 & 0.377655293 & 3.0963E-05 & 0.67471543 & 5.217418487 \\ \hline
        ACC & 3.530649396 & 0.014606984 & 0.598013102 & 0.015041169 & 0.376008709 & 3.08365E-05 & 0.670808044 & 5.20515824 \\ \hline
        ACA & 3.547081244 & 0.056553781 & 0.603720023 & 0.014863167 & 0.379043209 & 0.000212758 & 0.677167634 & 5.278641815 \\ \hline
        ACG & 3.536899878 & 0.056815037 & 0.599052985 & 0.014927169 & 0.376668611 & 0.000211774 & 0.671974969 & 5.256550423 \\ \hline
        GCT & 3.455800391 & 0.013267099 & 0.378471754 & 0.012296292 & 0.468179328 & 3.80069E-05 & 0.892568987 & 5.220621856 \\ \hline
        GCC & 3.453214682 & 0.013297913 & 0.376621232 & 0.012349382 & 0.466376206 & 3.78684E-05 & 0.888279279 & 5.210176563 \\ \hline
        GCA & 3.468405986 & 0.054328136 & 0.379818438 & 0.012159746 & 0.469862316 & 0.000256088 & 0.895747559 & 5.28057827 \\ \hline
        GCG & 3.459197838 & 0.054592209 & 0.377262787 & 0.012233116 & 0.467177181 & 0.000254964 & 0.889793419 & 5.260511515 \\ \hline
        TAT & 3.580009225 & 0.078661292 & 0.137062285 & 0.079276979 & 0.615768193 & 0.000339179 & 0.129872643 & 4.620989794 \\ \hline
        TAC & 3.558910001 & 0.084534442 & 0.117352892 & 0.085856045 & 0.60312524 & 0.00033595 & 0.115054219 & 4.565168789 \\ \hline
        TAA & 3.592729562 & 0.119257359 & 0.137108341 & 0.078635387 & 0.61800361 & 0.002219653 & 0.130019363 & 4.677973275 \\ \hline
        TAG & 3.564959579 & 0.125425203 & 0.117382972 & 0.085245061 & 0.604190233 & 0.002199747 & 0.115129973 & 4.614532768 \\ \hline
        CAT & 3.47554165 & 0.066933194 & 0.098662294 & 0.066318145 & 0.664868967 & 0.000357515 & 0.158241731 & 4.530923496 \\ \hline
        CAC & 3.460196028 & 0.07297319 & 0.085550585 & 0.073084663 & 0.651903149 & 0.000354213 & 0.142999782 & 4.48706161 \\ \hline
        CAA & 3.487610167 & 0.106514787 & 0.09866 & 0.065665766 & 0.667202238 & 0.002332711 & 0.158418355 & 4.586404025 \\ \hline
        CAG & 3.465937091 & 0.112852761 & 0.085557307 & 0.072468642 & 0.653016926 & 0.002312328 & 0.143092587 & 4.535237642 \\ \hline
        AAT & 3.466634722 & 0.079007995 & 0.143084121 & 0.079969876 & 0.227238407 & 2.41752E-05 & 0.141246471 & 4.137205767 \\ \hline
        AAC & 3.445283728 & 0.08493272 & 0.12321236 & 0.086607315 & 0.214067911 & 2.36041E-05 & 0.126315586 & 4.080443225 \\ \hline
        AAA & 3.478422673 & 0.118204276 & 0.143122459 & 0.079327627 & 0.227738455 & 0.000168968 & 0.14139622 & 4.188380677 \\ \hline
        AAG & 3.450891733 & 0.124419172 & 0.123238791 & 0.085995231 & 0.214310813 & 0.000165483 & 0.126392876 & 4.1254141 \\ \hline
        GAT & 3.419105357 & 0.067170398 & 0.101895586 & 0.06673599 & 0.254398442 & 2.53857E-05 & 0.170985619 & 4.080316778 \\ \hline
        GAC & 3.403628857 & 0.073248622 & 0.088706663 & 0.073545654 & 0.240888618 & 2.47999E-05 & 0.155667245 & 4.035710458 \\ \hline
        GAA & 3.430713065 & 0.106053997 & 0.101889794 & 0.066083481 & 0.254926719 & 0.000176354 & 0.171166327 & 4.131009737 \\ \hline
        GAG & 3.409151711 & 0.112427943 & 0.088711741 & 0.072929222 & 0.241147493 & 0.000172776 & 0.155762148 & 4.080303032 \\ \hline
        TGT & 3.532781661 & 0.041719863 & 0.138315863 & 0.079909085 & 0.565305216 & 0.000296488 & 0.131346977 & 4.489675153 \\ \hline
        TGC & 3.522030886 & 0.044501036 & 0.118438105 & 0.086503331 & 0.559166539 & 0.000294916 & 0.116380526 & 4.44731534 \\ \hline
        TGA & 3.545642361 & 0.082872626 & 0.138359422 & 0.079255059 & 0.567482572 & 0.001958008 & 0.131492424 & 4.547062472 \\ \hline
        TGG & 3.528138678 & 0.085906189 & 0.118467291 & 0.085881395 & 0.560201318 & 0.001947976 & 0.116455901 & 4.496998748 \\ \hline
        CGT & 3.442331837 & 0.036044011 & 0.099492218 & 0.0668442 & 0.602689543 & 0.000312047 & 0.160148386 & 4.407862242 \\ \hline
        CGC & 3.43435307 & 0.03890563 & 0.086273692 & 0.073630037 & 0.596390228 & 0.000310439 & 0.144746433 & 4.374609531 \\ \hline
        CGA & 3.454566934 & 0.076220295 & 0.099488146 & 0.066180152 & 0.604951712 & 0.002053693 & 0.160323955 & 4.463784888 \\ \hline
        CGG & 3.440164668 & 0.079334772 & 0.086279808 & 0.073003855 & 0.597466385 & 0.002043411 & 0.144839032 & 4.423131929 \\ \hline
        AGA & 3.426492466 & 0.081448878 & 0.144600196 & 0.079977056 & 0.164332376 & 0.000120047 & 0.143257037 & 4.040228056 \\ \hline
        AGG & 3.409413445 & 0.084503191 & 0.12453643 & 0.086662076 & 0.157801902 & 0.000118376 & 0.128089819 & 3.991125239 \\ \hline
        AGT & 3.414577378 & 0.041742042 & 0.144560966 & 0.080628226 & 0.163894506 & 1.6233E-05 & 0.143104322 & 3.988523672 \\ \hline
        AGC & 3.403753499 & 0.04454797 & 0.124509408 & 0.087281958 & 0.157591459 & 1.59575E-05 & 0.128011019 & 3.945711271 \\ \hline
        GGT & 3.383286221 & 0.036086653 & 0.10285348 & 0.067278908 & 0.177483525 & 1.68599E-05 & 0.173354437 & 3.940360085 \\ \hline
        GGC & 3.375268028 & 0.038966723 & 0.089551705 & 0.074108929 & 0.171015104 & 1.65773E-05 & 0.157861195 & 3.906788262 \\ \hline
        GGA & 3.395053354 & 0.075540226 & 0.102847791 & 0.066616576 & 0.177936161 & 0.000123833 & 0.173539149 & 3.99165709 \\ \hline
        GGG & 3.380858003 & 0.078671119 & 0.089557006 & 0.073483978 & 0.171233801 & 0.000122117 & 0.157958172 & 3.951884196 \\ \hline
    \end{tabular}
		\caption{Theoretically calculated residence times at all the seven intermediate states and the total residence times of all 64 codons. Note that the cognate codons of the smallest amino acid Glycine, GGN, have the shortest residence times.}
		\label{residencetimestable}
\end{table}

\end{document}